\documentclass[11pt,english,onecolumn]{IEEEtran}
\usepackage[T1]{fontenc}
\usepackage[latin9]{inputenc}
\usepackage{units}
\usepackage{mathtools}
\usepackage{enumitem}
\usepackage{amsmath}
\usepackage{amsthm}
\usepackage{amssymb}
\usepackage{cancel}
\usepackage{mathdots}
\usepackage{stmaryrd}
\usepackage{stackrel}
\usepackage{graphicx}
\usepackage{setspace}
\PassOptionsToPackage{version=3}{mhchem}
\usepackage{mhchem}
\PassOptionsToPackage{normalem}{ulem}
\usepackage{ulem}
\makeatletter
\theoremstyle{plain}
\newtheorem{thm}{\protect\theoremname}
\theoremstyle{plain}
\newtheorem{prop}[thm]{\protect\propositionname}
\theoremstyle{plain}
\newtheorem{cor}[thm]{\protect\corollaryname}
\theoremstyle{definition}
\newtheorem{defn}[thm]{\protect\definitionname}
\theoremstyle{definition}
\newtheorem{example}[thm]{\protect\examplename}
\theoremstyle{plain}
\newtheorem{lem}[thm]{\protect\lemmaname}
\newlist{casenv}{enumerate}{4}
\setlist[casenv]{leftmargin=*,align=left,widest={iiii}}
\setlist[casenv,1]{label={{\itshape\ \casename} \arabic*.},ref=\arabic*}
\setlist[casenv,2]{label={{\itshape\ \casename} \roman*.},ref=\roman*}
\setlist[casenv,3]{label={{\itshape\ \casename\ \alph*.}},ref=\alph*}
\setlist[casenv,4]{label={{\itshape\ \casename} \arabic*.},ref=\arabic*}
\theoremstyle{remark}
\newtheorem{rem}[thm]{\protect\remarkname}

\usepackage{hyperref}
\usepackage{breqn}

\usepackage{calligra}
\usepackage[T1]{fontenc}

\DeclareMathOperator*{\argmin}{arg\,min}
\DeclareMathOperator{\supp}{supp} 
 
\DeclareMathOperator{\dotgeq}{\dot{\geq}}
\DeclareMathOperator{\dotleq}{\dot{\leq}}
\allowdisplaybreaks

\global\long\def\s[#1]{\mathrm{\scriptsize #1}}

\global\long\def\P{\mathbb{P}}
\global\long\def\E{\mathbb{E}}

\global\long\def\I{\mathbb{I}}

\global\long\def\binomial{\mathrm{Binomial}}

\global\long\def\dfn{\stackrel{\mathrm{def}}{=}}
\global\long\def\trre[#1,#2]{\overset{{\scriptstyle (#2)}}{#1}} 

\author{Nir Weinberger, \IEEEmembership{Member, IEEE} and Yuval Kochman, \IEEEmembership{Member, IEEE}}

\makeatother

\usepackage{babel}
\providecommand{\casename}{Case}
\providecommand{\corollaryname}{Corollary}
\providecommand{\definitionname}{Definition}
\providecommand{\examplename}{Example}
\providecommand{\lemmaname}{Lemma}
\providecommand{\propositionname}{Proposition}
\providecommand{\remarkname}{Remark}
\providecommand{\theoremname}{Theorem}

\begin{document}

\title{On the Reliability Function of Distributed Hypothesis Testing Under
Optimal Detection\thanks{The material in this paper was presented in part at the Information
Theory and Applications (ITA) Workshop, San Diego, California, USA,
February 2017, and at the IEEE International Symposium on Information
Theory (ISIT), Vail, Colorado, U.S.A., June 2018. }}

\maketitle
\renewcommand\[{\begin{equation}}
\renewcommand\]{\end{equation}}
\thispagestyle{empty}
\begin{abstract}
The distributed hypothesis testing problem with full side-information
is studied. The trade-off (reliability function) between the two types
of error exponents under limited rate is studied in the following
way. First, the problem is reduced to the problem of determining the
reliability function of channel codes designed for detection (in analogy
to a similar result which connects the reliability function of distributed
lossless compression and ordinary channel codes). Second, a single-letter
random-coding bound based on a hierarchical ensemble, as well as a
single-letter expurgated bound, are derived for the reliability of
channel-detection codes. Both bounds are derived for a system which
employs the optimal detection rule. We conjecture that the resulting
random-coding bound is ensemble-tight, and consequently optimal within
the class of quantization-and-binning schemes.
\end{abstract}


\section{Introduction\label{sec:Introduction}}

The exponential decay of error probabilities in the \emph{hypothesis
testing }(HT) problem is well-understood, with known sharp results
such as \emph{Stein's exponent} - the optimal type 2 exponent given
that the type 1 error probability is bounded away from one, and the\emph{
reliability function} - the optimal trade-off between the two types
of exponents (typically obtained via\emph{ Sanov's theorem} \cite{blahut1974hypothesis},
\cite[Ch. 1]{csiszar2011information}, \cite[Sec. 2]{csiszar2004information},\cite[Ch. 11]{Cover:2006:EIT:1146355}.
However, a similar characterization for the problem of \emph{distributed
hypothesis testing} (DHT) problem \cite{berger1979decentralized,Ahlswede_Csiszar1986}
is much more challenging. The reliability function of the DHT problem
is the topic of this paper.

We consider a model, in which the observations are memoryless realizations
of a pair of discrete random variables $(X,Y)$. We focus on the asymmetric
case (also referred to as the\emph{ side-information} \emph{case}),
for which the $X$-observations are required to be compressed at a
rate $R$, while the $Y$-observations are fully available to the
detector. For this problem, Ahlswede and Csisz{\'a}r \cite[Th. 2]{Ahlswede_Csiszar1986}
have used entropy characterization and strong converse results from
\cite{Ahlswede1976bounds,Ahlswede_Korner195source} to fully characterize
Stein's exponent in the \emph{testing against independence} case (i.e.,
when the null hypothesis states that $(X,Y)\sim P_{XY}$, whereas
the alternative hypothesis states that $(X,Y)\sim P_{X}\times P_{Y}$).
Further, they have used quantization-based encoding to derive an achievable
Stein's exponent for a general pair of memoryless hypotheses \cite[Th. 5]{Ahlswede_Csiszar1986},
but without a converse bound. Consecutive progress on this problem,
as well as on the symmetric case (in which the $Y$-observations must
also be compressed) is summarized in \cite[Sec. IV]{Han_Amari1998},
with notable contributions from \cite{Han1987,Han_kobayashi1989,Shimokawa1994}.
The the \emph{zero-rate} case was also considered, for which \cite{Han1987,Han_kobayashi1989,Shalaby1992}
and \cite[Th. 5.5]{Han_Amari1998} derived matching achievable and
converse bounds under various kind of assumptions on the distributions
induced each of the hypotheses.

In the last decade, a renewed interest in the problem arose, aimed
both at tackling more elaborate models, as well as at improving the
results on the basic model. As for the former, notable examples include
the following. Stein's exponents under positive rates were explored
in successive refinement models \cite{Tian_Chan2008}, for multiple
encoders \cite{Rahman_Wagner2012}, for interactive models \cite{katz2016collaborative,Xiang_Kim2013},
under privacy constraints \cite{Mhanna_2015}, combined with lossy
compression \cite{Katz2017}, over noisy channels \cite{Sreekumar_2017,Tuncel_error_exponents},
for multiple decision centers \cite{salehkalaibar2017hypothesis_multiple},
as well as over multi-hop networks \cite{salehkalaibar2017hypothesis_multihop}.
Exponents for the zero-rate problem were studied under restricted
detector structure \cite{polyanskiy2012} and for multiple encoders
\cite{Zhao_Lai2015}. The finite blocklength and second-order regimes
were addressed in \cite{watanabe2016neyman}. 

Notwithstanding the foregoing progress, the encoding approach proposed
in \cite{Shimokawa1994} is still the best known in general for the
basic model we study in this paper. It is based on \emph{quantization
and binning}, just as used, e.g., for \emph{distributed lossy compression}
(the\emph{ Wyner-Ziv} \emph{problem} \cite[Ch. 11]{El-Gamal_Kim2011network}
\cite{Wyner_Ziv1976}). First, the encoding rate is reduced by \emph{quantizing
}the source vector to a reproduction vector chosen from a limited-size
codebook. Second, the rate is further reduced by \emph{binning} of
the reproduction vectors. The detection is a \emph{two stage} process:
In the first stage, the detector attempts to decode the reproduction
vector with high probability using the side information. In the second
stage, the detector assumes that its reproduced source vector was
actually emitted from the distribution induced by one of the hypothesis
and the test channel of the quantization. It then uses an ordinary
hypothesis test of some kind for the reproduced-vector/side-information
pair. In \cite{Rahman_Wagner2012}, it was shown that the quantization-and-binning
scheme achieves the optimal Stein's exponent in a \emph{testing against
conditional independence }problem, in a model inspired by the Gel'fand-Pinsker
problem \cite{GelfandPinsker}, as well as in a Gaussian model. In
\cite{Katz2015}, the quantization-and-binning scheme was shown to
be necessary for the case of DHT with \emph{degraded hypotheses}.
In \cite{Haim2016}, a full achievable exponent trade-off was presented
for symmetric sources in the side-information case, and K{\"o}rner-Marton
coding \cite{Korner_Marton1979} was used in order to extend the analysis
to the symmetric-rate case. In \cite{Katz_asilomar2016}, an improved
detection rule was suggested, in which the reproduction vectors in
the bin are exhausted one by one, and the null hypothesis is declared
if a single vector is jointly typical with the side-information vector.

The two stage process used for detection (and its improvements) are
in general suboptimal for any given encoder. Intuitively, this is
because the decoding of the source vector (or the reproduction vector)
is totally superfluous for the DHT system, as the system is only required
to distinguish between the hypotheses. Consequently, unless some special
situation occurs (as, e.g., in Stein's exponent for testing against
independence \cite{Ahlswede_Csiszar1986}), there is no reason to
believe that the reliability function will be achieved for such detectors.
In this work, we investigate the performance of the \emph{optimal
}detector for any given encoder,\footnote{As an exception, in the zero-rate regime, \cite{watanabe2016neyman}
recently considered the use of an optimal Neyman-Pearson-like detector,
rather than the possibly suboptimal Hoeffeding-like detector \cite{hoeffding1965asymptotically}
that was used in \cite{Han_kobayashi1989}. } which, in fact, directly follows from the standard Neyman-Pearson
lemma (see Section \ref{sec:The-Reliability-Function}). Nonetheless,
the error exponents achieved for the optimal detector were not previously
analyzed.

To address the asymmetric DHT problem under optimal detection we apply
a methodology inspired by the analysis of distributed lossless compression
(DLC) systems (also known as the \emph{Slepian-Wolf} \emph{problem}
\cite[Ch. 10]{El-Gamal_Kim2011network} \cite{slepian1973noiseless}),
where the $X$-observations are required to be compressed at a rate
$R$, while the decoder uses the received message index and the $Y$-observations
to decode $X$. A direct analysis of the reliability function of the
DLC problem, namely, the optimal exponential decrease of the error
probability as a function of the compression rate, was made in \cite{gallager1976source,csiszar1980towards,csiszar1981graph}.
Nonetheless, an ``indirect'' analysis method was also suggested,
which is based on the intuition that the sets of $X$-vectors which
are mapped to the same message index (called \emph{bins}) should constitute
a good channel code for the memoryless channel $P_{Y|X}$. This intuition
was made precise in \cite[Th. 1]{Ahlswede_permutations}\cite{Chen2007reliability,Weinberger15},
by linking the reliability function of the DLC problem to that of
channel coding problem. With this link established, any bound on the
reliability function of channel decoding - e.g., the \emph{random-coding}
bound \cite[Th. 10.2]{csiszar2011information}, the \emph{expurgated}
bound \cite[Problem 10.18]{csiszar2011information} and the \emph{sphere-packing}
bound \cite[Th. 10.3]{csiszar2011information} - leads immediately
to a corresponding bound on the DLC reliability function. Furthermore,
any prospective result on the reliability function of the channel
coding problem may be immediately translated to the DLC problem. We
briefly mention that this link is established by constructing DLC
systems which use \emph{structured binning},\footnote{Structured binning was also proposed in \cite{Rahman_Wagner2012}
for the DHT problem, but there it was recognized as inessential.} obtained by a \emph{permutation }technique \cite{Ahlswede_permutations,ahlswede1979coloring}. 

Adapting this idea to the DHT problem, we introduce the problem of
\emph{channel detection} (CD), and show that it is serves as a reduction
of the DHT problem. Specifically, in the CD problem one has to construct
a code of a given cardinality, that would enable to distinguish between
two hypotheses on a channel distribution. It is related to problems
studied in \cite{TCW09,WCCW11,CoN,Weinberger_merhav15}, but unlike
all of these works, has no requirement to convey message (communicate)
over the channel. For the CD problem, we derive both \emph{random-coding
}bounds and \emph{expurgated} bounds on the reliability function of
CD under the optimal detector. Our analysis bears similarity to \cite{Weinberger_merhav15},
yet it goes beyond that work in two senses: First, it is based on
a Chernoff distance characterization of the optimal exponents, which
leads to simpler bounds; Second, the analysis is performed for a \emph{hierarchical
ensemble }\footnote{Yielding\emph{ superposition codes} \cite{Bergmans1973}, used for
the Wyner-Ziv problem \cite{Wyner_Ziv1976} as well as for the \emph{broadcast
}channel, see, e.g. \cite[Ch. 5]{El-Gamal_Kim2011network}.}. We note in passing that the choice of a hierarchical ensemble for
deriving the random-coding bound on the reliability function of CD
is related to the fact that the best known ensemble for bounding the
reliability function of DHT systems is based on the quantization-and-binning
method described above.

The outline of the rest of the paper is as follows. System model and
preliminaries such as notation conventions and background on ordinary
HT, will be given in Section \ref{sec:System model}. The main result
of the paper - an achievable bound on the reliability function of
DHT under optimal detection - will be stated in Section \ref{sec:The-Reliability-Function},
along with some consequences. For the sake of proving these bounds,
the reduction of the DHT reliability problem to the CD reliability
problem will be considered in Section \ref{sec:Channel-Detection-Codes}.
While only achievability bounds on the DHT reliability function will
ultimately be derived in this paper, the reduction to CD has both
an achievability part as well as a converse part. Derivation of single-letter
achievable bounds on the reliability of CD will be considered in Section
\ref{sec:Bounds CD}. Using these bounds, the achievability bounds
on the DHT reliability function will immediately follow. Afterwards,
a discussion on computational aspects along with a numerical example
will be given in Section \ref{sec:Computational-Aspects-and}. Several
directions for further research will be highlighted in Section \ref{sec:Conclusion}.

\section{System Model \label{sec:System model}}

\subsection{Notation Conventions}

Throughout the paper, random variables will be denoted by capital
letters, specific values they may take will be denoted by the corresponding
lower case letters, and their alphabets will be denoted by calligraphic
letters. Random vectors and their realizations will be superscripted
by their dimension. For example, the random vector $X^{n}=(X_{1},\ldots,X_{n})$
(where $n$ is a positive integer), may take a specific vector value
$x^{n}=(x_{1},\ldots,x_{n})\in{\cal X}^{n}$, the $n$th order Cartesian
power of ${\cal X}$, which is the alphabet of each component of this
vector. The Cartesian product of ${\cal X}$ and ${\cal Y}$ (finite
alphabets) will be denoted by ${\cal X}\times{\cal Y}$. 

We will follow the standard notation conventions for probability distributions,
e.g., $P_{X}(x)$ will denote the probability of the letter $x\in{\cal X}$
under the distribution $P_{X}$. The arguments will be omitted when
we address the entire distribution, e.g., $P_{X}$. Similarly, generic
distributions will be denoted by $Q$, $\overline{Q}$, and in other
similar forms, subscripted by the relevant random variables/vectors/conditionings,
e.g., $Q_{XY}$, $Q_{X|Y}$. The composition of a $Q_{X}$ and $Q_{Y|X}$
will be denoted by $Q_{X}\times Q_{Y|X}$. 

In what follows, we will extensively utilize the \emph{method of types
}\cite{csiszar2011information,csiszar1998method} and the following
notations. The \emph{type class} of a type $Q_{X}$ at blocklength
$n$, i.e., the set of all $x^{n}\in{\cal X}^{n}$ with empirical
distribution $Q_{X}$, will be denoted by ${\cal T}_{n}(Q_{X})$.
The set of all type classes of vectors of length $n$ from ${\cal X}^{n}$
will be denoted by ${\cal P}_{n}({\cal X})$, and the set of all possible
types over ${\cal X}$ will be denoted by ${\cal P}({\cal X})\dfn\bigcup_{n=1}^{\infty}{\cal P}_{n}({\cal X})$.
Similar notations will be used for pairs of random variables (and
larger collections), e.g., ${\cal P}_{n}({\cal U}\times{\cal X})$,
and ${\cal T}_{n}(Q_{UXY})\subseteq{\cal U}^{n}\times{\cal X}^{n}\times{\cal Y}^{n}$.
The conditional type class of $x^{n}$ for a conditional type $Q_{Y|X}$,
namely, the subset of ${\cal T}_{n}(Q_{Y})$ such that the joint type
of $(x^{n},y^{n})$ is $Q_{XY}$ (sometimes called the \emph{Q-shell}
of $x^{n}$ \cite[Definition 2.4]{csiszar2011information}), will
be denoted by ${\cal T}_{n}(Q_{Y|X},x^{n})$. For a given $Q_{X}\in{\cal P}_{n}({\cal X})$,
the set of conditional types $Q_{Y|X}$ such that ${\cal T}_{n}(Q_{Y|X},x^{n})$
is not empty when $x^{n}\in{\cal T}_{n}(Q_{X})$ will be denoted by
${\cal P}_{n}({\cal Y},Q_{X})$. The probability simplex for an alphabet
${\cal X}$ will be denoted by ${\cal S}({\cal X})$.

The probability of the event ${\cal A}$ will be denoted by $\P({\cal A})$,
and its indicator function will be denoted by $\I({\cal A})$. The
expectation operator with respect to a given distribution $Q$ will
be denoted by $\E_{Q}[\cdot]$ where the subscript $Q$ will be omitted
if the underlying probability distribution is clear from the context.
The variational distance (${\cal L}_{1}$ norm) of $P_{X},Q_{X}\in{\cal S}({\cal X})$
will be denoted by$\|P_{X}-Q_{X}\|\dfn\sum_{x\in{\cal X}}|P_{X}(x)-Q_{X}(x)|.$
In general, information-theoretic quantities will be denoted by the
standard notation \cite{Cover:2006:EIT:1146355}, with subscript indicating
the distribution of the relevant random variables, e.g. $H_{Q}(X|Y),I_{Q}(X;Y),I_{Q}(X;Y|U)$,
under $Q=Q_{UXY}$. As an exception, the entropy of $X$ under $Q$
will be denoted by $H(Q_{X})$. The binary entropy function will be
denoted by $h_{\s[b]}(q)$ for $0\leq q\leq1$. The Kullback\textendash Leibler
divergence between $Q_{X}$ and $P_{X}$ will be denoted by $D(Q_{X}||P_{X})$,
and the conditional Kullback\textendash Leibler divergence between
$Q_{X|U}$ and $P_{X|U}$ averaged over $Q_{U}$ will be denoted by
$D(Q_{X|U}||P_{X|U}|Q_{U})$. \textbf{}

The Hamming distance between $x^{n},\overline{x}^{n}\in{\cal X}^{n}$
will be denoted by $d_{\s[H]}(x^{n},\overline{x}^{n})$. The complement
of a multiset ${\cal A}$ will be denoted by ${\cal A}^{c}$. The
number of \emph{distinct} elements of a finite multiset ${\cal A}$
will be denoted by $|{\cal A}|$. In optimization problem over the
simplex, the explicit display of the simplex constraint will be omitted,
i.e., $\min_{Q}f(Q)$ will be used instead of $\min_{Q\in{\cal S}({\cal X})}f(Q)$. 

For two positive sequences $\{a_{n}\}$ and $\{b_{n}\}$, the notation
$a_{n}\doteq b_{n}$, will mean asymptotic equivalence in the exponential
scale, that is, $\lim_{n\to\infty}\frac{1}{n}\log(\frac{a_{n}}{b_{n}})=0$.
Similarly, $a_{n}\dotleq b_{n}$ will mean $\limsup_{n\to\infty}\frac{1}{n}\log(\frac{a_{n}}{b_{n}})\leq0$,
and so on. The ceiling function will be denoted by $\lceil\cdot\rceil$.
The notation $|t|_{+}$ will stand for $\max\{t,0\}$. Logarithms
and exponents will be understood to be taken to the natural base.
Throughout, for the sake of brevity, we will ignore integer constraints
on large numbers. For example, $\lceil e^{nR}\rceil$ will be written
as $e^{nR}$. The set $\{1,\ldots,n\}$ for $n\in\mathbb{N}$ will
be denoted by $[n]$. 

\subsection{Ordinary Hypothesis Testing\label{subsec:Ordinary-Hypothesis-Testing}}

Before getting into the distributed scenario, we shortly review the
ordinary binary HT problem. Consider a random variable $Z\in{\cal Z}$,
whose distribution under the hypothesis $H$ (respectively, $\overline{H}$)
is $P$ (respectively, $\overline{P}$). It is common in the literature
to refer to $H$ (respectively, $\overline{H}$) as the \emph{null
hypothesis} (respectively, the \emph{alternative hypothesis}). However,
we will refrain from using such terminology, and the two hypotheses
will be considered to have an equal stature. 

Given $n$ independent and identically distributed (i.i.d.) observations
$Z^{n}$, a (possibly \emph{randomized}) \emph{detector} 
\[
\phi:{\cal Z}^{n}\to{\cal S}\left\{ H,\overline{H}\right\} ,
\]
has \emph{type 1 and type 2 error probabilities}\footnote{Also called the \emph{false-alarm probability }and \emph{misdetection}
\emph{probability} in engineering applications.}\emph{ }given by
\begin{equation}
p_{1}(\phi)\dfn P\left[\phi(Z^{n})=\overline{H}\right],\label{eq: FA ordinary def}
\end{equation}
and
\begin{equation}
p_{2}(\phi)\dfn\overline{P}\left[\phi(Z^{n})=H\right].\label{eq: MD ordinary def}
\end{equation}
For brevity, the probability of an event ${\cal A}$ under $H$ (respectively,
$\overline{H}$) is denoted by $P({\cal A})$ {[}respectively, $\overline{P}({\cal A})${]}. 

The \emph{Neyman-Pearson lemma} \cite[Prop. II.D.1]{Poor}, \cite[Th. 11.7.1]{Cover:2006:EIT:1146355}
states that the family of detectors $\{\phi_{n,T,\eta}^{*}\}_{T\in\mathbb{R},\;\eta\in[0,1]}$
which optimally trades between the two types of error probabilities
is given by
\begin{equation}
\P\left[\phi_{n,T,\eta}^{*}(z^{n})=H\right]\dfn\begin{cases}
1, & P(z^{n})>e^{nT}\cdot\overline{P}(z^{n})\\
0, & P(z^{n})<e^{nT}\cdot\overline{P}(z^{n})\\
\text{\ensuremath{\eta}}, & \text{otherwise}
\end{cases},\label{eq: ordinary HT detector one sample}
\end{equation}
where $T\in\mathbb{R}$ is a \emph{threshold} parameter. The parameter
$T$ controls the trade-off between the two types of error probabilities
- if $T$ is increased then the type 1 error probability also increases,
while the type 2 error probability decreases (and vice versa). The
parameters $T$ and $\eta$ may be tuned to obtain any desired type
1 error probability constraint, while providing the optimal type 2
error probability. 

To describe bounds on the error probabilities of the optimal detector,
let us define the \emph{hypothesis-testing reliability function} \cite[Section II]{blahut1974hypothesis}
as 
\begin{equation}
D_{2}(D_{1};P,\overline{P})\dfn\min_{Q:\;D(Q||P)\leq D_{1}}D(Q||\overline{P}).\label{eq: ordinary hypothesis testing reliability function}
\end{equation}
For brevity, we shall omit the dependence on $P,\overline{P}$ as
they remain fixed and can be understood from context. As is well known
\cite[Th. 3]{blahut1974hypothesis}, for a given $D_{1}\in(0,D(\overline{P}||P))$,
there exists a $T$ such that\footnote{These bounds were only proved in \cite{blahut1974hypothesis} for
a deterministic Neyman-Pearson detector, i.e., $\phi_{n,T,\eta}^{*}$
with $\eta\in\{0,1,\}$. Nonetheless, they also hold verbatim when
$\eta\in(0,1)$.}
\begin{equation}
p_{1}(\phi_{n,T,\eta}^{*})\leq\exp(-n\cdot D_{1}),\label{eq: FA exponent ordinary vector}
\end{equation}
\begin{equation}
p_{2}(\phi_{n,T,\eta}^{*})\leq\exp\left[-n\cdot D_{2}(D_{1})\right].\label{eq: MD exponent ordinary vector}
\end{equation}
Furthermore, it is also known that this exponential behavior is optimal
\cite[Corollary 2]{blahut1974hypothesis}, in the sense that if 
\[
\liminf_{n\to\infty}-\frac{1}{n}\log p_{1}(\phi_{n,T,\eta}^{*})\geq D_{1}
\]
then 
\[
\limsup_{n\to\infty}-\frac{1}{n}\log p_{2}(\phi_{n,T,\eta}^{*})\leq D_{2}(D_{1}).
\]
It should be noted, however, that the detector \eqref{eq: ordinary HT detector one sample}
is optimal and the bounds on its error probability \eqref{eq: FA exponent ordinary vector}-\eqref{eq: MD exponent ordinary vector}
hold for any given $n$. In fact, in what follows, we will use this
detector and bounds when $n=1$. Furthermore, since these bounds do
not depend on $\eta$, we shall henceforth assume an arbitrary value,
and omit the dependence on $\eta$. 

The function $D_{2}(D_{1})$ is known to be a convex function of $D_{1}$,
continuous on $(0,\infty)$ and strictly decreasing up to a critical
point for which it remains constant above it \cite[Th. 3]{blahut1974hypothesis}.
Furthermore, it is known \cite[Th. 7]{blahut1974hypothesis} that
up to the critical point, it can be represented as 
\begin{equation}
D_{2}(D_{1})=\sup_{\tau\geq0}\left\{ -\tau\cdot D_{1}+(\tau+1)\cdot d_{\tau}\right\} ,\label{eq: ordinary hypothesis testing Chernoff information}
\end{equation}
where 
\begin{equation}
d_{\tau}\dfn-\log\left[\sum_{z\in{\cal Z}}P^{\nicefrac{\tau}{\tau+1}}(z)\overline{P}^{\nicefrac{1}{\tau+1}}(z)\right],\label{eq: Chernoff distance HT}
\end{equation}
is the \emph{Chernoff distance} \emph{between distributions.} The
representation \eqref{eq: ordinary hypothesis testing Chernoff information}
will be used in the sequel to derive bounds on the reliability of
DHT systems. We also note in passing that \emph{Stein's exponent }is
defined as the largest type 2 error exponent that can be achieved
under the constraint $p_{1}(\phi_{n.T}^{*})\leq\epsilon$ for $\epsilon>0$.
It turns out \cite[Th. 2.2]{csiszar2004information} that this exponent
is independent of $\epsilon$, and given by $D(P||\overline{P})$
(which agrees with $\lim_{D_{1}\downarrow0}D_{2}(D_{1})$).

\subsection{Distributed Hypothesis Testing\label{sec:Distributed-Hypothesis-Testing}}

Let $\{(X_{i},Y_{i})\}_{i=1}^{n}$ be i.i.d. realizations of a pair
of random variables $(X,Y)\in({\cal X},{\cal Y})$, where $|{\cal X}|,|{\cal Y}|<\infty$,
where under $H$, the joint distribution of $(X,Y)$ is given by $P_{XY}$,
whereas under $\overline{H}$, this distribution is given by $\overline{P}_{XY}$.
To avoid trivial cases of an infinite exponent at zero rate, we will
assume throughout that $\supp(P_{X})\cap\supp(\overline{P}_{X})\neq\phi$
and $\supp(P_{Y})\cap\supp(\overline{P}_{Y})\neq\phi$.

\begin{figure}
\begin{centering}
\includegraphics[scale=1.2]{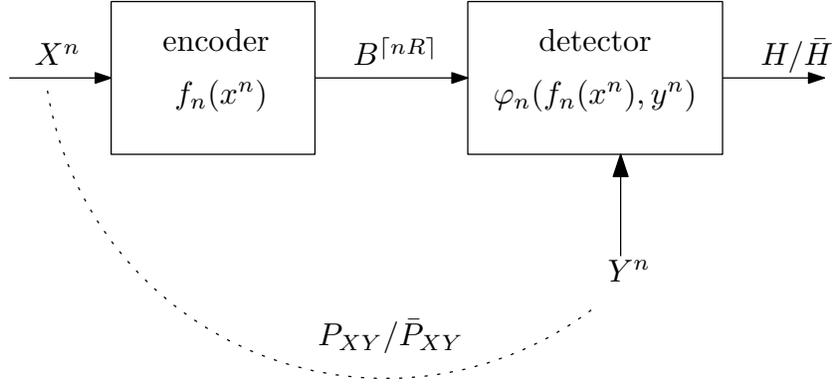}
\par\end{centering}
\caption{A DHT system. \label{fig: DHT system}}
\end{figure}

A DHT system ${\cal H}_{n}\dfn(f_{n},\varphi_{n})$, as depicted in
Fig. \ref{fig: DHT system}, is defined by an \emph{encoder} 
\[
f_{n}:{\cal X}^{n}\to[m_{n}],
\]
which maps a source vector into an index $i=f_{n}(x^{n})$, and a
\emph{detector }(possibly \emph{randomized}\footnote{Randomized encoding can also be defined. In this case, the encoder
takes the form $f_{n}:{\cal X}^{n}\to{\cal S}([m_{n}])$, where $f_{n}(x^{n})$
is a probability vector whose $i$th entry is the probability of mapping
$x^{n}$ to the index $i\in[m_{n}]$. In the sequel, we will also
use a rather simple form of randomized encoding, which does not require
this general definition. There, the source vector $x^{n}$ will be
used to randomly generate a new source vector $\tilde{X}^{n}$, and
the latter will be encoded by a deterministic encoder (see the proof
of the achievability part of Theorem \ref{thm: Redcution to CD codes}
in Appendix \ref{subsec: Achievability reduction}).})
\[
\varphi_{n}:[m_{n}]\times{\cal Y}^{n}\to{\cal S}\{H,\overline{H}\}.
\]
 The inverse image of $f_{n}$ for $i\in[m_{n}]$, i.e., 
\begin{equation}
f_{n}^{-1}(i)\dfn\left\{ x^{n}\in{\cal X}^{n}:f_{n}(x^{n})=i\right\} ,\label{eq: bin definition}
\end{equation}
is called the \emph{bin} associated with index $i$.\footnote{Traditionally, the term ``binning\textquotedblright{} refers to mapping
multiple \textquotedblleft distant\textquotedblright{} sequences to
a single index. For example, in quantization-and-binning schemes,
this term refers to sets of quantized source vectors. However, we
use it in a more general sense, referring to all source sequences
mapped to a single index. Thus in these terms, the whole \textquotedblleft quantization-and-binning\textquotedblright{}
process merely produces bins.} The \emph{rate} of ${\cal H}_{n}$ is defined as $\frac{1}{n}\log m_{n}$.
The \emph{type 1 error probability }of\emph{ ${\cal H}_{n}$} is defined
as
\[
p_{1}({\cal H}_{n})\dfn P\left[\varphi_{n}(f_{n}(X^{n}),Y^{n})=\overline{H}\right],
\]
and the\emph{ type 2 error probability }is defined as
\[
p_{2}({\cal H}_{n})\dfn\overline{P}\left[\varphi_{n}(f_{n}(X^{n}),Y^{n})=H\right].
\]
In the sequel,\footnote{Mainly in Appendix B.} conditional error
probabilities given an event ${\cal A}$ will be abbreviated as, e.g.,
\[
p_{1}({\cal H}_{n}|{\cal A})\dfn P\left[\varphi_{n}(f_{n}(X^{n}),Y^{n})=\overline{H}\vert{\cal A}\right].
\]
A sequence of DHT systems will be denoted by ${\cal H}\dfn\{{\cal H}_{n}\}_{n\geq1}$.
A sequence ${\cal H}$ is associated with two different error exponents
for each of the two error probabilities defined above. The \emph{infimum
type 1 exponent} of ${\cal H}$ is defined by 
\begin{equation}
\liminf_{n\to\infty}-\frac{1}{n}\log p_{1}({\cal H}_{n}),\label{eq: inf FA exponent definition given system}
\end{equation}
and the \emph{supremum type 1 exponent} is defined by 

\begin{equation}
\limsup_{n\to\infty}-\frac{1}{n}\log p_{1}({\cal H}_{n}).\label{eq: sup FA exponent definition given system}
\end{equation}
Analogous exponents can be defined for the type 2 error probability.

The \emph{reliability function }of a DHT system is the optimal trade-off
between the two types of exponents achieved by any encoder-detector
pair under a rate $R$. Specifically, the \emph{infimum DHT reliability
function} is defined by 
\[
E_{2}^{-}(R,E_{1};P_{XY},\overline{P}_{XY})\dfn\sup_{{\cal H}}\left\{ \liminf_{n\to\infty}-\frac{1}{n}\log p_{2}({\cal H}_{n}):\forall n,\;m_{n}\leq e^{nR},\;p_{1}({\cal H}_{n})\leq e^{-n\cdot E_{1}}\right\} ,
\]
and the \emph{supremum DHT reliability function} $E_{2}^{+}(R,E_{1};P_{XY},\overline{P}_{XY})$
is analogously defined, albeit with a $\limsup$. For brevity, the
dependence on $P_{Y|X},\overline{P}_{Y|X}$ will be omitted henceforth
whenever it is understood from context. While the focus of this paper
is the reliability function, one may also define \emph{Stein's exponent}
for some $\epsilon>0$ as
\[
\sup_{{\cal H}}\left\{ \liminf_{n\to\infty}-\frac{1}{n}\log p_{2}({\cal H}_{n}):\forall n,\;m_{n}\leq e^{nR},\;p_{1}({\cal H}_{n})\leq\epsilon\right\} .
\]
Unlike in ordinary HT, it is not assured that Stein's exponent is
independent of $\epsilon$. However, one can obtain an achievable
bound on Stein's exponent by taking the limit $E_{1}\downarrow0$
of an achievable bound on $E_{2}^{-}(R,E_{1})$. 

\section{Main Result: Bounds on The Reliability Function of DHT\label{sec:The-Reliability-Function}}

Our main result (Theorem \ref{thm: DHT RC and EX}) is an achievable
bound on the reliability function of DHT. Before that, we state the
trivial converse bound, obtained when $X^{n}$ is not compressed,
or alternatively, when $R=\log|{\cal X}|$ (immediately deduced from
the discussion in Section \ref{subsec:Ordinary-Hypothesis-Testing}).
\begin{prop}
\label{prop: DHT converse}The\emph{ }supremum DHT reliability function
is bounded as
\[
E_{2}^{+}(R,E_{1})\leq\min_{Q_{XY}:D(Q_{XY}||P_{XY})\leq E_{1}}D(Q_{XY}||\overline{P}_{XY}).
\]
\end{prop}
To state our achievability bound, we will need several additional
notations. We denote the \emph{Chernoff parameter} \emph{for a pair
of symbols }$(x,\tilde{x})$ by
\begin{equation}
d_{\tau}(x,\tilde{x})\dfn-\log\sum_{y\in{\cal Y}}P_{Y|X}^{\nicefrac{\tau}{\tau+1}}(y|x)\overline{P}_{Y|X}^{\nicefrac{1}{\tau+1}}(y|\tilde{x}),\label{eq: Chernoff distance}
\end{equation}
and \emph{for a pair of vectors }$(x^{n},\tilde{x}^{n})$ by
\begin{equation}
d_{\tau}(x^{n},\tilde{x}^{n})\dfn\frac{1}{n}\sum_{i=1}^{n}d_{\tau}(x_{i},\tilde{x}_{i}).\label{eq: Chernoff distance vectors}
\end{equation}
Further, when $(X,\tilde{X})$ are distributed according to $Q_{X\tilde{X}}$
we define the \emph{average} \emph{Chernoff parameter} as
\begin{equation}
d_{\tau}(Q_{X\tilde{X}})\dfn\E_{Q}\left[d_{\tau}(X,\tilde{X})\right],\label{eq: Chernoof distance type}
\end{equation}
and when $X$ is distributed according to $Q_{X}$, we denote, for
brevity,
\begin{equation}
d_{\tau}(Q_{X})\dfn\E_{Q}\left[d_{\tau}(X,X)\right].\label{eq: Chernoff distance type - one RV}
\end{equation}
Next, we denote the random-coding exponent\textbf{
\begin{equation}
B_{\s[rc]}(R,R_{\s[b]},Q_{UX},\tau)\dfn\min\left\{ B_{\s[rc]}'(R,R_{\s[b]},Q_{UX},\tau),\;B_{\s[rc]}''(R,R_{\s[b]},Q_{UX},\tau)\right\} ,\label{eq: B_rc}
\end{equation}
}where\textbf{
\begin{align}
 & B_{\s[rc]}'(R,R_{\s[b]},Q_{UX},\tau)\nonumber \\
 & \dfn\min_{(Q_{UXY},\overline{Q}_{UXY}):Q_{UX}=\overline{Q}_{UX},\;Q_{Y}=\overline{Q}_{Y}}\Bigg\{\tau\cdot D(Q_{Y|UX}||P_{Y|X}|Q_{UX})+D(\overline{Q}_{Y|UX}||\overline{P}_{Y|X}|\overline{Q}_{UX})\nonumber \\
 & \hphantom{=}+\max\left\{ \left|I_{Q}(U;Y)-R_{\s[b]}\right|_{+},\;I_{Q}(U,X;Y)-H(Q_{X})+R\right\} \nonumber \\
 & \hphantom{=}+\tau\cdot\max\left\{ \left|I_{\overline{Q}}(U;Y)-R_{\s[b]}\right|_{+},\;I_{\overline{Q}}(U,X;Y)-H(\overline{Q}_{X})+R\right\} \Bigg\},\label{eq: B_rc'}
\end{align}
}and\textbf{
\begin{align}
 & B_{\s[rc]}''(R,R_{\s[b]},Q_{UX},\tau)\nonumber \\
 & \dfn\min_{(Q_{UXY},\overline{Q}_{UXY}):Q_{UX}=\overline{Q}_{UX},\;Q_{UY}=\overline{Q}_{UY},\;I_{Q}(U;Y)>R_{\s[b]}}\Bigg\{\tau\cdot D(Q_{Y|UX}||P_{Y|X}|Q_{UX})+D(\overline{Q}_{Y|UX}||\overline{P}_{Y|X}|\overline{Q}_{UX})\nonumber \\
 & \hphantom{=}+\left|I_{Q}(X;Y|U)-H(Q_{X})+R+R_{\s[b]}\right|_{+}\nonumber \\
 & \hphantom{=}+\tau\cdot\left|I_{\overline{Q}}(X;Y|U)-H(\overline{Q}_{X})+R+R_{\s[b]}\right|_{+}\Bigg\},\label{eq: B_rc''}
\end{align}
}as well as\textbf{ }the expurgated exponent 
\begin{equation}
B_{\s[ex]}(R,Q_{X},\tau)\dfn(\tau+1)\cdot\min_{Q_{X\tilde{X}}:\;Q_{X}=Q_{\tilde{X}},\;H_{Q}(X|\tilde{X})\geq R}\left\{ d_{\tau}(Q_{X\tilde{X}})+R-H_{Q}(X|\tilde{X})\right\} .\label{eq: B_ex}
\end{equation}
Finally, we denote
\begin{equation}
B(R,Q_{X},\tau)\dfn\max\left\{ \sup_{Q_{U|X}}\sup_{R_{\s[b]}:\:R_{\s[b]}\geq\left|I_{Q}(U;X)-R\right|_{+}}B_{\s[rc]}(R,R_{\s[b]},Q_{UX},\tau),\;B_{\s[ex]}(R,Q_{X},\tau)\right\} .\label{eq: B}
\end{equation}
For brevity, arguments such as $(R,R_{\s[b]},Q_{UX},\tau)$ will sometimes
be omitted henceforth. 
\begin{thm}
\label{thm: DHT RC and EX}The\emph{ }infimum DHT reliability function
bounded as
\begin{multline}
E_{2}^{-}(R,E_{1};P_{XY},\overline{P}_{XY})\\
\geq\min_{Q_{X}}\sup_{\tau\geq0}\Bigg[-\tau\cdot E_{1}+D(Q_{X}||\overline{P}_{X})+\tau\cdot D(Q_{X}||P_{X})+\min\Big\{(\tau+1)\cdot d_{\tau}(Q_{X}),\;B(R,Q_{X},\tau)\Big\}\Bigg].\label{eq: DHT achievability bound}
\end{multline}
\end{thm}
The rest of the paper is mainly devoted to the proof of Theorem \ref{thm: DHT RC and EX},
which is based on two main steps. In the first step (Section \ref{sec:Channel-Detection-Codes})
we will reduce the DHT problem to an auxiliary problem of CD. In the
second step (Section \ref{sec:Bounds CD}), we will derive single-letter
achievable bounds for the CD problem. The bound of Theorem \ref{thm: DHT RC and EX}
on the DHT reliability function then follow as easy corollary to these
results, and its proof appears at the end of Section \ref{sec:Bounds CD}. 

Before stating a few implications of Theorem \ref{thm: DHT RC and EX}
and delving into its proof, we would like to describe several features
of the bound \eqref{eq: DHT achievability bound}. In general, any
bound relies on the choice of the encoder (or the random ensemble
from which it is drawn), the detector, and the analysis method of
the error probabilities. The bound of Theorem \ref{thm: DHT RC and EX}
is based on the following choices: 
\begin{itemize}
\item \emph{\uline{Encoder ensemble:}} The achieving ensemble for the
random-coding bound is based on quantization-and-binning. For any
$Q_{X}$ (with $H(Q_{X})>R$), the conditional type $Q_{U|X}$ is
the \emph{test channel} for quantizing $|{\cal T}_{n}(Q_{X})|\doteq e^{nH(Q_{X})}$
source vectors into one of $e^{nR_{\s[q]}}$ possible reproduction
vectors, where the \emph{quantization rate} \emph{$R_{\s[q]}$} satisfies
$R_{\s[q]}>R$. These reproduction vectors are grouped to bins of
size (at most) $e^{nR_{\s[b]}}$ each, such that the \emph{binning
rate }$R_{\s[b]}$ satisfies $R_{\s[b]}=R_{\s[q]}-R$. Both $Q_{U|X}$
and $R_{\s[b]}$ may be separately optimized for any given $Q_{X}$
to obtain the best type 2 exponent. The achievable ensemble for the
expurgated bound is based on binning, without quantization.
\item \emph{\uline{Detector:}} The bound is derived under the optimal
detector $\varphi_{n,T,\eta}^{*}(i,y^{n})$, which, following \eqref{eq: ordinary HT detector one sample},
is given by 
\begin{equation}
\P\left[\varphi_{n,T,\eta}^{*}(i,y^{n})=H\right]\dfn\begin{cases}
1, & \sum_{x^{n}:\;f_{n}(x^{n})=i}P_{XY}(x^{n},y^{n})>e^{nT}\cdot\sum_{x^{n}:\;f_{n}(x^{n})=i}\overline{P}_{XY}(x^{n},y^{n})\\
0, & \sum_{x^{n}:\;f_{n}(x^{n})=i}P_{XY}(x^{n},y^{n})<e^{nT}\cdot\sum_{x^{n}:\;f_{n}(x^{n})=i}\overline{P}_{XY}(x^{n},y^{n})\\
\eta, & \text{otherwise}
\end{cases}\label{eq: optimal NP detector for DHT}
\end{equation}
for some $T\in\mathbb{R}$ and $\eta\in[0,1]$. 
\item \emph{\uline{Analysis method:}} As apparent from \eqref{eq: B},
for any given input type $Q_{X}$, the best of a random-coding bound
{[}as defined in \eqref{eq: B_rc}{]} and an expurgated bound {[}as
defined in \eqref{eq: B_ex}{]} can be chosen. The bounds on the error
probabilities are derived using a Chernoff type bound, and the random
coding analysis, in particular, is based on analyzing the Chernoff
parameter using the \emph{type-enumeration method} \cite[Sec. 6.3]{Merhav09}.
This method avoids any use of bounds such as Jensen's inequality,
and leads to ensemble-tight random coding exponents in many scenarios.
We conjecture that our random coding bounds are ensemble-tight, and
thus cannot be improved. 
\end{itemize}
Besides the detector which clearly cannot be improved, to the best
of our knowledge, both the analysis method and the encoder ensemble
are the tightest known for providing exponential bounds. It should
be mentioned though, that these features are only implicit in the
proof, since following the reduction from DHT to CD, we will only
address the CD problem. 

We further discuss several implications of Theorem \ref{thm: DHT RC and EX}.
First, simpler bounds, perhaps at the cost of worse exponents, can
be obtained by considering two extermal choices. To obtain a binning-based
scheme, without quantization, we choose $U$ to be a degenerated random
variable (deterministic, i.e., $|{\cal U}|=1$) and $R_{\s[b]}=H(Q_{X})-R$.
We then get that $B_{\s[rc]}'$ dominates the minimization in \eqref{eq: B_rc},
and
\begin{align}
B_{\s[rc]}(R,H(Q_{X})-R,Q_{UX},\tau) & =B_{\s[rc,b]}(R,Q_{X},\tau)\\
 & \dfn\min_{(Q_{XY},\overline{Q}_{XY}):Q_{X}=\overline{Q}_{X},\;Q_{Y}=\overline{Q}_{Y}}\Bigg\{\tau\cdot D(Q_{Y|X}||P_{Y|X}|Q_{X})+D(\overline{Q}_{Y|X}||\overline{P}_{Y|X}|\overline{Q}_{X})\nonumber \\
 & \hphantom{=}+\left|R-H_{Q}(X|Y)\right|_{+}+\tau\cdot\left|R-H_{\overline{Q}}(X|Y)\right|_{+}\Bigg\}.\label{eq: B_rc pure binning}
\end{align}
To obtain a quantization-based scheme, without binning, we choose
$R_{\s[b]}=0$, and limit $Q_{U|X}$ to satisfy $R\geq I_{Q}(U;X)$. 

Second, if the rate is large enough then no loss is expected in the
reliability function of DHT compared to the ordinary-HT bound of Proposition
\ref{prop: DHT converse}. We can deduce from Theorem \ref{thm: DHT RC and EX}
an upper bound on the minimal rate required, as follows.
\begin{cor}
\label{corr: no-loss rate}Suppose that $R$ is sufficiently large
such that 
\begin{equation}
B(R,Q_{X},\tau)\geq d_{\tau}(Q_{X})\label{eq: no-loss rate condition}
\end{equation}
 for all $Q_{X}\in{\cal S}({\cal X})$ and $\tau\geq0$. Then, 
\[
E_{2}^{-}(R,E_{1})=E_{2}^{-}(\infty,E_{1})=D_{2}(E_{1}),
\]
\end{cor}
where $D_{2}(\cdot)$ is the ordinary HT reliability function \eqref{eq: ordinary hypothesis testing reliability function}.
The proof of this corollary appears in Appendix \ref{sec:Proofs-of-Corollaries}. 

Third, by setting $E_{1}=0$, Theorem \ref{thm: DHT RC and EX} yields
an achievable bound on Stein's exponent, as follows. 
\begin{cor}
\label{corr: Stein's exponent}Stein's exponent is lower bounded by
$E_{2}^{-}(R,0)$, which satisfies
\begin{align}
 & E_{2}^{-}(R,0)\nonumber \\
 & \geq D(P_{X}||\overline{P}_{X})+\sup_{\tau\geq0}\min\left\{ (\tau+1)\cdot d_{\tau}(P_{X}),\;B(R,P_{X},\tau)\right\} \label{eq: Stein's exponent corollary}\\
 & \geq\min\left\{ D(P_{XY}||P_{X}\times\overline{P}_{Y|X}),\;D(P_{X}||\overline{P}_{X})+\sup_{Q_{U|X}}\sup_{R_{\s[b]}:\:R_{\s[b]}\geq\left|I_{P_{X}\times Q_{U|X}}(U;X)-R\right|_{+}}\lim_{\tau\to\infty}B_{\s[rc]}(R,R_{\s[b]},P_{X}\times Q_{U|X},\tau)\right\} .\label{eq: Stein's exponent corollary only RC}
\end{align}
\end{cor}
The first term in \eqref{eq: Stein's exponent corollary only RC}
can be identified as Stein's exponent when the rate is not constrained
at all. The proof of this corollary also appears in Appendix \ref{sec:Proofs-of-Corollaries}.
It is worth to note, however, that the resulting bound is quite different
from the bound of \cite[Th. 4.3]{Han_Amari1998}, \cite{Shimokawa1994}
(and its refinement in \cite{Haim2016}). Nonetheless, our bound is
presumably tighter simply because it was derived for the optimal Neyman-Pearson
detector, using the type-enumeration method. 

Fourth, it is interesting to examine the case $R=0$. Using analysis
similar to the proof of Corollary \ref{corr: Stein's exponent}, it
is easy to verify that using a binning-based scheme {[}i.e., substituting
\eqref{eq: B_rc pure binning} in \eqref{eq: DHT achievability bound}
for $B(R,Q_{X},\tau)${]} achieves the lower bound
\[
E_{2}^{-}(R=0,E_{1})\geq\min_{(Q_{XY},\overline{Q}_{XY}):Q_{X}=\overline{Q}_{X},\;Q_{Y}=\overline{Q}_{Y},\;D(Q_{XY}||P_{XY})\leq E_{1}}D(\overline{Q}_{XY}||\overline{P}_{XY}).
\]
As expected, this is the same type 2 error exponent obtained when
$y^{n}$ is not fully available to the detector and also must be encoded
at zero rate, as obtained in \cite[Th. 5.4]{Han_Amari1998}, \cite[Th. 6]{Han1987}.
For this bound, a matching converse is known \cite[Th. 5.5]{Han_Amari1998}.
When $E_{1}=0$ then $Q_{XY}=P_{XY}$, and then Stein's exponent is
given by 
\[
E_{2}^{-}(R=0,E_{1}=0)\geq\min_{\overline{Q}_{XY}:\overline{Q}_{X}=P_{X},\;\overline{Q}_{Y}=P_{Y}}D(\overline{Q}_{XY}||\overline{P}_{XY}).
\]
In \cite[Th. 2]{Shalaby1992} it was determined that this exponent
is optimal (even when $y^{n}$ is not encoded and given as side information
to the detector).

\section{A Reduction of Distributed Hypothesis Testing to Channel-Detection
Codes\label{sec:Channel-Detection-Codes}}

In this section, we formulate the CD problem which is relevant to
the characterization of the DHT reliability function. To motivate
their definition, let us assume that the detector knows the type of
$x^{n}$ (notice that sending this information requires zero rate),
or equivalently, that each DHT bin only contains source vectors of
the same type class. Then, conditioned on the message index $f_{n}(X^{n})=i$,
$X^{n}$ is distributed uniformly over $f_{n}^{-1}(y^{n})\dfn{\cal C}_{n,i}\subseteq{\cal T}_{n}(Q_{X})$,
and consequently, $Y^{n}$ is distributed according to the \emph{induced
distribution}
\begin{equation}
P_{Y^{n}}^{({\cal C}_{n,i})}(y^{n})\dfn\frac{1}{|{\cal C}_{n,i}|}\sum_{x^{n}\in{\cal C}_{n,i}}P_{Y|X}(y^{n}|x^{n}).\label{eq: induced distribution}
\end{equation}
under $H$, and according to $\overline{P}_{Y^{n}}^{({\cal C}_{n,i})}(y^{n})$
(defined similarly with $\overline{P}$ replacing $P$) under $\overline{H}$.
The detector thus may assume the following model. First, $X^{n}$
is chosen randomly and uniformly over ${\cal C}_{n,i}$. Second, the
chosen codeword $X^{n}$ is transmitted either over a channel $P_{Y|X}$
or a channel $\overline{P}_{Y|X}$. The detector should decide on
the hypothesis given the output of this channel. Following this observation,
we will henceforth refer to ${\cal C}_{n,i}$ as a CD code for the
channels $P_{Y|X}$ and $\overline{P}_{Y|X}$. 

Now, if there exists a set of CD codes ${\cal C}_{n,i}\subseteq{\cal T}_{n}(Q_{X})$
such that $\cup_{i=1}^{e^{nR}}{\cal C}_{n,i}={\cal T}_{n}(Q_{X})$,
and each ${\cal C}_{n,i}$ has low error probabilities in the CD problem
described above, then a DHT system can be constructed for $x^{n}\in{\cal T}_{n}(Q_{X})$
by setting $f_{n}(x^{n})=i$ if $x^{n}\in{\cal C}_{n,i}$. Thus, trivially,
a ``good'' DHT system is a ``good'' \emph{set} of CD codes and
vice versa. The main idea of the reduction in this section is to show
that a \emph{single} ``good'' CD code suffice, say ${\cal C}_{n,1}$.
All other CD codes $\{{\cal C}_{n,i}\}_{i=2}^{e^{nR}}$ may be generated
from ${\cal C}_{n,1}$ in a structured way, based on a permutation
idea \cite{ahlswede1979coloring,Ahlswede_permutations} which we will
shortly describe after stating the theorem.

It should be noted, however, that unlike \cite{TCW09,WCCW11,CoN,Weinberger_merhav15},
${\cal C}_{n}$ should be designed solely for attaining low error
probabilities in the detection problem between $P_{Y^{n}}^{({\cal C}_{n})}(y^{n})$
and $\overline{P}_{Y^{n}}^{({\cal C}_{n})}(y^{n})$, without any communication
goal. In this case, if the codewords of ${\cal C}_{n}$ are allowed
to be identical, then that indeed would be the optimal choice. However,
since ${\cal C}_{n}$ is to be used as a bin $f_{n}^{-1}(y^{n})$
of a DHT system, its codewords are\emph{ }unique, by definition. With
this in mind, we next define CD codes, which are required to have
a prescribed number of \emph{unique} codewords. The required definitions
are quite similar to the ones required for DHT systems, but as some
differences do exist, we explicitly outline them in what follows. 

A CD code for a type class $Q_{X}\in{\cal P}_{n}({\cal X})$ is given
by ${\cal C}_{n}\subseteq{\cal T}_{n}(Q_{X})$. An input $X^{n}\in{\cal C}_{n}$
to the channel is chosen with a uniform distribution over ${\cal C}_{n}$,
and sent over $n$ uses of a DMC which may be either $P_{Y|X}$ when
$H$ is active or $\overline{P}_{Y|X}$ when $\overline{H}$ is. The
random channel output is given by $Y^{n}\in{\cal Y}^{n}$. The detector
has to decide based on $y^{n}$ whether the DMC conditional probability
distribution is $P_{Y|X}$ or $\overline{P}_{Y|X}$. A detector (possibly
randomized) for ${\cal C}_{n}$ is given by
\[
\phi_{n}:{\cal Y}^{n}\to{\cal S}\{H,\overline{H}\}.
\]
In accordance, two error probabilities can be defined, namely, the
type 1 error probability
\begin{equation}
p_{1}({\cal {\cal C}}_{n},\phi_{n})\dfn P\left[\phi_{n}(Y^{n})=\overline{H}\right],\label{eq: FA CD}
\end{equation}
and the type 2 error probability
\begin{equation}
p_{2}({\cal {\cal C}}_{n},\phi_{n})\dfn\overline{P}\left[\phi_{n}(Y^{n})=H\right].\label{eq: MD CD}
\end{equation}
As for the DHT problem, the Neyman-Pearson lemma implies that the
optimal detector $\phi_{n,T,\eta}^{*}$ is given by 
\begin{equation}
\P\left[\phi_{n,T,\eta}^{*}(y^{n})=1\right]\dfn\begin{cases}
1, & \sum_{x^{n}\in{\cal C}_{n}}P_{Y|X}(y^{n}|x^{n})>e^{nT}\cdot\sum_{x^{n}\in{\cal C}_{n}}\overline{P}_{Y|X}(y^{n}|x^{n})\\
0, & \sum_{x^{n}\in{\cal C}_{n}}P_{Y|X}(y^{n}|x^{n})<e^{nT}\cdot\sum_{x^{n}\in{\cal C}_{n}}\overline{P}_{Y|X}(y^{n}|x^{n})\\
\eta, & \text{otherwise}
\end{cases},\label{eq: optimal detector CD}
\end{equation}
for some threshold $T\in\mathbb{R}$ and $\eta\in[0,1]$. 

Let $Q_{X}\in{\cal P}({\cal X})$ be a given type, and let $\{n_{l}\}_{l=1}^{\infty}$
be the subsequence of blocklengths such that ${\cal P}_{n}(Q_{X})$
is not empty. As for a DHT sequence of systems ${\cal H}$, a sequence
of CD codes ${\cal C}\dfn\{{\cal C}_{n_{l}}\}_{l=1}^{\infty}$ is
associated with two exponents. The \emph{infimum type 1 exponent}
of a sequence of codes ${\cal C}$ and detector $\{\phi_{n_{l}}\}_{l=1}^{\infty}$
is defined as
\begin{equation}
\liminf_{l\to\infty}-\frac{1}{n_{l}}\log p_{1}({\cal C}_{n_{l}},\phi_{n_{l}}),\label{eq: inf FA exponent definition given code}
\end{equation}
and the \emph{supremum type 1 exponent} is similarly defined, albeit
with a $\limsup$. Analogous exponents are defined for the type 2
error probability. In the sequel, we will construct DHT systems whose
bins are good CD codes, for each $Q_{X}\in{\cal P}({\cal X})$. Since
to obtain an achievability bound for a DHT system, good performance
of CD codes of all types of the source vectors will be simultaneously
required, the blocklengths of the components CD codes must match.
Thus, the limit inferior definition of exponents must be used, as
it assures convergence for all sufficiently large blocklength. For
the converse bound, we will use the limit superior definition. 

For a given type $Q_{X}\in{\cal P}({\cal X})$, rate $\rho\in[0,H(Q_{X}))$,
and type 1 constraint $F_{1}>0$, we define the \emph{infimum CD reliability
function} as 
\begin{multline}
F_{2}^{-}(\rho,Q_{X},F_{1};P_{Y|X},\overline{P}_{Y|X})\\
\dfn\sup_{{\cal C},\{\phi_{n_{l}}\}_{l=1}^{\infty}}\left\{ \liminf_{l\to\infty}-\frac{1}{n_{l}}\log p_{2}({\cal C}_{n_{l}},\phi_{n_{l}}):\forall l,\;{\cal C}_{n_{l}}\subseteq{\cal T}_{n_{l}}(Q_{X}),\;|{\cal C}_{n_{l}}|\geq e^{n_{l}\rho},\;p_{1}({\cal C}_{n_{l}},\phi_{n_{l}})\leq e^{-n_{l}\cdot F_{1}}\right\} ,\label{eq: CD reliability function}
\end{multline}
and the \emph{supremum CD reliability function} $F_{2}^{+}(\rho,Q_{X},F_{1};P_{Y|X},\overline{P}_{Y|X})$
is analogously defined, albeit with a $\limsup$. For brevity, the
dependence on $P_{Y|X},\overline{P}_{Y|X}$ will be omitted whenever
it is understood from context. Thus, the only difference in the reliability
function of CD codes from ordinary HT, is that in CD codes the distributions
are to be optimally designed under the rate constraint $|{\cal C}_{n}|\geq e^{n\rho}$.
Indeed, for $|{\cal C}_{n}|=1$ symmetry implies that any $x^{n}\in{\cal T}_{n}(Q_{X})$
is an optimal CD code. Basic properties of $F_{2}^{\pm}(\rho,Q_{X},F_{1})$
are given as follows. 
\begin{prop}
\label{pro: monotonic and continuous F_FA}As a function of $F_{1}$,
$F_{2}^{\pm}(\rho,Q_{X},F_{1})$ are nonincreasing and have both limit
from the right and from the left at every point. They have no discontinuities
of the second kind and the set of first kind discontinuities (i.e.,
jump discontinuity points) is at most countable. Similar properties
hold as a function of $\rho\in[0,H(Q_{X}))$.
\end{prop}
\begin{IEEEproof}
It follows from their definition that $F_{2}^{\pm}(\rho,Q_{X},F_{1})$
are nonincreasing in $F_{1}$. The continuity statements follow from
properties of monotonic functions \cite[Th. 4.29 and its Corollary, Th. 4.30]{rudin1964principles}
(Darboux-Froda's theorem).
\end{IEEEproof}
With the above, we can state the main result of this section, which
is a characterization of the reliability of DHT systems using the
reliability of CD codes.
\begin{thm}
\label{thm: Redcution to CD codes}The DHT reliability functions $E_{2}^{\pm}(R,E_{1})$
satisfy:
\begin{itemize}
\item Achievability part: 
\[
E_{2}^{-}(R,E_{1})\geq\lim_{\delta\downarrow0}\inf_{Q_{X}\in{\cal P}({\cal X})}\left\{ D(Q_{X}||\overline{P}_{X})+F_{2}^{-}(H(Q_{X})-R,Q_{X},E_{1}-D(Q_{X}||P_{X})+\delta)\right\} .
\]
\item Converse part: 
\[
E_{2}^{+}(R,E_{1})\leq\lim_{\delta\downarrow0}\inf_{Q_{X}\in{\cal P}({\cal X})}\left\{ D(Q_{X}||\overline{P}_{X})+F_{2}^{+}(H(Q_{X})-R+\delta,Q_{X},E_{1}-D(Q_{X}||P_{X})-\delta)\right\} .
\]
\end{itemize}
\end{thm}
The proof of Theorem \ref{thm: Redcution to CD codes} appears in
Appendix \ref{sec:Proof-of-Theorem-Reduction-To-CD }, and its achievability
part is based on the following idea. To begin, let us define for a
given permutation $\pi$ of $[n]$, the permutation of $x^{n}=(x_{1},x_{2},\ldots,x_{n})$
as
\begin{equation}
\pi(x^{n})\dfn\left(x_{\pi(1)},x_{\pi(2)},\ldots,x_{\pi(n)}\right),\label{eq: permutation of vector}
\end{equation}
and let us define the permutation of a set ${\cal D}_{n}\dfn\{x^{n}(0),\ldots,x^{n}(|{\cal D}_{n}|-1)\}$
as
\begin{equation}
\pi({\cal D}_{n})\dfn\left\{ \pi(x^{n}(0)),\ldots,\pi(x^{n}(|{\cal D}_{n}|-1))\right\} .\label{eq: permutation of set defintion}
\end{equation}
Given a single CD codes ${\cal C}_{n}\in{\cal T}_{n}(Q_{X})$, we
can construct a DHT system for $x^{n}\in{\cal T}_{n}(Q_{X})$ by setting
the first bin as $f_{n}^{-1}(1)={\cal C}_{n}.$ Then, the second bin
is set to $f_{n}^{-1}(2)=\pi_{n,2}({\cal C}_{n})\backslash f_{n}^{-1}(1)$
for some permutation $\pi_{n,2}$, and so on. The construction continues
in the same manner until for each $x^{n}\in{\cal T}_{n}(Q_{X})$ there
exists a permutation $\pi_{n,i}$ such that $x^{n}\in\pi_{n,i}({\cal C}_{n})$.
Since the number of required permutations determines the number of
bin, or the encoding rate, such a construction is useful only if the
required number of permutations is not ``too large'', i.e., equal
$\frac{|{\cal T}_{n}(Q_{X})|}{|{\cal C}_{n}|}$ on the exponential
scale. Furthermore, the error probabilities of the DHT systems should
be related to that of the CD code. The proof of the achievability
part of Theorem \ref{thm: Redcution to CD codes} establishes these
properties. 

The achievability and converse part match up to two discrepancies.
First, in the achievability (respectively, converse) part the infimum
(supremum) reliability function appears. This seems unavoidable, as
it is not known if the infimum and supremum reliability functions
are equal even for ordinary channel codes \cite[Problem 10.7]{csiszar2011information}.
Second, the bounds include left and right limits of $E_{2}^{+}(R,E_{1})$
at rate $R$ and exponent $E_{1}$. Nonetheless, due to monotonicity,
$E_{2}^{+}(R,E_{1})$ is continuous function of $R$ and $E_{1}$
for all rates and exponents, perhaps excluding a countable set (Proposition
\ref{pro: monotonic and continuous F_FA}). Thus, for any given $(R,E_{1})$
there exists an arbitrarily close $(\tilde{R},\tilde{E}_{1})$ such
that Theorem \ref{thm: Redcution to CD codes} holds with $\delta=0$. 

\begin{figure}
\begin{centering}
\includegraphics[scale=0.8]{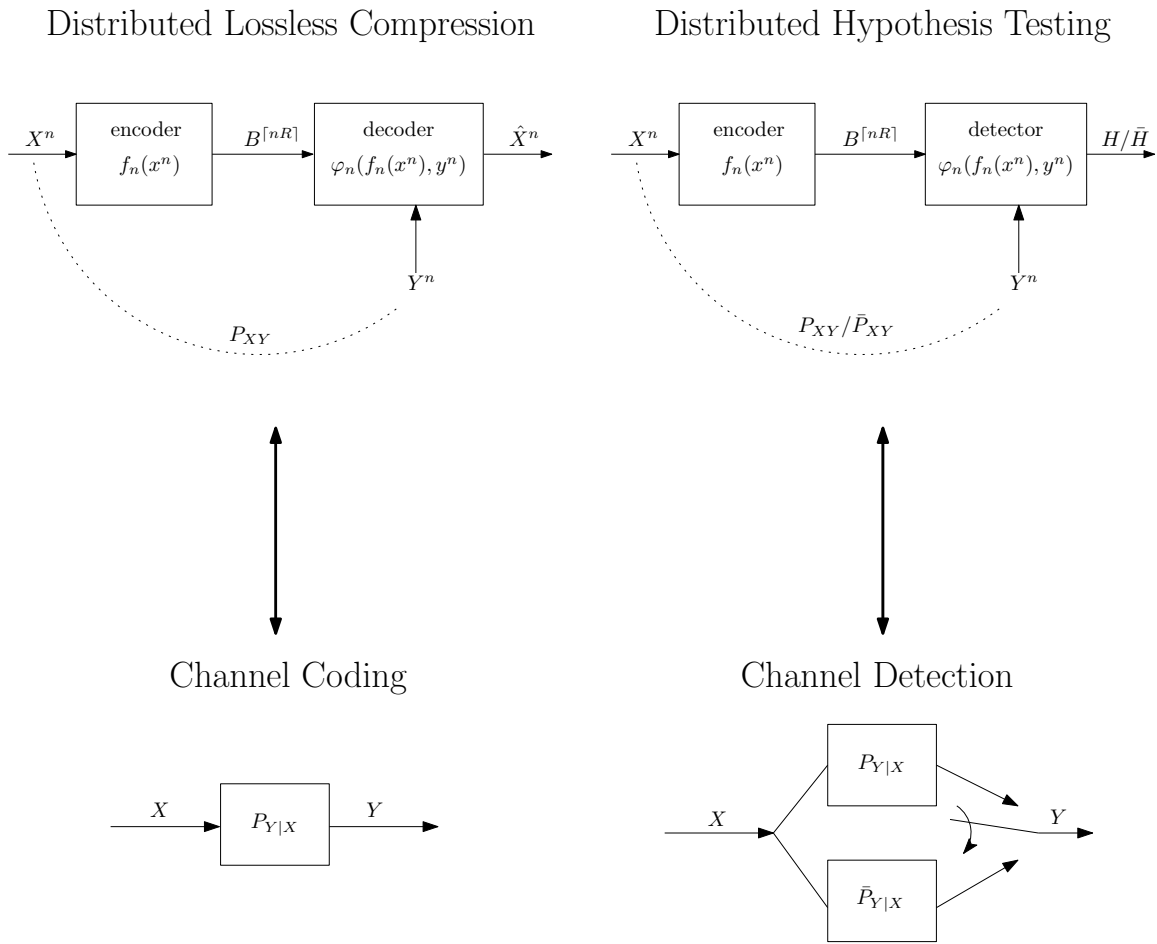}
\par\end{centering}
\caption{An analogy between distributed compression and DHT. \label{fig: analogy to SW}}
\end{figure}

As illustrated in Fig. \ref{fig: analogy to SW}, this theorem parallels
a similar result of\textbf{ }\cite{Chen2007reliability,Weinberger15},
 which characterizes the reliability function of DLC with that of
channel ordinary channel codes. While the reliability function of
the latter is itself not fully known, bounds such as the \emph{random-coding}
and \emph{expurgated }achievability bounds, and the \emph{sphere-packing},
\emph{zero-rate}, and \emph{straight-line} converse bounds \cite[Ch. 10]{csiszar2011information}
may be used to obtain analogous bounds for DLC. Similarly, Theorem
\ref{thm: Redcution to CD codes} reveals that the DHT problem is
characterized by the reliability function of CD codes. For the latter,
we derive in the next section a random-coding bound and an expurgated
bound on its reliability. Using Theorem \ref{thm: Redcution to CD codes},
these bounds directly lead to bounds on the reliability of the DHT
problem, as stated in Theorem \ref{thm: DHT RC and EX}.

\section{Bounds on the Reliability of CD Codes\label{sec:Bounds CD}}

\begin{figure}
\begin{centering}
\includegraphics[scale=0.5]{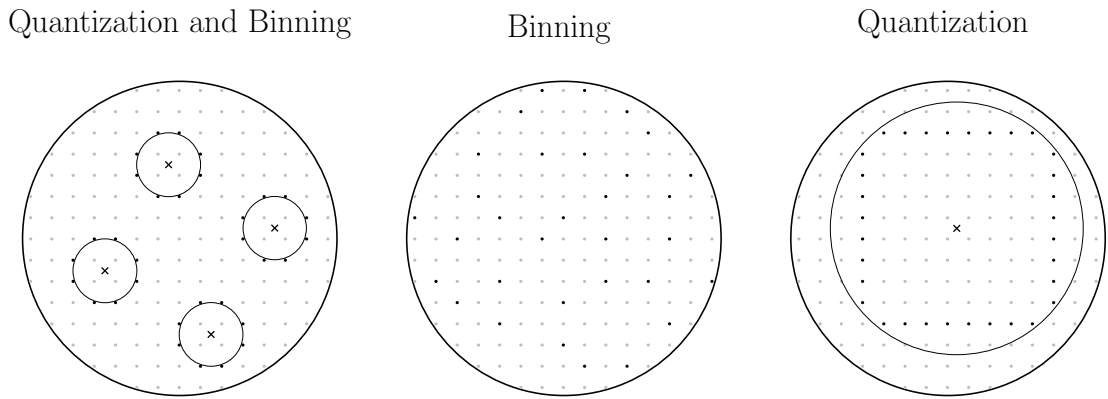}
\par\end{centering}
\caption{An illustration of various types of CD codes which pertain to a bin
of a DHT system. The grey dots within the large circle represent the
members of ${\cal T}_{n}(Q_{X})$. In a quantization based scheme,
a bin corresponds to a single reproduction cell of the quantization
scheme, and thus all the codewords of the CD code share a common ``center''
(reproduction vector). In a binning-based scheme, the codewords of
the CD code are scattered over the type class with no particular structure.
In a quantization-and-binning scheme, the codewords are partitioned
to ``distant'' clouds, where the black dots within one of the small
circles represent the satellite codebook pertaining to one of the
cloud centers. \label{fig: types of codes}}
\end{figure}

In the previous section, we have linked the DHT reliability function
to that of CD. In this section, we derive bounds on the latter using
random coding arguments, to wit, choosing ${\cal C}_{n}\subseteq{\cal T}_{n}(Q_{X})$
of size $2^{nR}$ at random, and analyzing the average error probabilities.
To obtain good bounds for the DHT problem, however, the random ensemble
should be chosen with some attention. We now list the encoding schemes
typically analyzed for DHT systems, and state the corresponding CD
ensemble for each of them, allowing a DHT system to be constructed
(for vectors of the given input type) via its permutations:
\begin{enumerate}
\item \emph{Binning} - meaning assigning source vectors to bins uniformly
at random. This corresponds to an \emph{ordinary ensemble}, i.e.,
choosing the codewords uniformly at random over ${\cal T}_{n}(Q_{X})$. 
\item \emph{Quantization} - meaning assigning ``close'' source vectors
to the same bin. This corresponds to a \emph{conditional ensemble},
i.e., choosing the codewords uniformly at random in some ${\cal T}_{n}(Q_{X|U},u^{n})$
given a ``cloud center'' $u^{n}$.
\item \emph{Quantization-and-binning} which combines both. This corresponds
to an hierarchical ensemble - a combination of the ordinary and conditional
ensembles - i.e., choosing cloud centers from the ordinary ensemble
over ${\cal T}_{n}(Q_{U})$ uniformly at random, and then draw \textquotedblleft satellite\textquotedblright{}
codewords for each center from the conditional ensemble ${\cal T}_{n}(Q_{X|U},u^{n})$
uniformly at random (independently over clouds).
\end{enumerate}
See Fig. \ref{fig: types of codes} for illustration. 

In what follows, we will analyze the hierarchical ensemble since,
as discussed in the introduction, the best known achievable bounds
for DHT systems are obtained via quantization-and-binning-based schemes.
Furthermore, it generalizes both the ordinary ensemble and the conditional
ensemble. We next rigorously define the specific \emph{hierarchical
ensemble} used:
\begin{defn}
\label{def:A-fixed-composition-hierarchical-Codebook}A \emph{fixed-composition}
\emph{hierarchical ensemble} for an input type $Q_{X}\in{\cal P}_{n}({\cal X})$
and rate $\rho$ is defined by a conditional type $Q_{U|X}\in{\cal P}_{n}({\cal U},Q_{X})$,
where $U\in{\cal U}$ is an auxiliary random variable $|{\cal U}|<\infty$,
a \emph{cloud-center rate} $\rho_{\s[c]}$ and a \emph{satellite rate}
$\rho_{\s[s]}$, such that $\rho=\rho_{\s[c]}+\rho_{\s[s]}$. A random
codebook $\mathfrak{C}_{n}$ from this ensemble is drawn in two stages.
First, $e^{n\rho_{\s[c]}}$ cloud centers ${\cal C}_{\s[c],n}$ are
drawn, independently and uniformly over ${\cal T}_{n}(Q_{U})$. Second,
for each of the cloud centers $u^{n}\in{\cal C}_{\s[c],n}$, $e^{n\rho_{\s[s]}}$
satellites are drawn independently and uniformly over ${\cal T}_{n}(Q_{X|U},u^{n})$.
\end{defn}
Evidently, codewords which pertain to the same cloud are dependent,
whereas codewords from different clouds are independent. Further,
the ordinary ensemble is obtained as a special case by choosing $U=X$
and $\rho_{\s[c]}=\rho$, and the conditional ensemble is obtained
by setting $\rho_{\s[s]}=\rho$. Whenever the CD code is to be used
as bins of a DHT system for source vectors of type $Q_{X}$ the correspondence
between the parameters is as follows: A quantization-and-binning DHT
system of rate $R$, binning rate $R_{\s[b]}$, and quantization rate
$R_{\s[q]}=R+R_{\s[b]}$, requires hierarchical CD codes of rate $\rho=H(Q_{X})-R$,
cloud-center rate $\rho_{\s[c]}=R_{\s[b]}$ and satellite rate $\rho_{\s[s]}=H(Q_{X})-R_{\s[q]}$.
The cloud centers ${\cal C}_{\s[c],n}$ are the reproduction vectors
of the DHT system, where the joint distribution of any source vector
and its reproduction vector is exactly $Q_{UX}$, and the choice of
the \emph{test channel }$Q_{U|X}$ is used to control the distortion
of the quantization.\footnote{Usually in rate-distortion theory, the test channel is used to control
the average distortion $\E_{Q}[d(X,U)]$ for some distortion measure
$d:{\cal X}\times{\cal U}\to\mathbb{R}_{+}$. Here, $x^{n}$ is always
chosen from ${\cal T}_{n}(Q_{X|U},u^{n})$ and therefore the distortion
between $u^{n}$ and $x^{n}$ is constant depending on $Q_{UX}$. } 

In this section, it will be more convenient to use the parameter $\lambda\dfn\frac{1}{\tau+1}\in\lambda\in[0,1]$
instead of $\tau$. Using this convention, we will use, e.g., $d_{\lambda}$
instead of $d_{\tau}$ for the Chernoff parameter.

To state a random-coding bound on the reliability of CD codes, we
denote
\begin{align}
 & A_{\s[rc]}'(\rho,\rho_{\s[c]},Q_{UX},\lambda)\nonumber \\
 & \dfn\min_{(Q_{UXY},\overline{Q}_{UXY}):Q_{U|X}=\overline{Q}_{U|X},\;Q_{Y}=\overline{Q}_{Y}}\left\{ (1-\lambda)\cdot D(Q_{Y|UX}||P_{Y|X}|Q_{UX})+\lambda\cdot D(\overline{Q}_{Y|UX}||\overline{P}_{Y|X}|\overline{Q}_{UX})\vphantom{\left[I_{\overline{Q}}(X;Y|U)-\left[\rho_{\s[c]}-I_{\overline{Q}}(U;Y)\right]_{+}-\rho_{\s[s]}\right]_{+}}\right.\nonumber \\
 & \left.\hphantom{=}+\lambda\cdot\max\left\{ \left|I_{Q}(U;Y)-\rho_{\s[c]}\right|_{+},\;I_{Q}(U,X;Y)-\rho\right\} +(1-\lambda)\cdot\max\left\{ \left|I_{\overline{Q}}(U;Y)-\rho_{\s[c]}\right|_{+},\;I_{\overline{Q}}(U,X;Y)-\rho\right\} \right\} ,\label{eq: A_rc_tag}
\end{align}
and
\begin{align}
 & A_{\s[rc]}''(\rho,\rho_{\s[c]},Q_{UX},\lambda)\nonumber \\
 & \dfn\min_{(Q_{UXY},\overline{Q}_{UXY}):Q_{U|X}=\overline{Q}_{U|X},\;Q_{UY}=\overline{Q}_{UY},\;I_{Q}(U;Y)>\rho_{\s[c]}}\left\{ (1-\lambda)\cdot D(Q_{Y|UX}||P_{Y|X}|Q_{UX})+\lambda\cdot D(\overline{Q}_{Y|UX}||\overline{P}_{Y|X}|\overline{Q}_{UX})\vphantom{\left[I_{\overline{Q}}(X;Y|U)-\left[\rho_{\s[c]}-I_{\overline{Q}}(U;Y)\right]_{+}-\rho_{\s[s]}\right]_{+}}\right.\nonumber \\
 & \left.\hphantom{=}+\lambda\cdot\left|I_{Q}(X;Y|U)-\rho+\rho_{\s[c]}\right|_{+}+(1-\lambda)\cdot\left|I_{\overline{Q}}(X;Y|U)-\rho+\rho_{\s[c]}\right|_{+}\right\} ,\label{eq: A_rc_tag2}
\end{align}
as well as\textbf{
\begin{equation}
A_{\s[rc]}(\rho,\rho_{\s[c]},Q_{UX},\lambda)\dfn\min\left\{ A_{\s[rc]}'(\rho,\rho_{\s[c]},Q_{UX},\lambda),\;A_{\s[rc]}''(\rho,\rho_{\s[c]},Q_{UX},\lambda)\right\} .\label{eq: A_rc}
\end{equation}
}For brevity, when can be understood from context, the dependency
on $(\rho,\rho_{\s[c]},Q_{UX},\lambda)$ will be omitted. Our random-coding
bound is as follows.
\begin{thm}
\label{thm: random coding bound}The infimum CD reliability function
is bounded as 
\begin{equation}
F_{2}^{-}(\rho,Q_{X},F_{1})\geq\sup_{0\leq\lambda\leq1}\left\{ -\frac{1-\lambda}{\lambda}\cdot F_{1}+\frac{1}{\lambda}\cdot\min\left[d_{\lambda}(Q_{X}),\;\sup_{Q_{U|X}}\sup_{\rho_{\s[c]}:\;\rho_{\s[c]}\geq\left|\rho-H_{Q}(X|U)\right|_{+}}A_{\s[rc]}(\rho,\rho_{\s[c]},Q_{UX},\lambda)\right]\right\} .\label{eq: CD random coding bound}
\end{equation}
\end{thm}
The proof of Theorem \ref{thm: random coding bound} appears in Appendix
\ref{sec:Proof-of-Theorem-Random-Coding}. We make the following comments:
\begin{enumerate}
\item Loosely speaking, in \eqref{eq: A_rc}, the exponent $A_{\s[rc]}'$
corresponds to the contribution to the error probability from codewords
which belong to different cloud centers, whereas the exponent $A_{\s[rc]}''$
corresponds to the contribution to the error probability from codewords
which belong to the same cloud center as the transmitted codeword.
Thus, for a given rate $\rho$, $A_{\s[rc]}'$ is monotonically nonincreasing
with $\rho_{\s[c]}$, while $A_{\s[rc]}''$ is monotonically non-decreasing
with $\rho_{\s[c]}$ (or, monotonically nonincreasing with $\rho_{\s[s]}$).
The cloud-center rate $\rho_{\s[c]}$ and the test channel $Q_{U|X}$
therefore should be chosen to optimally balance between these two
contributions to the error probability. 
\item In comparison to \cite{Weinberger_merhav15}, we have generalized
the random coding analysis of the detection error exponents to hierarchical
ensembles, and also obtained simpler expressions using the ensemble
average of the exponent of the Chernoff parameter.
\item In fact, a stronger claim than the one appears in Theorem \ref{thm: random coding bound}
can be made. It can be shown that there exists a \emph{single} sequence
of CD codes $\{{\cal C}_{n_{l}}^{*}\}_{l=1}^{\infty}$ such that 
\begin{multline}
\liminf_{l\to\infty}-\frac{1}{n_{l}}\log\min_{T:\;p_{1}({\cal C}_{n_{l}}^{*},\phi_{n_{l},T}^{*})\leq e^{-nF_{1}}}p_{2}({\cal C}_{n_{l}}^{*},\phi_{n_{l},T}^{*})\\
\geq\max_{0\leq\lambda\leq1}\left\{ -\frac{1-\lambda}{\lambda}\cdot F_{1}+\frac{1}{\lambda}\cdot\min\left[d_{\lambda}(Q_{X}),\;\sup_{Q_{U|X}}\sup_{\rho_{\s[c]}:\;\rho_{\s[c]}\geq\left[\rho-H_{Q}(X|U)\right]_{+}}A_{\s[rc]}\right]\right\} ,
\end{multline}
\emph{simultaneously} for all $F_{1}$. Thus, when using such a CD
code, the operating point along the trade-off curve between the two
exponents can be determined solely by the detector, and can be arbitrarily
changed from block to block. For details regarding the proof of this
claim, see Remark \ref{rem:good for all type 1 exponent}. 
\end{enumerate}
Next, we state our \emph{expurgated} exponent, and to this end we
denote 
\[
A_{\s[ex]}(\rho,Q_{X},\lambda)\dfn\min_{Q_{X\tilde{X}}:\;Q_{X}=Q_{\tilde{X}},\;I_{Q}(X;\tilde{X})\leq\rho}\left\{ d_{\lambda}(Q_{X\tilde{X}})+I_{Q}(X;\tilde{X})-\rho\right\} .
\]

\begin{thm}
\label{thm: Expurgated bound}The infimum CD reliability function
is bounded as 
\[
F_{2}^{-}(\rho,Q_{X},F_{1})\geq\max_{0\leq\lambda\leq1}\left\{ -\frac{1-\lambda}{\lambda}\cdot F_{1}+\frac{1}{\lambda}\cdot\min\left\{ d_{\lambda}(Q_{X}),\;A_{\s[ex]}(\rho,Q_{X},\lambda)\right\} \right\} .
\]
\end{thm}
The proof appears in Appendix \ref{sec:Proof-of-Theorem-Random-Coding}.
We make the following comments:
\begin{enumerate}
\item A similar expurgated bound can be derived for hierarchical ensembles.
However, when optimizing the rates $(\rho_{\s[s]},\rho_{\s[c]})$
for this expurgated bound, it turns out that choosing $\rho_{\s[s]}=0$
is optimal. Thus, the resulting bound exactly equals the bound of
Theorem \ref{thm: Expurgated bound}, which corresponds to an ordinary
ensemble. 
\item Since the expurgated exponent only improves the random-coding exponent
of the ordinary ensemble (which is inferior in performance to the
hierarchical ensemble), it is anticipated that expurgation does not
play a significant role in this problem, compared to the channel coding
problem. This might be due to the fact that the aim of expurgation
is to remove codewords which have ``close'' neighbors, while this
is not actually required in the DHT problem. This can also be attributed
to the bounding technique of the expurgated bound, which is based
on pairwise Chernoff parameters.
\end{enumerate}
After deriving the bounds on the reliability of CD, we return to the
DHT problem, and conclude the section with a short proof of Theorem
\ref{thm: DHT RC and EX}. 
\begin{IEEEproof}[Proof of Theorem \ref{thm: DHT RC and EX}]
Up to the arbitrariness of $\delta>0$, Theorem \ref{thm: Redcution to CD codes}
states that
\begin{equation}
E_{2}^{-}(R,E_{1})=\inf_{Q_{X}\in{\cal P}({\cal X})}\left\{ D(Q_{X}||\overline{P}_{X})+F_{2}^{-}\left(H(Q_{X})-R,Q_{X},E_{1}-D(Q_{X}||P_{X})\right)\right\} .\label{eq: proof of DHT bound - CD charactarization}
\end{equation}
Further, the random-coding bound of Theorem \ref{thm: random coding bound}
and the expurgated bound of Theorem \ref{thm: Expurgated bound} both
imply that 
\begin{align}
 & F_{2}^{-}\left(H(Q_{X})-R,Q_{X},E_{1}-D(Q_{X}||P_{X})\right)\nonumber \\
 & \geq\max\Bigg\{\max_{0\leq\lambda\leq1}\Bigg[-\frac{1-\lambda}{\lambda}\cdot E_{1}+\frac{1-\lambda}{\lambda}\cdot D(Q_{X}||P_{X})\nonumber \\
 & \hphantom{=}+\frac{1}{\lambda}\cdot\min\left\{ d_{\lambda}(Q_{X}),\;\sup_{Q_{U|X}}\sup_{R_{\s[b]}:\;R_{\s[b]}\geq\left|I_{Q}(U;X)-R\right|_{+}}A_{\s[rc]}\left(H(Q_{X})-R,R_{\s[b]},Q_{UX},\lambda\right)\right\} \Bigg],\nonumber \\
 & \hphantom{=}\max_{0\leq\lambda\leq1}\Bigg[-\frac{1-\lambda}{\lambda}\cdot E_{1}+\frac{1-\lambda}{\lambda}\cdot D(Q_{X}||P_{X})+\nonumber \\
 & \hphantom{=}+\frac{1}{\lambda}\cdot\min\left\{ d_{\lambda}(Q_{X}),\;A_{\s[ex]}\left(H(Q_{X})-R,Q_{X},\lambda\right)\right\} \Bigg]\Bigg\}\\
 & =\sup_{Q_{U|X}}\sup_{R_{\s[b]}:\;R_{\s[b]}\geq\left|I_{Q}(U;X)-R\right|_{+}}\max_{0\leq\lambda\leq1}\Bigg\{-\frac{1-\lambda}{\lambda}\cdot E_{1}+\frac{1-\lambda}{\lambda}\cdot D(Q_{X}||P_{X})\nonumber \\
 & \hphantom{=}+\frac{1}{\lambda}\cdot\min\bigg[d_{\lambda}(Q_{X}),\;\max\left\{ A_{\s[rc]}\left(H(Q_{X})-R,R_{\s[b]},Q_{UX},\lambda\right),\;A_{\s[ex]}\left(H(Q_{X})-R,Q_{X},\lambda\right)\right\} \bigg]\Bigg\}.\label{eq: combined rc and ex for CD}
\end{align}
The bound of Theorem \ref{thm: DHT RC and EX} is obtained by substituting
in \eqref{eq: combined rc and ex for CD} in \eqref{eq: proof of DHT bound - CD charactarization},
while changing variables from $\lambda$ to $\tau\dfn\frac{1-\lambda}{\lambda}$
and from $\rho_{\s[c]}$ to $R_{\s[b]}$, as well as using the definitions
of $B_{\s[rc]}$ \eqref{eq: B_rc}, $B_{\s[ex]}$ \eqref{eq: B_ex}
and $B$ \eqref{eq: B}.
\end{IEEEproof}

\section{Computational Aspects and a Numerical Example\label{sec:Computational-Aspects-and}}

The bound of Theorem \ref{thm: DHT RC and EX} is rather involved,
and therefore it is of interest to discuss how to compute it efficiently.
Evidently, the main computational task is the computation of $B_{\s[rc]}'$
and $B_{\s[rc]}''$ for a given $(R,R_{\s[b]},Q_{UX},\tau)$. To this
end, it can be seen that the objective functions of both $B_{\s[rc]}'$
and $B_{\s[rc]}''$ are convex functions of $(Q_{Y|UX},\overline{Q}_{Y|UX})$
(and strictly convex, if $P_{Y|X}\ll\gg\overline{P}_{Y|X}$ ).\footnote{Indeed, the divergence terms and $I(U,X;Y)$ are convex in $Q_{Y|UX}$.
The term $I(U;Y)$ is also convex in $Q_{Y|UX}$, as a composition
of a linear function which maps $Q_{Y|UX}$ to $Q_{Y|U}$ and the
mutual information $I(U;Y)$. The pointwise maximum of two convex
functions is also a convex function (note that $|f(t)|_{+}=\max\{0,f(t)\}$). } Furthermore, the feasible set of $B_{\s[rc]}'$ is a convex set (only
has linear constraints) and thus the computation of $B_{\s[rc]}'$
is a convex optimization problem, which can be solved efficiently
\cite{boyd2004convex}. However, the feasible set of $B_{\s[rc]}''$
is not convex, due to the additional constraint $I_{Q}(U;Y)>R_{\s[b]}$
beyond the linear constraints. Nevertheless, the value of $B_{\s[rc]}$
can be computed efficiently, by only solving \emph{convex} optimization
problems, according to the following algorithm:
\begin{enumerate}
\item Solve the optimization problem \eqref{eq: B_rc'} defining $B_{\s[rc]}'$,
and let the optimal value be $v'$.
\item Solve the optimization problem \eqref{eq: B_rc''} defining $B_{\s[rc]}''$,
but without the constraint $I_{Q}(U;Y)>R_{\s[b]}$ (this is a convex
optimization problem). Let the solution be $(Q_{UXY}^{*},\overline{Q}_{UXY}^{*})$
and the optimal value be $v''$. If $I_{Q^{*}}(U;Y)>R_{\s[b]}$ then
set $B_{\s[rc]}''=v''$, and otherwise, set $B_{\s[rc]}''=\infty$.
\item The result is $B_{\s[rc]}=\min\{v',v''\}$.
\end{enumerate}
The correctness of the algorithm follows from the following argument.
It is easily verified that if $I_{Q^{*}}(U;Y)>R_{\s[b]}$ then the
constraint $I_{Q}(U;Y)>R_{\s[b]}$ is inactive, and therefore can
be omitted. Thus, in this case $B_{\s[rc]}''=v''$. However, if this
is not the case, then the solution must be on the boundary, i.e.,
must satisfy $I_{Q}(U;Y)=R_{\s[b]}$. This is because the objective
in $B_{\s[rc]}''$ is a convex function. In the latter case, it can
be easily seen that $B_{\s[rc]}'\leq B_{\s[rc]}''$, and as $B_{\s[rc]}$
is the minimum between the two, $B_{\s[rc]}''=\infty$ can be set. 

Given the value of $B_{\s[rc]}$, the next step is to optimize over
$Q_{U|X}$ and $\rho_{\s[c]}$. While this should be done exhaustively,\footnote{Or by any other general-purpose global optimization algorithm.}
any specific choice of $Q_{U|X}$ and $\rho_{\s[c]}$ (or a restricted
optimization set for them) leads to an achievable bound on $E_{2}^{-}(R,E_{1})$.
It should be mentioned, however, that it is not clear to us how to
apply standard cardinality-bounding techniques (i.e., those based
on the support lemma \cite[p. 310]{csiszar2011information}) to bound
$|{\cal U}|$ in this problem. Thus, in principle, the cardinality
of ${\cal U}$ is unrestricted, improved bounds are possible. Computing
$B_{\s[ex]}(R,Q_{X},\tau)$ of \eqref{eq: B_ex} is a convex optimization
problem, over $Q_{\tilde{X}|X}$. 

Finally, both $\tau$ and $Q_{X}$ should be optimized, which is feasible
when ${\cal X}$ is not very large and exhausting the simplex ${\cal S}({\cal X})$
in search of the minimizer $Q_{X}$ is possible. Furthermore, as we
have seen in Corollary \ref{corr: Stein's exponent}, when only Stein's
exponent is of interest, i.e., $E_{1}=0$, the minimal value in \eqref{eq: DHT achievability bound}
must be attained for $Q_{X}=P_{X}$. Thus, there is no need to minimize
over $Q_{X}\in{\cal S}({\cal X})$, but rather only on $Q_{X}=P_{X}$.
We can also set $\tau\to\infty$ if the weak version \eqref{eq: Stein's exponent corollary only RC}
of Corollary \ref{corr: Stein's exponent} is used as a bound. More
generally, the minimizer of $Q_{X}$ must satisfy $D(Q_{X}||P_{X})\leq E_{1}$,
and this can decrease the size of the feasible set of $Q_{X}$ whenever
the required $E_{1}$ is not very large. 

\begin{figure}
\begin{centering}
\includegraphics[scale=0.5]{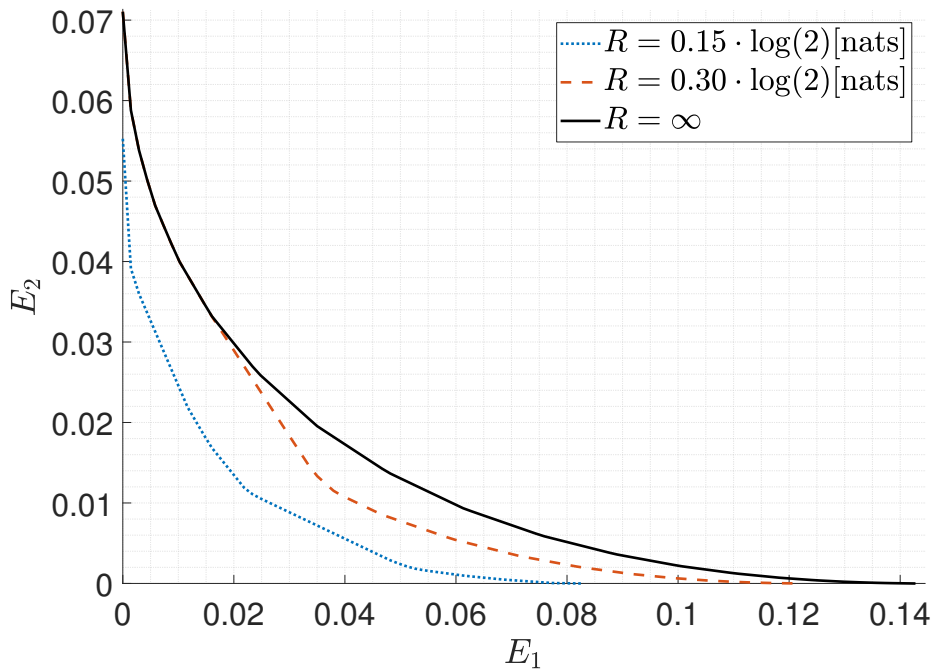}
\par\end{centering}
\caption{A binary example. \label{fig: Binary example}}
\end{figure}

A simple example for using the above methods to compute bounds on
the DHT reliability function is given as follows.
\begin{example}
Consider the case ${\cal X}={\cal Y}=\{0,1\}$, and $P_{X}=\overline{P}_{X}=(\nicefrac{1}{2},\nicefrac{1}{2})$,
where $P_{Y|X}$ and $\overline{P}_{Y|X}$ are binary symmetric channels
with crossover probabilities $10^{-1}$ and $10^{-2}$, respectively.
We have used an auxiliary alphabet of size $|{\cal U}|=|{\cal X}|+1=3$,
and due to the symmetry in the problem, we have only optimized over
symmetric $Q_{U|X}$. The random-coding bounds on the reliability
of the DHT is shown in Fig. \ref{fig: Binary example} for two different
rates. The convex optimization problems were solved using CVX, a Matlab
package for disciplined convex programming \cite{cvx}. 
\end{example}

\section{Conclusion and Further Research\label{sec:Conclusion}}

We have considered the trade-off between the two types of error exponents
of a DHT system, with full side information. We have shown that its
reliability is intimately related to the reliability of CD, and thus
the latter simpler problem should be considered. Achievable bounds
on the reliability of CD were derived, under the optimal Neyman-Pearson
detector. 

There are multiple directions in which our understanding of the problem
can be broadened: 
\begin{enumerate}
\item \emph{Variable-rate coding}: The DLC reliability may be increased
when variable rate\emph{ }is allowed, either with an average rate
constraint \cite{Chen2007reliability}, or under excess-rate exponent
constraint \cite{Weinberger15}. It is interesting to use the techniques
developed for the DLC to the DHT problem (see also the discussion
in \cite[Appendix]{Weinberger17_secrecy}). 
\item \emph{Computation of the bounds}: The main challenge in the random-coding
bound computation is the optimization over the test channel $Q_{U|X}$.
First, deriving cardinality bounds on the auxiliary random variable
alphabet ${\cal U}$ is of interest. Second, finding an efficient
algorithm to optimize the test channel, perhaps an alternating-maximization
algorithm in the spirit of the Csisz{\'a}r-Tusn\'{a}dy \cite{csiszar_information_1984}
and the Blahut-Arimoto algorithms \cite{arimoto1972algorithm,arimoto_exponents_computation,blahut1972computation}.
As was noted in \cite{Tian_Chan2008,Katz2017}, Stein's exponent in
the DHT problem of testing against independence is identical to the
\emph{information bottleneck }problem \cite{tishby2000information},
for which such alternating-maximization algorithm was developed. 
\item \emph{Converse bounds}: We have shown converse bounds on the reliability
of DHT systems, it suffices to obtain converse bounds on the reliability
of CD codes, no concrete bounds were derived. To obtain converse bounds
which explicitly depend on the rate (in contrast to Proposition \ref{prop: DHT converse}),
two challenges are visible. First, it is tempting to conjecture that
the Chernoff characterization \eqref{eq: ordinary hypothesis testing Chernoff information}
characterizes the reliability of CD codes, in the sense that\footnote{As usual $\{n_{l}\}_{l=1}^{\infty}$ is the subsequence of blocklength
such that ${\cal T}_{n}(Q_{X})$ is not empty. } 
\begin{multline}
F_{2}^{+}(\rho,Q_{X},F_{1};P_{Y|X},\overline{P}_{Y|X})=\limsup_{l\to\infty}\max_{{\cal C}_{n_{l}}\subseteq{\cal T}_{n_{l}}(Q_{X}):\;|{\cal C}_{n_{l}}|\geq e^{n_{l}\rho}}\sup_{\tau\geq0}\\
\left\{ -\tau\cdot F_{1}-(\tau+1)\cdot\frac{1}{n_{l}}\log\left\{ \sum_{y^{n_{l}}\in{\cal Y}^{n_{l}}}\left[P_{Y^{n_{l}}}^{({\cal C}_{n_{l}})}(y^{n_{l}})\right]^{\nicefrac{\tau}{1+\tau}}\cdot\left[\overline{P}_{Y^{n_{l}}}^{({\cal C}_{n_{l}})}(y^{n_{l}})\right]^{\nicefrac{1}{1+\tau}}\right\} \right\} ,
\end{multline}
just as a similar quantity was used to derive the random-coding and
expurgated bounds. Second, even if this conjecture holds, the value
of 
\[
\left\{ \sum_{y^{n_{l}}\in{\cal Y}^{n_{l}}}\left[P_{Y^{n_{l}}}^{({\cal C}_{n_{l}})}(y^{n_{l}})\right]^{\nicefrac{\tau}{1+\tau}}\cdot\left[\overline{P}_{Y^{n_{l}}}^{({\cal C}_{n_{l}})}(y^{n_{l}})\right]^{\nicefrac{1}{1+\tau}}\right\} 
\]
should be lower bounded for all CD codes whose size is larger than
$e^{n\rho}$. As this term can be identified as a \emph{R\'{e}nyi
divergence }\cite{renyi1961measures,van2014renyi}\emph{, }the problem
of bounding its value is a \emph{R\'{e}nyi divergence characterization}.
This problem seems formidable, as the methods developed in \cite{Ahlswede1976bounds}
for the entropy characterization problem rely heavily on the chain
rule of mutual information; a property which is not naturally satisfied
by R\'{e}nyi entropies and divergences. Hence, the problem of obtaining
a non trivial converse bound for the reliability of DHT systems with
general hypotheses and a positive encoding rate remains an elusive
open problem.
\item \emph{Rate constraint on the side information}: The reliability of
a DHT systems in which the side-information vector $y^{n}$ is also
encoded at a limited rate should be studied under optimal detection.
Such systems will naturally lead to \emph{multiple-access} CD codes,
as studied for ordinary channel coding (see \cite{Nazari_lower,Nazari_upper}
and references therein). However, for such a scenario, it was shown
in \cite{Haim2016} that the use of linear codes (a-la K{\"o}rner-Marton
coding) dramatically improves performance. Thus, it is of interest
to analyze DHT systems with both linear codes and optimal detection.
However, it is not yet known how to apply the type-enumeration method,
used here for analysis of optimal detection, to linear codes. Hence,
either the type-enumeration method should be refined, or an alternative
approach should be sought after.
\item \emph{Generalized hypotheses}: Hypotheses regarding the distributions
of continuous random variables, or regarding the distributions of
sources with memory can be considered. Furthermore, the case of composite
hypotheses, in which the distribution under each hypotheses is not
exactly known (e.g., belongs to a subset of a given parametric family),
and finding universal detectors which operate as well as the for simple
hypotheses can also be considered. For preliminary results along this
line see \cite{Shalaby1992,Weinberger_merhav15}.
\end{enumerate}

\section*{Acknowledgment}

Discussions with N. Merhav, the comments of the associate editor,
A. Tchamkerten, and the comments of the anonymous reviewers, are acknowledged
with thanks. 

\appendices{\numberwithin{equation}{section}}

\section{Proofs of Corollaries to Theorem \ref{thm: DHT RC and EX}\label{sec:Proofs-of-Corollaries}}
\begin{IEEEproof}[Proof of Corollary \ref{corr: no-loss rate}]
 Suppose that the inner minimization in the bound of Theorem \ref{thm: DHT RC and EX}
is dominated by $(\tau+1)\cdot d_{\tau}(Q_{X})$ for all $Q_{X}\in{\cal P}({\cal X})$.
Then, \eqref{eq: DHT achievability bound} reads
\begin{align}
 & E_{2}^{-}(R,E_{1};P_{XY},\overline{P}_{XY})\nonumber \\
 & \geq\min_{Q_{X}}\sup_{\tau\geq0}\left\{ -\tau\cdot E_{1}+D(Q_{X}||\overline{P}_{X})+\tau\cdot D(Q_{X}||P_{X})+(\tau+1)\cdot d_{\tau}(Q_{X})\right\} \\
 & \trre[=,a]\min_{Q_{X}}\sup_{\tau\geq0}\min_{Q_{Y|X}}\big\{-\tau\cdot E_{1}+D(Q_{X}||\overline{P}_{X})+\tau\cdot D(Q_{X}||P_{X})\nonumber \\
 & \hphantom{=}+\tau\cdot D(Q_{Y|X}||P_{Y|X}|Q_{UX})+D(Q_{Y|UX}||\overline{P}_{Y|X}|Q_{UX})\big\}\\
 & =\min_{Q_{X}}\sup_{\tau\geq0}\min_{Q_{Y|X}}\left\{ -\tau\cdot E_{1}+D(Q_{X}||\overline{P}_{X})+\tau\cdot D(Q_{XY}||P_{XY})+D(Q_{XY}||\overline{P}_{XY})\right\} \\
 & \trre[=,b]\min_{Q_{XY}}\sup_{\tau\geq0}\left\{ \tau\cdot\left[D(Q_{XY}||P_{XY})-E_{1}\right]+D(Q_{XY}||\overline{P}_{XY})\right\} \\
 & =\min_{Q_{XY}:D(Q_{XY}||P_{XY})\leq E_{1}}D(Q_{XY}||\overline{P}_{XY}),\label{eq: distributed hypothesis testing reduced to ordinary}
\end{align}
where: $(a)$ follows since (see \eqref{eq: divergence charctarization of Chernoff}
in the proof of Lemma \ref{lem: average of Chernoff})
\begin{align}
(\tau+1)\cdot d_{\tau}(Q_{X}) & =\min_{Q_{Y|UX}}\left[\tau\cdot D(Q_{Y|UX}||P_{Y|X}|Q_{UX})+D(Q_{Y|UX}||\overline{P}_{Y|X}|\overline{Q}_{UX})\right]\\
 & =\min_{Q_{Y|X}}\left[\tau\cdot D(Q_{Y|X}||P_{Y|X}|Q_{X})+D(Q_{Y|X}||\overline{P}_{Y|X}|Q_{X})\right].\label{eq: divergence charctarization of Chernoff tau}
\end{align}
$(b)$ follows since the objective function is linear in $\tau$ (and
hence concave) and convex in $Q_{Y|X}$, and therefore the minimization
and maximization order can be interchanged \cite{Sion1958minimax}.
Thus, the achievability bound of Theorem \ref{thm: DHT RC and EX}
coincides with the converse bound of Proposition \ref{prop: DHT converse},
where the latter is obtained when the rate of the DHT system is not
constrained at all, and given by the reliability function of the ordinary
HT problem between $P_{XY}$ and $\overline{P}_{XY}$.
\end{IEEEproof}
\begin{IEEEproof}[Proof of Corollary \ref{corr: Stein's exponent}]
 It can be seen that the outermost minimum in \eqref{eq: DHT achievability bound}
must be attained for $Q_{X}=P_{X}$. Intuitively, since we are only
interested in negligible type 1 exponent, any event with $Q_{X}\neq P_{X}$
has exponentially decaying probability $\exp[-nD(Q_{X}||P_{X})]$,
and does not affect the exponent. More rigorously, if $Q_{X}\neq P_{X}$
then by taking $\tau\to\infty$ the objective function becomes unbounded.
Hence \eqref{eq: Stein's exponent corollary} immediately follows.
Further simplifications are possible if the bound is weakened by ignoring
the expurgated term, i.e., setting $B_{\s[ex]}(R,Q_{X},\tau)=0$ in
\eqref{eq: B}. In this case, since 
\[
(\tau+1)\cdot d_{\tau}(P_{X})=\min_{Q_{Y|X}}\left[\tau\cdot D(Q_{Y|X}||P_{Y|X}|P_{X})+D(Q_{Y|X}||\overline{P}_{Y|X}|P_{X})\right],
\]
{[}see \eqref{eq: divergence charctarization of Chernoff tau}{]},
and since $\tau$ only multiplies positive terms in the objective
functions of $B_{\s[rc]}',B_{\s[rc]}''$ {[}see \eqref{eq: B_rc'}
and \eqref{eq: B_rc''}{]}, it is evident that the supremum in \eqref{eq: Stein's exponent corollary}
is obtained as $\tau\to\infty$. Hence, the supremum and minimum in
\eqref{eq: Stein's exponent corollary} can be interchanged to yield
the bound
\begin{align}
 & E_{2}^{-}(R,0)\nonumber \\
 & \geq D(P_{X}||\overline{P}_{X})+\min\left\{ \lim_{\tau\to\infty}(\tau+1)\cdot d_{\tau}(P_{X}),\;\sup_{Q_{U|X}}\sup_{R_{\s[b]}:\:R_{\s[b]}\geq\left|I_{P_{X}\times Q_{U|X}}(U;X)-R\right|_{+}}\lim_{\tau\to\infty}B_{\s[rc]}(R,R_{\s[b]},P_{X}\times Q_{U|X},\tau)\right\} \\
 & \trre[=,a]\min\left\{ D(P_{XY}||P_{X}\times\overline{P}_{Y|X},\;D(P_{X}||\overline{P}_{X})+\sup_{Q_{U|X}}\sup_{R_{\s[b]}:\:R_{\s[b]}\geq\left|I_{P_{X}\times Q_{U|X}}(U;X)-R\right|_{+}}\lim_{\tau\to\infty}B_{\s[rc]}(R,R_{\s[b]},P_{X}\times Q_{U|X},\tau)\right\} ,\label{eq: Stein's exponent corollary proof only RC}
\end{align}
where $(a)$ follows since 
\begin{align}
\sup_{\tau\geq0}(\tau+1)\cdot d_{\tau}(P_{X}) & =\sup_{\tau\geq0}\min_{Q_{Y|X}}\left[\tau\cdot D(Q_{Y|X}||P_{Y|X}|P_{X})+D(Q_{Y|X}||\overline{P}_{Y|X}|P_{X})\right]\\
 & =\sup_{\tau\geq0}\min_{Q_{Y|X}}\sup_{\tau\geq0}\left[\tau\cdot D(Q_{Y|X}||P_{Y|X}|P_{X})+D(Q_{Y|X}||\overline{P}_{Y|X}|P_{X})\right]\\
 & =D(P_{Y|X}||\overline{P}_{Y|X}|P_{X}).
\end{align}
\end{IEEEproof}

\section{Proof of Theorem \ref{thm: Redcution to CD codes}\label{sec:Proof-of-Theorem-Reduction-To-CD }}

\subsection{Proof of the Achievability Part\label{subsec: Achievability reduction}}

In the course of the proof , we will use subcodes of CD codes, and
would like to claim that the error probabilities of these subcodes
is not significantly different than the code itself. The following
lemma establish such a property. 
\begin{lem}
\label{ lem: expurgating to maximal probabilities}Let ${\cal C}_{n}$
be a CD code, and $\phi_{n}$ be a detector. Then, there exists a
CD code $\tilde{{\cal C}}_{n}$ with $|\tilde{{\cal C}}_{n}|\geq\frac{1}{3}\cdot|{\cal C}_{n}|$
which satisfies the following: For any subcode $\overline{{\cal C}}_{n}\subseteq\tilde{{\cal C}}_{n}$,
there exists a of detector $\overline{\phi}_{n}$ such that
\[
p_{i}({\cal \overline{{\cal C}}}_{n},\overline{\phi}_{n})\leq3\cdot p_{i}({\cal {\cal C}}_{n},\phi_{n})
\]
holds for both $i=1,2$. 
\end{lem}
\begin{IEEEproof}
Note that the error probabilities in \eqref{eq: FA CD} and \eqref{eq: MD CD}
are averaged over the transmitted codeword $X^{n}\in{\cal C}_{n}$.
We first prove that by expurgating enough codewords from a codebook
with good \emph{average} error probabilities, a codebook with \emph{maximal}
(over the codewords) error probabilities can be obtained (for both
types of error). The proof follows the standard expurgation argument
from average error probability to maximal error probability (which
in turn follows from Markov's inequality).\textbf{ }Denoting the conditional
type 1 error probability by\footnote{Note that conditioned on $X^{n}=x^{n}$, $p_{1}({\cal C}_{n},\phi_{n}|X^{n}=x^{n})$
depends on the code ${\cal C}_{n}$ only if the detector $\phi_{n}$
depends on the code. In this lemma, the detector $\phi_{n}$ is arbitrary,
and the use of this notation is therefore just for the sake of consistency.} 
\[
p_{1}\left({\cal C}_{n},\phi_{n}|X^{n}=x^{n}\right)\dfn P\left[\phi_{n}(Y^{n})=\overline{H}\vert X^{n}=x^{n}\right],
\]
we may write 
\begin{align}
p_{1}({\cal {\cal C}}_{n},\phi_{n}) & =\sum_{x^{n}\in{\cal C}_{n}}\P\left(X^{n}=x^{n}\right)\cdot p_{1}\left({\cal C}_{n},\phi_{n}|X^{n}=x^{n}\right)\\
 & =\frac{1}{|{\cal C}_{n}|}\sum_{x^{n}\in{\cal C}_{n}}p_{1}\left({\cal C}_{n},\phi_{n}|X^{n}=x^{n}\right).
\end{align}
Thus, at least $\nicefrac{2}{3}$ of the codewords in $x^{n}\in{\cal C}_{n}$
satisfy 
\begin{equation}
p_{1}\left({\cal C}_{n},\phi_{n}|X^{n}=x^{n}\right)\leq3\cdot p_{1}({\cal {\cal C}}_{n},\phi_{n}).\label{eq: expurgating bad codewords FA}
\end{equation}
Using a similar notation for the conditional type 2 error probability,
and repeating the same argument, we deduce that there exists $\tilde{{\cal C}}_{n}\subset{\cal C}_{n}$
such that $|\tilde{{\cal C}}_{n}|\geq\nicefrac{|{\cal C}_{n}|}{3}$
and both \eqref{eq: expurgating bad codewords FA} as well as
\begin{equation}
p_{2}\left({\cal C}_{n},\phi_{n}|X^{n}=x^{n}\right)\leq3\cdot p_{2}({\cal {\cal C}}_{n},\phi_{n}),\label{eq: expurgating bad codewords MD}
\end{equation}
hold for any $x^{n}\in\tilde{{\cal C}}_{n}$. Let us now consider
any $\overline{{\cal C}}_{n}\subseteq\tilde{{\cal C}}_{n}$ . For
the code $\overline{{\cal C}}_{n}$, the detector $\phi_{n}$ is possibly
suboptimal, and thus might be improved. Using the standard Neyman-Pearson
lemma \cite[Prop. II.D.1]{Poor}, one can find a detector $\overline{\phi}_{n}$
(perhaps randomized) to match any prescribed type 1 error probability
 value, which is optimal in the sense that if any other detector $\hat{\phi}_{n}$
satisfies $p_{1}(\overline{{\cal C}}_{n},\hat{\phi}_{n})\leq p_{1}(\overline{{\cal C}}_{n},\overline{\phi}_{n})$
then $p_{2}(\overline{{\cal C}}_{n},\hat{\phi}_{n})\geq p_{2}(\overline{{\cal C}}_{n},\overline{\phi}_{n})$.
Specifically, let us require that
\begin{equation}
p_{1}(\overline{{\cal C}}_{n},\overline{\phi}_{n})=3\cdot p_{1}(\mathcal{C}_{n},\phi_{n}),\label{eq: avg to max expurgation - FA equality}
\end{equation}
and choose $\hat{\phi}_{n}=\phi_{n}$. Then, as \eqref{eq: expurgating bad codewords FA}
holds for any $x^{n}\in\overline{{\cal C}}_{n}$, 
\begin{align}
p_{1}(\overline{{\cal C}}_{n},\phi_{n}) & =\sum_{x^{n}\in\overline{{\cal C}}_{n}}\P\left(X^{n}=x^{n}\right)\cdot p_{1}\left(\overline{{\cal C}}_{n},\phi_{n}|X^{n}=x^{n}\right)\\
 & \leq3\cdot p_{1}(\mathcal{C}_{n},\phi_{n})\\
 & =p_{1}(\overline{{\cal C}}_{n},\overline{\phi}_{n}),
\end{align}
and as $\overline{\phi}_{n}$ is optimal in the Neyman-Pearson sense,
\eqref{eq: expurgating bad codewords MD} implies that 
\begin{align}
p_{2}(\overline{{\cal C}}_{n},\overline{\phi}_{n}) & \leq p_{2}(\overline{{\cal C}}_{n},\phi_{n})\\
 & =\sum_{x^{n}\in\overline{{\cal C}}_{n}}\P\left(X^{n}=x^{n}\right)\cdot p_{2}\left({\cal C}_{n},\phi_{n}|X^{n}=x^{n}\right)\\
 & \leq3\cdot p_{2}({\cal C}_{n},\phi_{n}).\label{eq: avg to max expurgation - MD equality}
\end{align}
The result follows from \eqref{eq: avg to max expurgation - FA equality}
and \eqref{eq: avg to max expurgation - MD equality}. Note that $\overline{{\cal C}}_{n}=\tilde{\mathcal{C}}_{n}$
is a valid choice. 
\end{IEEEproof}
Next, we focus on encoding a single type class of the source, say
${\cal T}_{n}(Q_{X})$. Given an optimal sequence of CD codes $\{{\cal C}_{n}\}_{n=1}^{\infty}$
for the input type $Q_{X}$, we construct a DHT system which has the
same conditional error probabilities (given that $X^{n}\in{\cal T}_{n}(Q_{X})$)
by permutations of the CD code, as described in Section \ref{sec:Channel-Detection-Codes}.
\begin{lem}
\label{lem: achievability conditional exponents}Let $\delta>0$ and
$Q_{X}\in{\cal P}({\cal X})$ be given, such that $\supp(Q_{X})\subseteq\supp(P_{X})\cap\supp(\overline{P}_{X})$,
and let $\{n_{l}\}$ the subsequence of blocklengths such that ${\cal T}_{n}(Q_{X})$
is not empty. Further, let ${\cal C}$ be a sequence of CD codes of
type $Q_{X}$ and rate $\rho$, and $\{\phi_{n_{l}}\}_{l=1}^{\infty}$
be a sequence of detectors. Then, there exist a sequence of DHT systems
${\cal H}$ of rate $H(Q_{X})-\rho$ such that 
\[
\liminf_{l\to\infty}-\frac{1}{n_{l}}\log p_{i}\left[{\cal H}_{n_{l}}|X^{n_{l}}\in{\cal T}_{n_{l}}(Q_{X})\right]\geq\liminf_{l\to\infty}-\frac{1}{n_{l}}\log p_{i}({\cal C}_{n_{l}},\phi_{n_{l}})-\delta,
\]
holds for both $i=1,2$.
\end{lem}
\begin{IEEEproof}
We only need to focus on $x^{n}\in{\cal T}_{n}(Q_{X})$. For notational
simplicity, let us assume that $n$ is always such that ${\cal T}_{n}(Q_{X})$
is not empty. Let us first extract from ${\cal C}$ the sequence of
CD codes $\tilde{\mathcal{C}}$ whose existence is assured by Lemma
\ref{ lem: expurgating to maximal probabilities}. The rate of $\tilde{\mathcal{C}}$
is chosen to be larger than $\rho-\delta$ (for all sufficiently large
$n$), and for any given codeword, the error probability of each type
is assured to be up to a factor of $3$ of its average error probability. 

Recall the definition of permutation of $\pi(\tilde{{\cal C}}_{n})$
in \eqref{eq: permutation of vector} and \eqref{eq: permutation of set defintion}.
As $\tilde{{\cal C}}_{n}\in{\cal T}_{n}(Q_{X})$, clearly so is $\pi(\tilde{{\cal C}}_{n})\in{\cal T}_{n}(Q_{X})$
for any $\pi$, and thus, there exists a set of permutations $\{\pi_{n,i}\}_{i=i}^{\kappa_{n}}$
such that
\begin{equation}
\bigcup_{i=1}^{\kappa_{n}}\pi_{n,i}(\tilde{{\cal C}}_{n})={\cal T}_{n}(Q_{X}).\label{eq: permutations cover the entire type}
\end{equation}
By a simple counting argument, the minimal number of permutations
required $\kappa_{n}$ is at least $\nicefrac{|{\cal T}_{n}(Q_{X})|}{|\tilde{\mathcal{C}}_{n}|}$.
This is achieved when the permuted sets are pairwise disjoint, i.e.,
$\pi_{n,i}(\tilde{{\cal C}}_{n})\cap\pi_{n,i'}(\tilde{{\cal C}}_{n})=\phi$,
for all $i\neq i'$. While this property is difficult to assure, Ahlswede's
covering lemma \cite[Section 6, Covering Lemma 2]{ahlswede1979coloring}
(see also \cite[Sec. 3, Covering Lemma]{Ahlswede_Csiszar1986}) implies
that up to the first order in the exponent, this minimal number can
be achieved. In particular, there exists a set of permutations $\{\pi_{n,i}^{*}\}_{i=1}^{\kappa_{n}^{*}}$
such that 
\[
\kappa_{n}^{*}\leq\frac{|{\cal T}_{n}(Q_{X})|}{|\tilde{\mathcal{C}}_{n}|}\cdot e^{n\delta}\leq\frac{e^{n[H(Q_{X})+\delta]}}{e^{n(\rho-\delta)}}\cdot e^{n\delta}=e^{n[H(Q_{X})-\rho+3\delta]},
\]
for all $n$ sufficiently large. Without loss of generality (w.l.o.g.),
we assume that $\pi_{n,1}^{*}$ is the identity permutation, and thus
${\cal C}_{n,1}^{*}=\tilde{{\cal C}}_{n}$. Further, for $2\leq i\leq\kappa_{n}^{*}$,
we let 
\[
{\cal C}_{n,i}^{*}\dfn\pi_{n,i}^{*}(\tilde{{\cal C}}_{n})\backslash\left\{ \bigcup_{j=1}^{i-1}\pi_{n,j}^{*}(\tilde{{\cal C}}_{n})\right\} .
\]
 In words, the code ${\cal C}_{n,i}^{*}$ is the permutation $\pi_{n,i}^{*}$
of the code $\tilde{{\cal C}}_{n}$, excluding codewords which belong
to a permutation of $\tilde{{\cal C}}_{n}$ with a smaller index.
Thus, $\{{\cal C}_{n,i}^{*}\}_{i=1}^{\kappa_{n}^{*}}$ forms a disjoint
partition of ${\cal T}_{n}(Q_{X})$. Moreover, Lemma \ref{ lem: expurgating to maximal probabilities}
implies that for any given 
\[
{\cal \overline{{\cal C}}}_{n,i}\dfn[\pi_{n,i}^{*}]^{-1}({\cal C}_{n,i}^{*})\subseteq\tilde{{\cal C}}_{n}
\]
(where $\pi^{-1}$ is the inverse permutation of $\pi$), one can
find a detector $\overline{\phi}_{n,i}$ such that 
\begin{equation}
p_{1}({\cal \overline{{\cal C}}}_{n,i},\overline{\phi}_{n,i})=3\cdot p_{1}({\cal {\cal C}}_{n},\phi_{n}),\label{eq: subset of main bin also have good FA}
\end{equation}
and 
\begin{equation}
p_{2}({\cal \overline{{\cal C}}}_{n,i},\overline{\phi}_{n,i})\leq3\cdot p_{2}({\cal {\cal C}}_{n},\phi_{n}).\label{eq: subset of main bin also have good MD}
\end{equation}
Now, let 
\[
\phi_{n,i}^{*}(y^{n})\dfn\overline{\phi}_{n,i}\left[[\pi_{n,i}^{*}]^{-1}(y^{n})\right].
\]
Since the hypotheses are memoryless, the permutation does not change
the probability distributions. Indeed, for an arbitrary CD code ${\cal C}_{n}'$,
a detector $\phi_{n}'$ and a permutation $\pi$,
\begin{align}
p_{1}({\cal C}_{n}',\phi_{n}') & =\sum_{x^{n}\in{\cal C}_{n}'}\P\left(X^{n}=x^{n}\right)\cdot p_{1}(x^{n},\phi_{n}')\\
 & =\sum_{x^{n}\in{\cal C}_{n}'}\P\left(X^{n}=x^{n}\right)\cdot\sum_{y^{n}:\;\phi_{n}'(y^{n})=\overline{H}}P\left(Y^{n}=y^{n}\vert X^{n}=x^{n}\right)\\
 & =\frac{1}{\left|{\cal C}_{n}'\right|}\sum_{x^{n}\in{\cal C}_{n}'}\sum_{y^{n}:\;\phi_{n}'(y^{n})=\overline{H}}P\left(Y^{n}=y^{n}\vert X^{n}=x^{n}\right)\\
 & =\frac{1}{\left|{\cal C}_{n}'\right|}\sum_{x^{n}\in{\cal C}_{n}'}\sum_{y^{n}:\;\phi_{n}'(y^{n})=\overline{H}}P\left[Y^{n}=\pi(y^{n})\vert X^{n}=\pi(x^{n})\right]\\
 & =\frac{1}{\left|{\cal C}_{n}'\right|}\sum_{x^{n}\in\pi^{-1}({\cal C}_{n}')}\sum_{y^{n}:\;\phi_{n}'[\pi^{-1}(y^{n})]=\overline{H}}P\left(Y^{n}=y^{n}\vert X^{n}=x^{n}\right)\\
 & =p_{1}\left(\pi^{-1}({\cal C}_{n}'),\phi_{n,\pi}'\right)
\end{align}
where $\phi_{n,\pi}'(y^{n})\dfn\phi_{n}'[\pi^{-1}(y^{n})].$ Hence,
\begin{equation}
p_{1}({\cal \overline{{\cal C}}}_{n,i},\overline{\phi}_{n,i})=p_{1}({\cal C}_{n,i}^{*},\phi_{n,i}^{*}),\label{eq: permutation does not change FA proability}
\end{equation}
and 
\begin{equation}
p_{2}({\cal \overline{{\cal C}}}_{n,i},\overline{\phi}_{n,i})=p_{2}({\cal C}_{n,i}^{*},\phi_{n,i}^{*}).\label{eq: permutation does not change MD proability}
\end{equation}

We thus construct $\mathcal{H}_{n}=(f_{n},\varphi_{n})$ as follows.
The codes $\{{\cal C}_{n,i}^{*}\}_{i=1}^{\kappa_{n}^{*}}$ will serve
as the bins of $f_{n}$, and detectors $\{\phi_{n,i}^{*}\}_{i=1}^{\kappa_{n}^{*}}$
as the decision function, given that the bin index is $i$. As said
above, only $x^{n}\in{\cal T}_{n}(Q_{X})$ will be encoded. More rigorously,
the encoding of $x^{n}\in{\cal T}_{n}(Q_{X})$ is given by $f_{n}(x^{n})=i$
whenever $x^{n}\in{\cal C}_{n,i}^{*}$, and by $f_{n}(x^{n})=0$ whenever
$x^{n}\not\in{\cal T}_{n}(Q_{X})$. Clearly, the rate of the code
is less than 
\[
\frac{1}{n}\log\kappa_{n}^{*}\leq H(Q_{X})-\rho+3\delta,
\]
for all $n$ sufficiently large. The detector $\varphi_{n}$ is given
by $\varphi_{n}(i,y^{n})=\phi_{n,i}^{*}(y^{n})$. The conditional
type 1 error probability of this DHT system is given by
\begin{align}
P\left[\varphi_{n}(f_{n}(X^{n}),Y^{n})=\overline{H}\vert X^{n}\in{\cal T}_{n}(Q_{X})\right] & =\sum_{i=1}^{\kappa_{n}^{*}}\P\left[f_{n}(X^{n})=i\right]\cdot P\left[\varphi_{n}(f_{n}(X^{n}),Y^{n})=\overline{H}\vert f_{n}(X^{n})=i\right]\\
 & =\sum_{i=1}^{\kappa_{n}^{*}}\P\left[f_{n}(X^{n})=i\right]\cdot P\left[\phi_{n,i}^{*}(Y^{n})=\overline{H}\vert f_{n}(X^{n})=i\right]\\
 & \trre[=,a]\sum_{i=1}^{\kappa_{n}^{*}}\P\left[f_{n}(X^{n})=i\right]\cdot p_{1}({\cal C}_{n,i}^{*},\phi_{n,i}^{*})\\
 & \trre[=,b]\sum_{i=1}^{\kappa_{n}^{*}}\P\left[f_{n}(X^{n})=i\right]\cdot p_{1}({\cal \overline{{\cal C}}}_{n,i},\overline{\phi}_{n,i})\\
 & \trre[\leq,c]3\cdot p_{1}({\cal {\cal C}}_{n},\phi_{n}),
\end{align}
where $(a)$ follows because given $f_{n}(X^{n})=i$ the source vector
$X^{n}$ is distributed uniformly over ${\cal C}_{n,i}^{*}$, $(b)$
follows from \eqref{eq: permutation does not change FA proability},
and $(c)$ follows from \eqref{eq: subset of main bin also have good FA}.
Similarly, the conditional type 2 error probability is upper bounded
as
\[
\overline{P}\left[\varphi_{n}(f_{n}(X^{n}),Y^{n})=H\vert X^{n}\in{\cal T}_{n}(Q_{X})\right]\leq3\cdot p_{2}({\cal {\cal C}}_{n},\phi_{n}).
\]
The factor $3$ in the error probabilities is negligible asymptotically. 
\end{IEEEproof}
The DHT system constructed in Lemma \ref{lem: achievability conditional exponents}
achieves asymptotically optimal error probabilities, only conditional
on $X^{n}\in{\cal T}_{n}(Q_{X})$, for a \emph{single} $Q_{X}\in{\cal P}({\cal X})$.
To construct a DHT system which achieves \emph{unconditional} asymptotically
optimal error probabilities, one can, in principle, construct a different
DHT subsystem ${\cal H}_{n,Q_{X}}$ for \emph{any} $Q_{X}\in{\cal P}({\cal X})$.
Then, the encoder will choose the appropriate system according to
the type of $x^{n}$, and then inform the detector of the actual system
utilized by a short header (for which the required rate is negligible
since the number of types only increases polynomially). However, as
clearly $|{\cal P}_{n}({\cal X})|\to\infty$ as $n\to\infty$, such
a method might fail since the convergence of the error probabilities
to their exponential bounds, may depend on the type. For example,
let $\{Q_{X}^{(n)}\}$ be a sequence of types which satisfies $Q_{X}^{(n)}\in{\cal P}_{n}({\cal X})$\emph{
}and \emph{$Q_{X}^{(n)}\notin{\cal P}_{n'}({\cal X})$ }for all $n'<n$\emph{.
}A priori, it might be that the error probabilities of ${\cal H}_{n,Q_{X}^{(n)}}$
are far from their asymptotic values, for all $n$\emph{. }In other
words, uniform convergence of the error probabilities to their asymptotic
exponential bounds is required. 

We solve this problem (see also \cite{Weinberger15,Weinberger17_secrecy})
by defining a \emph{finite} \emph{grid} of types ${\cal P}_{n_{0}}(\tilde{{\cal X}})$
for a fixed $n_{0}$, and construct DHT subsystems only for $Q_{X}\in{\cal P}_{n_{0}}(\tilde{{\cal X}})$,
where $\tilde{{\cal X}}\dfn\supp(P_{X})\cap\supp(\overline{P}_{X})$.
As $|{\cal P}_{n_{0}}(\tilde{{\cal X}})|<\infty$ uniform convergence
of the error probabilities of ${\cal H}_{n,Q_{X}}$ for $Q_{X}\in{\cal P}_{n_{0}}(\tilde{{\cal X}})$
is assured. Now, if the type of $x^{n}$ belongs to ${\cal P}_{n_{0}}(\tilde{{\cal X}})$,
it can be encoded using ${\cal H}_{n,Q_{X}}$. Otherwise, $x^{n}$
is slightly modified to a different vector $\tilde{X}^{n}$, where
the type of the latter does belong to ${\cal P}_{n_{0}}(\tilde{{\cal X}})$.
Then, $\tilde{X}^{n}$ is encoded using the DHT subsystem which pertain
to its type. Since the DHT subsystems are designed for $(X^{n},Y^{n})$,
rather than for $(\tilde{X}^{n},Y^{n})$, the side-information vector
$Y^{n}$ is also modified to a vector $\tilde{Y}^{n}$, using additional
information sent from the encoder. To analyze the effect of this modification
on the error probabilities, we will need the following \emph{partial}
\emph{mismatch }lemma.
\begin{lem}
\label{lem: mismatch lemma}Let ${\cal C}_{n}$ be a CD code, and
$\phi_{n}$ a detector. Also fix $\tilde{P}_{Y|X}$ which satisfies
both $\tilde{P}_{Y|X}\gg P_{Y|X}$ and $\tilde{P}_{Y|X}\gg\overline{P}_{Y|X}$
(for example, $\tilde{P}_{Y|X}=\frac{1}{2}\overline{P}_{Y|X}+\frac{1}{2}P_{Y|X}$)
and let 
\[
\Omega\dfn\max_{x\in{\cal X},\;y\in{\cal Y}}\left|\log\frac{P_{Y|X}(y|x)}{\tilde{P}_{Y|X}(y|x)}\right|<\infty
\]
and 
\[
\overline{\Omega}\dfn\max_{x\in{\cal X},\;y\in{\cal Y}}\left|\log\frac{\overline{P}_{Y|X}(y|x)}{\tilde{P}_{Y|X}(y|x)}\right|<\infty.
\]
Further, for $d\dfn\delta n\in[n]$, assume that $\tilde{Y}^{n}=(\tilde{Y}_{1}^{d},\tilde{Y}_{d+1}^{n})$
is drawn as follows: Given $x^{n}\in{\cal C}_{n}$, $\tilde{Y}_{1}^{d}\sim\tilde{P}_{Y|X}(\cdot|x_{1}^{d})$
under both hypotheses, $Y_{d+1}^{n}\sim P_{Y|X}(\cdot|x^{n})$ under
the hypothesis $H$, and $\tilde{Y}_{d+1}^{n}\sim\overline{P}_{Y|X}(\cdot|x_{d+1}^{n})$
under the hypothesis $\overline{H}$. Then, 
\[
P\left[\phi_{n}(\tilde{Y}^{n})=\overline{H}\right]\leq e^{n\Omega\delta}p_{1}({\cal {\cal C}}_{n},\phi_{n}),
\]
and 
\[
\overline{P}\left[\phi_{n}(\tilde{Y}^{n})=H\right]\leq e^{n\overline{\Omega}\delta}\cdot p_{2}({\cal {\cal C}}_{n},\phi_{n}).
\]
\end{lem}
\begin{IEEEproof}
We will show that the ``wrong'' distribution of $Y^{n}$ in the
first $d$ coordinates does not change likelihoods and probabilities
significantly. Indeed, for the type 1 error probability, conditioning
on $\tilde{Y}_{1}^{d}=y_{1}^{d}$
\[
P\left[\phi_{n}(\tilde{Y}^{n})=\overline{H}\vert\tilde{Y}_{1}^{d}=y_{1}^{d}\right]=P\left[\phi_{n}(Y^{n})=\overline{H}\vert Y_{1}^{d}=y_{1}^{d}\right].
\]
Then, since, 
\begin{align}
P\left(\tilde{Y}_{1}^{d}=y_{1}^{d}\right) & =\frac{1}{|{\cal C}_{n}|}\sum_{x^{n}\in{\cal C}_{n}}\tilde{P}_{Y|X}(y_{1}^{d}|x_{1}^{d})\\
 & \leq e^{d\Omega}\cdot\frac{1}{|{\cal C}_{n}|}\sum_{x^{n}\in{\cal C}_{n}}P_{Y|X}(y_{1}^{d}|x_{1}^{d})\\
 & =e^{d\Omega}\cdot P\left(Y_{1}^{d}=y_{1}^{d}\right),
\end{align}
we obtain
\begin{align}
P\left[\phi_{n}(\tilde{Y}^{n})=\overline{H}\right] & =\sum_{y_{1}^{d}\in{\cal Y}^{d}}P\left(\tilde{Y}_{1}^{d}=y_{1}^{d}\right)\cdot P\left[\phi_{n}(\tilde{Y}^{n})=\overline{H}\vert\tilde{Y}_{1}^{d}=y_{1}^{d}\right]\\
 & \leq\sum_{y_{1}^{d}\in{\cal Y}^{d}}e^{d\Omega}\cdot P\left(Y_{1}^{d}=y_{1}^{d}\right)\cdot P\left[\phi_{n}(Y^{n})=\overline{H}\vert Y_{1}^{d}=y_{1}^{d}\right]\\
 & =e^{d\Omega}\cdot p_{1}({\cal {\cal C}}_{n},\phi_{n}).
\end{align}
The statement regarding the type 2 error probability is similar.
\end{IEEEproof}
We will also use the following lemma whose simple proof is omitted.
\begin{lem}
\label{lem: hamming distance close types}Let $Q_{X},\tilde{Q}{}_{X}\in{\cal P}_{n}({\cal X})$
and assume that $||Q_{X}-\tilde{Q}{}_{X}||=\frac{2d}{n}$ where $d>0$.
If $x^{n}\in{\cal T}_{n}(Q_{X})$ then 
\[
\min_{\tilde{x}^{n}\in{\cal T}_{n}(\tilde{Q}{}_{X})}d_{\s[H]}(\tilde{x}^{n},x^{n})\leq d.
\]
\end{lem}
We are now ready to prove the achievability part.
\begin{IEEEproof}[Proof of the achievability part of the Theorem \ref{thm: Redcution to CD codes}]
 We will describe the construction of the sequence of DHT systems.
Then we will describe the encoder and show that satisfies the rate
constraint. Finally, we will describe the detector and show that it
satisfies the type 1 error exponent constraint, and prove that the
achieved type 2 error exponent is good as the bound stated in the
theorem.

\emph{\uline{Construction of a sequence of DHT systems:}}
\begin{enumerate}
\item Choose a finite \emph{grid} of types: Given $\epsilon>0$ (to be specified
later), choose $n_{0}\in\mathbb{N}$ such that $\Phi_{\epsilon}(Q_{X})\leq\frac{\epsilon}{2}$
for any $Q_{X}\in{\cal P}(\tilde{{\cal X}})$, where\footnote{If the minimizer is not unique, one of the minimizers can be arbitrarily
and consistently chosen.}
\[
\Phi_{\epsilon}(Q_{X})\dfn\argmin_{\tilde{Q}_{X}\in{\cal P}_{n_{0}}(\tilde{{\cal X}})}||Q_{X}-\tilde{Q}_{X}||.
\]
\item Let $\delta>0$ be given. For any $Q_{X}\in{\cal P}_{n_{0}}(\tilde{{\cal X}})$,
construct the optimal CD code ${\cal C}_{n,Q_{X}}^{*}$ of rate $\rho=H(Q_{X})-R$,
and its optimal detector $\phi_{n,Q_{X}}^{*}$ such that 
\[
p_{1}({\cal C}_{n,Q_{X}}^{*},\phi_{n}^{*})\leq\exp[-n\cdot F_{1}]
\]
where $F_{1}=E_{1}-D(Q_{X}||P_{X})$. By definition of the CD reliability
function, there exists $n_{1}(Q_{X},\delta)$ such that for all $n>n_{1}(Q_{X},\delta)$
\[
p_{2}({\cal C}_{n,Q_{X}}^{*},\phi_{n}^{*})\leq\exp\left[-n\cdot\left(F_{2}^{-}(R,Q{}_{X},F_{1})-\delta/2\right)\right].
\]
\item For any $Q_{X}\in{\cal P}_{n_{0}}(\tilde{{\cal X}})$, construct a
DHT subsystem ${\cal H}_{n,Q_{X}}=(f_{n,Q_{X}},\varphi_{n,Q_{X}})$
such that 
\begin{equation}
p_{1}(f_{n,Q_{X}},\varphi_{n,Q_{X}})\leq\exp\left[-n\cdot\left(E_{1}-D(Q_{X}||P_{X})-\delta\right)\right],\label{eq: uniform convergence of FA probability to exponent over grid}
\end{equation}
and
\begin{equation}
p_{2}(f_{n,Q_{X}},\varphi_{n,Q_{X}})\leq\exp\left\{ -n\cdot\left[F_{2}^{-}(R,Q_{X},E_{1}-D(Q_{X}||P_{X}))-\delta\right]\right\} .\label{eq: uniform convergence of MD probability to exponent over grid}
\end{equation}
for \emph{all} $n>n_{2}(Q_{X},\delta)$. The existence of such construction
is assured by Lemma \ref{lem: achievability conditional exponents},
using the CD codes ${\cal C}_{n,Q_{X}}^{*}$. 
\end{enumerate}
\emph{\uline{The encoder operation and rate analysis:}} Upon observing
$X^{n}$ of type $Q_{X}$, the encoder: 
\begin{enumerate}
\item Sends the detector a description of $Q_{X}$. As $|{\cal P}_{n}({\cal X})|\leq(n+1)^{|{\cal X}|}$,
this description requires no more than $\lceil|{\cal X}|\cdot\log(n+1)\rceil$
nats. If $Q_{X}\notin{\cal P}_{n}(\tilde{{\cal X}})$ then no additional
bits are sent (otherwise further bits are sent as follows).
\item Finds $\Phi_{\epsilon}(Q_{X})$ and generates $\tilde{X}^{n}\in{\cal T}_{n}(\Phi_{\epsilon}(Q_{X}))$
with a uniform distribution over the set
\[
\left\{ \tilde{x}^{n}\in{\cal T}_{n}(\Phi_{\epsilon}(Q_{X}))\colon d_{\s[H]}(x^{n},\tilde{x}^{n})=\frac{n\epsilon}{4}\right\} .
\]
Note that Lemma \ref{lem: hamming distance close types} assures that
this set is not empty, and that if $X^{n}$ is distributed uniformly
over ${\cal T}_{n}(Q_{X})$ then $\tilde{X}^{n}$ is distributed uniformly
over ${\cal T}_{n}(\Phi_{\epsilon}(Q_{X}))$ (due to the permutation
symmetry of type classes). 
\item Sends to the detector a description of the set ${\cal I}(x^{n},\tilde{x}^{n})\dfn\{i\colon x_{i}\neq\tilde{x}_{i}\}$.
Since $|{\cal I}(x^{n},\tilde{x}^{n})|=\frac{n\epsilon}{4}$, the
number possible sets is less than
\[
{n \choose \frac{n\epsilon}{4}}\leq e^{nh_{\s[b]}(\frac{\epsilon}{4})},
\]
and so sending its description requires no more than $nh_{\s[b]}(\frac{\epsilon}{4})$
nats. 
\item Sends the detector the value of $\tilde{x}^{n}$ for $i\in{\cal I}(x^{n},\tilde{x}^{n})$.
Each letter can be encoded using $\lceil\log|{\cal X}|\rceil$ nats,
and so this requires no more than $\frac{n\epsilon}{4}\lceil\log|{\cal X}|\rceil$
nats.
\item Sends the message index $\tilde{i}\dfn f_{n,\Phi_{\epsilon}(Q_{X})}(\tilde{X}^{n})$
to the detector. This requires $nR$ nats. 
\end{enumerate}
The required rate is therefore no more than 

\[
\frac{1}{n}\log\lceil|{\cal X}|\cdot\log(n+1)\rceil+h_{\s[b]}\left(\frac{\epsilon}{4}\right)+\frac{\epsilon}{4}\lceil\log|{\cal X}|\rceil+R.
\]
By choosing $\epsilon>0$ sufficiently small, the required rate can
be made less than $R+\delta$ for all $n$ sufficiently large.

\emph{\uline{The detector operation and error probability analysis:}}
Upon receiving the message of the encoder and observing $Y^{n}$ the
detector:
\begin{enumerate}
\item Decodes $Q_{X}$ the type of $x^{n}$. If $Q_{X}\notin{\cal P}_{n}(\tilde{{\cal X}})$
then it decides on the hypothesis based on $Q_{X}$. Otherwise it
continuous.
\item Finds $\Phi_{\epsilon}(Q_{X})$ and generates $\tilde{Y}^{n}$ as
follows: For all $i\in[n]$, if $i\not\in{\cal I}(x^{n},\tilde{x}^{n})$
then $\tilde{Y}_{i}=Y_{i}$, and if $i\in{\cal I}(x^{n},\tilde{x}^{n})$
then $\tilde{Y}_{i}\sim\tilde{P}_{Y|X}(\cdot|\tilde{x}_{i})$, where
$\tilde{P}_{Y|X}$ is as chosen in Lemma \ref{lem: mismatch lemma}. 
\item Decides on the hypothesis as $\varphi_{n,\Phi_{\epsilon}(Q_{X})}(\tilde{i},\tilde{Y}^{n})$. 
\end{enumerate}
Note that the encoder alters $x^{n}$ to $\tilde{x}^{n}$ such that
$\tilde{x}^{n}$ has a type which matches one of the subsystems ${\cal H}_{n,Q_{X}}$,
$Q_{X}\in{\cal P}_{n_{0}}(\tilde{{\cal X}})$. Due to this modification,
a similar change is made to the side-information vector. The detector
generates a proper $\tilde{Y}^{n}$ by using the information sent
from the encoder (namely, ${\cal I}(x^{n},\tilde{x}^{n})$ and the
values of $\tilde{x}^{n}$ on this set). However, $\tilde{Y}_{i}\sim\tilde{P}_{Y|X}(\cdot|\tilde{x}{}_{i})$
for $i\in{\cal I}(x^{n},\tilde{x}^{n})$ rather than according to
the true distribution ($P$ or $\overline{P}$) . As we shall see,
Lemma \ref{lem: mismatch lemma} assures that this mismatch has small
effect on the error probabilities. Let us denote the constructed system
by ${\cal H}_{n}=(f_{n},\varphi_{n})$, and analyze the error probabilities
for for all $n>\max\{n_{0},\;\max_{Q_{X}\in{\cal P}_{n_{0}}(\tilde{{\cal X}})}n_{2}(Q_{X},\delta)\}$. 

For the type 1 error exponent, note that 
\begin{align}
p_{1}({\cal H}_{n}) & \trre[=,a]\sum_{Q_{X}\in{\cal P}_{n}(\tilde{{\cal X}})}\P\left[X^{n}\in{\cal T}_{n}(Q_{X})\right]\cdot P\left[\varphi_{n}(Y^{n})=\overline{H}\vert X^{n}\in{\cal T}_{n}(Q_{X})\right]\\
 & \trre[\leq,b]e^{n\delta}\cdot\max_{Q_{X}\in{\cal P}_{n}(\tilde{{\cal X}})}e^{-nD(Q_{X}||P_{X})}\cdot P\left[\varphi_{n}(Y^{n})=\overline{H}\vert X^{n}\in{\cal T}_{n}(Q_{X})\right]\\
 & \trre[=,c]e^{n\delta}\cdot\max_{Q_{X}\in{\cal P}_{n}(\tilde{{\cal X}})}e^{-nD(Q_{X}||P_{X})}\cdot P\left[\varphi_{n,\Phi_{\epsilon}(Q{}_{X})}(\tilde{Y}^{n})=\overline{H}\vert\tilde{X}^{n}\in{\cal T}_{n}(\Phi_{\epsilon}(Q_{X}))\right]\\
 & \trre[=,d]e^{n\delta}\cdot e^{n\Omega\epsilon/4}\cdot\max_{Q_{X}\in{\cal P}_{n}(\tilde{{\cal X}})}e^{-nD(Q_{X}||P_{X})}\cdot P\left[\varphi_{n,\Phi_{\epsilon}(Q{}_{X})}(Y^{n})=\overline{H}\vert X^{n}\in{\cal T}_{n}(\Phi_{\epsilon}(Q_{X}))\right]\\
 & \trre[\leq,e]\exp\left\{ -n\cdot\min_{Q_{X}\in{\cal P}_{n}(\tilde{{\cal X}})}\left[D(Q_{X}||P_{X})+E_{1}-D\left(\Phi_{\epsilon}(Q{}_{X})||P_{X}\right)-\delta-\frac{\Omega\epsilon}{4}\right]\right\} \\
 & \trre[\leq,f]\exp\left[-n\cdot\left(E_{1}-2\delta-\frac{\Omega\epsilon}{4}\right)\right],
\end{align}
where $(a)$ follows since if $Q_{X}\notin{\cal P}_{n}(\tilde{{\cal X}})$
the detector can decide on the hypothesis with zero error, $(b)$
follows since $|{\cal P}_{n}({\cal X})|\leq(n+1)^{|{\cal X}|}\leq e^{n\delta}$
and $\P[X^{n}\in{\cal T}_{n}(Q_{X})]\leq\exp[-n\cdot D(Q_{X}||P_{X})]$,
$(c)$ follows from the definition of the system ${\cal H}_{n}$,
$(d)$ follows from Lemma \ref{lem: mismatch lemma}, $(e)$ follows
from \eqref{eq: uniform convergence of FA probability to exponent over grid}
and as $n>n_{2}(Q_{X},\delta)$ for all $Q_{X}\in{\cal P}_{n}(\tilde{{\cal X}})$,
and $(f)$ follows from the fact that $D(Q_{X}||P_{X})$ is a continuous
function of $Q_{X}$ in ${\cal S}(\tilde{{\cal X}})$, and thus uniformly
continuous. 

For the type 2 error exponent, first note that, as for the type 1
error probability, 
\begin{align}
 & p_{2}({\cal H}_{n})\nonumber \\
 & =\sum_{Q_{X}\in{\cal P}_{n}(\tilde{{\cal X}})}\P\left[X^{n}\in{\cal T}_{n}(Q_{X})\right]\cdot\overline{P}\left[\varphi_{n}(Y^{n})=H\vert X^{n}\in{\cal T}_{n}(Q_{X})\right]\\
 & \leq\exp\left(-n\cdot\min_{Q_{X}\in{\cal P}_{n}(\tilde{{\cal X}})}\left\{ D(Q_{X}||P_{X})-\frac{1}{n}\log\overline{P}\left[\varphi_{n,\Phi_{\epsilon}(Q{}_{X})}(\tilde{Y}^{n})=H\vert\tilde{X}^{n}\in{\cal T}_{n}(\Phi_{\epsilon}(Q_{X}))\right]-\delta\right\} \right).
\end{align}
Now, 
\begin{align}
 & \liminf_{n\to\infty}-\frac{1}{n}\log p_{2}({\cal H}_{n})\nonumber \\
 & =\liminf_{n\to\infty}\min_{Q_{X}\in{\cal P}_{n}(\tilde{{\cal X}})}\left\{ D(Q_{X}||P_{X})-\frac{1}{n}\log\overline{P}\left[\varphi_{n,\Phi_{\epsilon}(Q{}_{X})}(\tilde{Y}^{n})=H\vert\tilde{X}^{n}\in{\cal T}_{n}(\Phi_{\epsilon}(Q_{X}))-\delta\right]\right\} \\
 & \trre[\geq,a]\liminf_{n\to\infty}\min_{Q_{X}\in{\cal P}_{n}(\tilde{{\cal X}})}\left\{ D(Q_{X}||P_{X})-\frac{1}{n}\log p_{2}\left[{\cal H}_{n,\Phi_{\epsilon}(Q_{X})}|X^{n}\in{\cal T}_{n}(\Phi_{\epsilon}(Q_{X}))\right]-\delta-\frac{\epsilon\overline{\Omega}}{4}\right\} \\
 & \trre[\geq,b]\liminf_{n\to\infty}\min_{Q_{X}\in{\cal P}_{n}(\tilde{{\cal X}})}\left\{ D(Q_{X}||P_{X})+F_{2}^{-}\left[H(\Phi_{\epsilon}(Q_{X}))-R,\Phi_{\epsilon}(Q_{X}),E_{1}-D(Q_{X}||P_{X})\right]\right\} \nonumber \\
 & \hphantom{=}-\frac{\epsilon\overline{\Omega}}{4}-2\delta\\
 & \trre[\geq,c]\liminf_{n\to\infty}\min_{Q_{X}\in{\cal P}_{n}(\tilde{{\cal X}})}\left\{ D\left(\Phi_{\epsilon}(Q_{X})||P_{X}\right)+F_{2}^{-}\left[H(\Phi_{\epsilon}(Q_{X}))-R,\Phi_{\epsilon}(Q_{X}),E_{1}-D\left(\Phi_{\epsilon}(Q_{X})||P_{X}\right)+\delta_{1}\right]\right\} \nonumber \\
 & \hphantom{=}-2\delta-\frac{\epsilon\overline{\Omega}}{4}-\delta_{1}\\
 & =\min_{Q_{X}\in{\cal P}_{n_{0}}(\tilde{{\cal X}})}\left\{ D(Q_{X}||P_{X})+F_{2}^{-}\left[H(Q_{X})-R,Q_{X},E_{1}-D(Q_{X}||P_{X})+\delta_{1}\right]-2\delta-\frac{\epsilon\overline{\Omega}}{4}-\delta_{1}\right\} \\
 & \geq\inf_{Q_{X}\in{\cal P}_{n}(\tilde{{\cal X}})}\Biggr\{ D\left(Q_{X}||P_{X}\right)+F_{2}^{-}\left[H(Q_{X})-R,Q_{X},E_{1}-D(Q_{X}||P_{X})+\delta_{1}\right]-2\delta-\frac{\epsilon\overline{\Omega}}{4}-\delta_{1}\Biggr\}\\
 & \geq\inf_{Q_{X}\in{\cal P}_{n}({\cal X})}\Biggr\{ D\left(Q_{X}||P_{X}\right)+F_{2}^{-}\left[H(Q_{X})-R,Q_{X},E_{1}-D(Q_{X}||P_{X})+\delta_{1}\right]-2\delta-\frac{\epsilon\overline{\Omega}}{4}-\delta_{1}\Biggr\}
\end{align}
where $(a)$ follows from Lemma \ref{lem: mismatch lemma} (and the
way $\tilde{Y}^{n}$ was generated), $(b)$ follows from \eqref{eq: uniform convergence of MD probability to exponent over grid}
and as $n>n_{2}(Q_{X},\delta)$ for all $Q_{X}\in{\cal P}_{n}(\tilde{{\cal X}})$.
Passage $(c)$ holds for some $\delta_{1}>0$ that satisfies $\delta_{1}\downarrow0$
as $\epsilon\downarrow0$, and follows from the fact that $D(Q_{X}||P_{X})$
is a continuous function of $Q_{X}$ over the compact set ${\cal S}(\tilde{{\cal X}})$,
and thus uniformly continuous. Finally, by choosing $\epsilon>0$
sufficiently small, and then $\delta>0$ sufficiently small, the loss
in exponents can be made arbitrarily small.
\end{IEEEproof}

\subsection{Proof of the Converse Part \label{subsec: Converse reduction}}

The proof of the converse part is based upon identifying for any sequence
of DHT systems $\{{\cal H}_{n}\}_{n=1}^{\infty}$ a sequence of bin
indices $i_{n}$ such that the size of $|f_{n}^{-1}(i_{n})|$ is ``typical''
to ${\cal H}_{n}$, and such that the conditional error probability
given $f_{n}(X^{n})=i_{n}$ is also ``typical'' to ${\cal H}_{n}$.
The sequence of bins $f_{n}^{-1}(i_{n})$ corresponds to a sequence
of CD codes, and thus clearly cannot have better exponents than the
ones dictated by the reliability function of CD codes. This restriction
is then translated back to bound the reliability of DHT systems.
\begin{IEEEproof}[Proof of the converse part of Theorem \ref{thm: Redcution to CD codes}]
 For a given DHT system ${\cal H}_{n}$, let us denote the (random)
bin index by $I_{n}\dfn f_{n}(X^{n})$. We will show that the converse
hold even if the detector of the DHT systems ${\cal H}_{n}$ is aware
of the type of $x^{n}$. Consequently, as an optimal Neyman-Pearson
detector will only average the likelihoods of source vectors from
the true type class, it can be assumed w.l.o.g. that each bin contains
only sequences from a unique type class.

Recall that $m_{n}$ is the number of possible bins, and let $m_{n,Q_{X}}$
be the number of bins associated with a specific type class $Q_{X}$,
i.e., $m_{n,Q_{X}}\dfn|{\cal M}_{n,Q_{X}}|$ where 
\[
{\cal M}_{n,Q_{X}}\dfn\left\{ i\in[m_{n}]\colon f_{n}^{-1}(i)\cap{\cal T}_{n}(Q_{X})=f_{n}^{-1}(i)\right\} .
\]
Further, conditioned on the type class $Q_{X}$ (note that $I_{n}$
is a function of $X^{n}$),
\begin{align}
\mu_{Q_{X}} & \dfn\E\left[\left|f_{n}^{-1}(I_{n})\right|^{-1}\vert X^{n}\in{\cal T}_{n}(Q_{X})\right]\\
 & =\sum_{i\in{\cal M}_{n,Q_{X}}}\left|f_{n}^{-1}(i)\right|^{-1}\cdot\P\left[I_{n}=i\vert X^{n}\in{\cal T}_{n}(Q_{X})\right]\\
 & =\sum_{i\in{\cal M}_{n,Q_{X}}}\left|f_{n}^{-1}(i)\right|^{-1}\cdot\frac{\left|f_{n}^{-1}(i)\right|}{\left|{\cal T}_{n}(Q_{X})\right|}\\
 & =\frac{m_{n,Q_{X}}}{\left|{\cal T}_{n}(Q_{X})\right|}\\
 & \leq\frac{e^{n(R+\delta)}}{e^{n[H(Q_{X})-\delta]}},\label{eq: expected value of inverse bin size}
\end{align}
as for all $n$ sufficiently large, $m_{n}\leq e^{n(R+\delta)}$,
and thus clearly $m_{n,Q_{X}}\leq e^{n(R+\delta)}$. Hence, for any
$\gamma>1$, Markov's inequality implies
\begin{align}
\P\left[I_{n}:\left|f_{n}^{-1}(I_{n})\right|\geq\frac{1}{\gamma\cdot\mu_{Q_{X}}}\Bigg|X^{n}\in{\cal T}_{n}(Q_{X})\right] & =\P\left[I_{n}:\;\left|f_{n}^{-1}(I_{n})\right|^{-1}\leq\gamma\cdot\mu_{Q_{X}}\Big|X^{n}\in{\cal T}_{n}(Q_{X})\right]\\
 & \geq1-\frac{1}{\gamma}.
\end{align}
Thus, using \eqref{eq: expected value of inverse bin size}, conditioned
on $X^{n}\in{\cal T}_{n}(Q_{X})$ 
\begin{equation}
\left|f_{n}^{-1}(I_{n})\right|\geq\frac{1}{\gamma\cdot\mu_{Q_{X}}}\geq\frac{1}{\gamma}\cdot e^{n\cdot[H(Q_{X})-R-2\delta]},\label{eq: large bin event}
\end{equation}
with probability larger than $1-\frac{1}{\gamma}>0$. Now, assume
by contradiction that the statement of the theorem does not hold.
This implies that there exists an increasing subsequence of blocklengths
$\{n_{k}\}_{k=1}^{\infty}$ and $\delta>0$ such that 
\[
p_{1}({\cal H}_{n_{k}})\leq\exp\left\{ -n_{k}\cdot\left[E_{1}-\delta\right]\right\} ,
\]
and 
\[
p_{2}({\cal H}_{k})\leq\exp\left\{ -n_{k}\cdot\left[E_{2}^{+}(R-3\delta,E_{1}-3\delta)+5\delta\right]\right\} ,
\]
for all $k$ sufficiently large. For brevity of notation, we assume
w.l.o.g. that these bounds hold for all $n$ sufficiently large, and
thus omit the subscript $k$. Let $\delta>0$ be given. Then, for
all $n$ sufficiently large {[}which only depends on $(\delta,|{\cal X}|)${]},
\begin{equation}
p_{1}\left[{\cal H}_{n}\vert X^{n}\in{\cal T}_{n}(Q_{X})\right]\leq\exp\left\{ -n\cdot\left[E_{1}-D(Q_{X}||P_{X})-2\delta\right]\right\} ,\label{eq: conditional type 1 prob upper bound}
\end{equation}
for all $Q_{X}$ such that ${\cal T}_{n}(Q_{X})$ is not empty. Indeed,
for all $n$ sufficiently large, it holds that
\begin{align}
\exp\left[-n\cdot(E_{1}-\delta)\right] & \geq p_{1}({\cal H}_{n})\\
 & =\sum_{Q_{X}\in{\cal P}_{n}({\cal X})}\P\left[X^{n}\in{\cal T}_{n}(Q_{X})\right]\cdot p_{1}\left[{\cal H}_{n}\vert X^{n}\in{\cal T}_{n}(Q_{X})\right]\\
 & \geq\sum_{Q_{X}\in{\cal P}_{n}({\cal X})}e^{-n\cdot\left[D(Q_{X}||P_{X})+\delta\right]}\cdot p_{1}\left[{\cal H}_{n}\vert X^{n}\in{\cal T}_{n}(Q_{X})\right]\\
 & \geq\max_{Q_{X}\in{\cal P}_{n}({\cal X})}e^{-n\cdot\left[D(Q_{X}||P_{X})+\delta\right]}\cdot p_{1}\left[{\cal H}_{n}\vert X^{n}\in{\cal T}_{n}(Q_{X})\right],
\end{align}
and \eqref{eq: conditional type 1 prob upper bound} is obtained by
rearranging. Writing
\[
p_{1}\left[{\cal H}_{n}\vert X^{n}\in{\cal T}_{n}(Q_{X})\right]=\sum_{i\in{\cal M}_{n,Q_{X}}}\P\left[I_{n}=i|X^{n}\in{\cal T}_{n}(Q_{X})\right]\cdot p_{1}\left({\cal H}_{n}\vert I_{n}=i\right),
\]
Markov's inequality implies
\begin{align}
 & \P\left\{ I_{n}:p_{1}\left({\cal H}_{n}\vert I_{n}\right)\geq e^{n\delta}\cdot p_{1}\left[{\cal H}_{n}\vert X^{n}\in{\cal T}_{n}(Q_{X})\right]\right\} \nonumber \\
 & \leq\frac{\E\left[p_{1}\left({\cal H}_{n}\vert I_{n}\right)\right]}{e^{n\delta}\cdot p_{1}\left[{\cal H}_{n}\vert X^{n}\in{\cal T}_{n}(Q_{X})\right]}\\
 & =\frac{\sum_{i\in{\cal M}_{n,Q_{X}}}\P\left[I_{n}=i|X^{n}\in{\cal T}_{n}(Q_{X})\right]\cdot p_{1}\left({\cal H}_{n}\vert I_{n}=i\right)}{e^{n\delta}\cdot p_{1}\left[{\cal H}_{n}\vert X^{n}\in{\cal T}_{n}(Q_{X})\right]}\\
 & =e^{-n\delta}.
\end{align}
The same arguments can be applied for the type 2 exponent. Thus, from
the above and \eqref{eq: large bin event}, with probability larger
than $1-\gamma^{-1}-2\cdot e^{-n\delta}$, which is strictly positive
for all sufficiently large $n$, the bin index satisfies \eqref{eq: large bin event},
\begin{align}
p_{1}\left({\cal H}_{n}\vert I_{n}\right) & \leq e^{n\delta}\cdot p_{1}\left[{\cal H}_{n}\vert X^{n}\in{\cal T}_{n}(Q_{X})\right]\\
 & \leq\exp\left\{ -n\cdot\left[E_{1}-D(Q_{X}||P_{X})-3\delta\right]\right\} ,\label{eq: bin has small FA prob}
\end{align}
as well as
\begin{equation}
p_{2}\left({\cal H}_{n}\vert I_{n}\right)\leq\exp\left\{ -n\cdot\left[E_{2}^{+}(R+3\delta,E_{1}-3\delta)-D(Q_{X}||\overline{P}_{X})+3\delta\right]\right\} .\label{eq: bin has small MD prob}
\end{equation}

Now, let $Q_{X}^{*}\in{\cal P}({\cal X})$ be chosen to achieve $E_{2}^{+}(R+3\delta,E_{1}-3\delta)$
up to $\delta$, i.e., to be chosen such that
\begin{equation}
E_{2}^{+}(R+3\delta,E_{1}-3\delta)-D(Q_{X}^{*}||\overline{P}_{X})\geq F_{2}^{+}\left(H(Q_{X}^{*})-R-3\delta,Q_{X}^{*},E_{1}-D(Q_{X}^{*}||P_{X})-3\delta\right)-\delta,\label{eq: Qx chosen to achieve sup reliability}
\end{equation}
and let $\{n_{l}\}_{l=1}^{\infty}$ be the subsequence of blocklengths
such that ${\cal T}_{n}(Q_{X}^{*})$ is not empty. From the above
discussion, there a sequence of bin indices $\{i_{n_{l}}^{*}\}_{l=1}^{\infty}$
such that \eqref{eq: large bin event}, \eqref{eq: bin has small FA prob}
and \eqref{eq: bin has small MD prob} hold for $Q_{X}^{*}$. Consider
the sequence of bins ${\cal C}_{n_{l}}^{*}=f_{n_{l}}^{-1}(i_{n_{l}}^{*})$
to be a sequence of CD codes, whose rate is larger than $H(Q_{X})-R-3\delta$,
its detectors are induced by the DHT system detector as $\phi_{n_{l}}^{*}(y^{n_{l}})=\varphi_{n_{l}}(i_{n_{l}}^{*},y^{n_{l}})$,
and such that 
\begin{equation}
p_{1}\left({\cal C}_{n_{l}}^{*},\phi_{n_{l}}^{*}\right)\leq\exp\left\{ -n_{l}\cdot\left[E_{1}-D(Q_{X}^{*}||P_{X})-3\delta\right]\right\} ,\label{eq: FA of induced CD code}
\end{equation}
and 
\begin{align}
p_{2}\left({\cal C}_{n_{l}}^{*},\phi_{n_{l}}^{*}\right) & \leq\exp\left\{ -n_{l}\cdot\left[E_{2}^{+}(R+3\delta,E_{1}-3\delta)-D(Q_{X}^{*}||\overline{P}_{X})+3\delta\right]\right\} \\
 & \leq\exp\left\{ -n_{l}\cdot\left[F_{2}^{+}\left(H(Q_{X}^{*})-R-3\delta,Q_{X}^{*},E_{1}-D(Q_{X}^{*}||P_{X})-3\delta\right)+2\delta\right]\right\} ,
\end{align}
where the last inequality follows from \eqref{eq: Qx chosen to achieve sup reliability}.
However, this is a contradiction, since whenever \eqref{eq: FA of induced CD code}
holds the definition of CD reliability function implies that 
\[
p_{2}\left({\cal C}_{n_{l}}^{*},\phi_{n_{l}}^{*}\right)\geq\exp\left\{ -n_{l}\cdot\left[F_{2}^{+}\left(H(Q_{X}^{*})-R-3\delta,Q_{X}^{*},E_{1}-D(Q_{X}^{*}||P_{X})-3\delta\right)+\delta\right]\right\} 
\]
for all $l$ sufficiently large.
\end{IEEEproof}

\section{Proof of Theorems \ref{thm: random coding bound} and \ref{thm: Expurgated bound}
\label{sec:Proof-of-Theorem-Random-Coding}}

We will prove the random-coding bound of Theorem \ref{thm: random coding bound}
by considering CD codes drawn from the fixed-composition hierarchical
ensemble defined in Definition \ref{def:A-fixed-composition-hierarchical-Codebook}.
In the course of the proof, we shall consider various types of the
form $Q_{UXY}$ and $\overline{Q}_{UXY}$. All of them are assume
to have $(U,X)$ marginal $Q_{UX}=\overline{Q}_{UX}$ even if it is
not explicitly stated. Furthermore, we shall assume that the blocklength
$n$ is such that ${\cal T}_{n}(Q_{UX})$ is not empty. In this case,
the notation for exponential equality (or inequality) needs to be
clarified as follows. We will say that $a_{n}\doteq b_{n}$ if 
\[
\lim_{l\to\infty}\frac{1}{n_{l}}\log\frac{a_{n_{l}}}{b_{n_{l}}}=1,
\]
where $\{n_{l}\}_{l=1}^{\infty}$ is the subsequence of blocklengths
such that ${\cal T}_{n}(Q_{UX})$ is not empty. 

The proof of Theorem \ref{thm: random coding bound} relies on the
following result, which is stated and proved by means of the \emph{type-enumeration
method }(see \cite[Sec. 6.3]{Merhav09}). Specifically, for a given
$y^{n}$, we define \emph{type-class enumerators }for a random CD
code $\mathfrak{C}_{n}$ by
\begin{equation}
M_{y^{n}}(Q_{UXY})\dfn\left|\left\{ x^{n}\in\mathfrak{C}_{n}:\exists u^{n}\text{ such that }x^{n}\in\mathfrak{C}_{n,\s[s]}(u^{n}),\;(u^{n},x^{n},y^{n})\in{\cal T}_{n}(Q_{UXY})\right\} \right|.\label{eq: hierarchical codewords enumerator}
\end{equation}
To wit, $M_{y^{n}}(Q_{UXY})$ counts the random number of codewords
whose joint type with their own cloud center $u^{n}$ and $y^{n}$
is $Q_{UXY}\in{\cal P}_{n}({\cal U}\times{\cal X}\times{\cal Y})$.
To derive a random-coding bound on the achievable CD exponents, we
will need to evaluate the exponential order of $\E[M_{y^{n}}^{1-\lambda}(Q_{UXY})M_{y^{n}}^{\lambda}(\overline{Q}_{UXY})]$
for an arbitrary sequence of $\{y^{n}\}$ taken from ${\cal T}_{n}(Q_{Y})={\cal T}_{n}(\overline{Q}_{Y})$. 

The result is summarized in the following proposition, interesting
on its own right. 
\begin{prop}
\label{prop: Correlation of hierarchical enumeartors}Let $Q_{UXY},\overline{Q}_{UXY}\in{\cal P}_{n_{0}}({\cal U}\times{\cal X}\times{\cal Y})$
be given for some $n_{0}$, with $Q_{UX}=\overline{Q}_{UX}$ and $Q_{Y}=\overline{Q}_{Y}$.
Also let $\{n_{l}\}$ be the subsequence of blocklengths such that
${\cal T}_{n_{l}}(Q_{UXY})$ and ${\cal T}_{n_{l}}(\overline{Q}_{UXY})$
are both not empty, and let $\{y^{n_{l}}\}_{l=1}^{\infty}$ satisfy
$y^{n_{l}}\in{\cal T}_{n_{l}}(Q_{Y})$ for all $l$. Then, for any
$\lambda\in(0,1)$
\begin{multline}
\lim_{n_{l}\to\infty}\frac{1}{n_{l}}\log\E\left[M_{y^{n_{l}}}^{1-\lambda}(Q_{UXY})M_{y^{n_{l}}}^{\lambda}(\overline{Q}_{UXY})\right]\\
=\begin{cases}
\rho-I_{Q}(U,X;Y), & Q_{UXY}=\overline{Q}_{UXY}\\
\Delta_{\lambda}(Q_{UXY},\overline{Q}_{UXY}), & Q_{UY}\neq\overline{Q}_{UY}\text{ and }Q_{UXY}\neq\overline{Q}_{UXY}\\
\Delta_{\lambda}(Q_{UXY},\overline{Q}_{UXY})-|I_{Q}(U;Y)-\rho_{\s[c]}|_{+}, & Q_{UY}=\overline{Q}_{UY}\text{ (and }Q_{UXY}\neq\overline{Q}_{UXY}\text{)}
\end{cases},\label{eq: correlation of hierarchical enumerator}
\end{multline}
where
\begin{align}
\Delta_{\lambda}(Q_{UXY},\overline{Q}_{UXY}) & \dfn(1-\lambda)\cdot\left[\rho-I_{Q}(U,X;Y)\right]-\lambda\cdot\max\left\{ \left|I_{Q}(U;Y)-\rho_{\s[c]}\right|_{+},\;I_{Q}(U,X;Y)-\rho\right\} \nonumber \\
 & \hphantom{=}+\lambda\left[\rho-I_{\overline{Q}}(U,X;Y)\right]-(1-\lambda)\cdot\max\left\{ \left|I_{\overline{Q}}(U;Y)-\rho_{\s[c]}\right|_{+},\;I_{\overline{Q}}(U,X;Y)-\rho\right\} .\label{eq: Delta functional}
\end{align}
\end{prop}
It is interesting to note that the expression \eqref{eq: correlation of hierarchical enumerator}
is not continuous, as, say, $\overline{Q}_{UXY}\to Q_{UXY}$. The
proof of Proposition \ref{prop: Correlation of hierarchical enumeartors}
is of technical nature, and thus relegated to Appendix \ref{sec:The-Type-Enumeration-Method}.
For the rest of the proof, no knowledge of the type-enumeration method
is required. 

As described in Section \ref{sec:Channel-Detection-Codes} the detector
of a CD code faces an ordinary HT problem between the $P_{Y^{n}}^{({\cal C}_{n})}$
and $\overline{P}_{Y^{n}}^{({\cal C}_{n})}$, and therefore the exponents
of this HT problem are simply given by \eqref{eq: FA exponent ordinary vector}
and \eqref{eq: MD exponent ordinary vector} (when letting ${\cal Z}={\cal Y}^{n}$).
In turn, using the characterization \eqref{eq: ordinary hypothesis testing Chernoff information},
the reliability function can be expressed using the Chernoff parameter
between $P_{Y^{n}}^{({\cal C}_{n})}$ and $\overline{P}_{Y^{n}}^{({\cal C}_{n})}$.
We thus next analyze the average exponent of the Chernoff parameter
over a random choice of CD codes.
\begin{lem}
\label{lem: average of Chernoff}Let $\mathfrak{C}_{n}$ be drawn
randomly from the hierarchical ensemble of Definition \ref{def:A-fixed-composition-hierarchical-Codebook},
with conditional distribution $Q_{U|X}$, cloud-center rate $\rho_{\s[c]}$,
and satellite rate $\rho_{\s[s]}$ (which satisfy $\rho=\rho_{\s[c]}+\rho_{\s[s]}$).
Then,
\begin{equation}
\E\left\{ \sum_{y^{n}\in{\cal Y}^{n}}\left[P_{Y^{n}}^{(\mathfrak{C}_{n})}(y^{n})\right]^{1-\lambda}\cdot\left[\overline{P}_{Y^{n}}^{(\mathfrak{C}_{n})}(y^{n})\right]^{\lambda}\right\} \dotleq\exp\left[-n\cdot\min\left\{ d_{\lambda}(Q_{X}),\;A_{\s[rc]}\right\} \right],\label{eq: expected Chernoff distance}
\end{equation}
where $d_{\lambda}(Q_{X})$ is defined in \eqref{eq: Chernoff distance type - one RV}
when setting $\tau=\frac{1-\lambda}{\lambda}$, and $A_{\s[rc]}$
is defined in \eqref{eq: A_rc}. 
\end{lem}
\begin{IEEEproof}
Let us denote the \emph{log-likelihood} of $(x^{n},y^{n})\in{\cal T}_{n}(Q_{XY})$
by 
\begin{align}
L(Q_{XY}) & \dfn-\frac{1}{n}\log P_{Y|X}(y^{n}|x^{n})\label{eq: log likelihood}\\
 & =-\sum_{x\in{\cal X},\;y\in\mathcal{Y}}Q_{XY}(x,y)\log P(y|x),
\end{align}
and the log-likelihood of $\overline{P}_{Y|X}$ by $\overline{L}(Q_{XY})$
(with $\overline{P}_{Y|X}$ replacing $P_{Y|X}$). For any given $n$,
\begin{align}
 & \E\left\{ \sum_{y^{n}\in{\cal Y}^{n}}\left[P_{Y^{n}}^{(\mathfrak{C}_{n})}(y^{n})\right]^{1-\lambda}\cdot\left[\overline{P}_{Y^{n}}^{(\mathfrak{C}_{n})}(y^{n})\right]^{\lambda}\right\} \nonumber \\
 & =\frac{1}{e^{n\rho}}\sum_{y^{n}\in{\cal Y}^{n}}\E\left\{ \left[\sum_{x^{n}\in\mathfrak{C}_{n}}P_{Y|X}(y^{n}|x^{n})\right]^{1-\lambda}\cdot\left[\sum_{\overline{x}^{n}\in\mathfrak{C}_{n}}\overline{P}_{Y|X}(y^{n}|\overline{x}^{n})\right]^{\lambda}\right\} \\
 & =\frac{1}{e^{n\rho}}\sum_{Q_{Y}\in{\cal P}_{n}({\cal Y})}\sum_{y^{n}\in{\cal T}_{n}(Q_{Y})}\E\left\{ \left[\sum_{x^{n}\in\mathfrak{C}_{n}}P_{Y|X}(y^{n}|x^{n})\right]^{1-\lambda}\cdot\left[\sum_{\overline{x}^{n}\in\mathfrak{C}_{n}}\overline{P}_{Y|X}(y^{n}|\overline{x}^{n})\right]^{\lambda}\right\} \\
 & \trre[=,a]\frac{1}{e^{n\rho}}\sum_{Q_{Y}\in{\cal P}_{n}({\cal Y})}\left|{\cal T}_{n}(Q_{Y})\right|\cdot\E\left\{ \left[\sum_{Q_{UXY}}M_{y^{n}}(Q_{UXY})e^{-nL(Q_{XY})}\right]^{1-\lambda}\cdot\left[\sum_{\overline{Q}_{UXY}}M_{y^{n}}(\overline{Q}_{UXY})e^{-n\overline{L}(\overline{Q}_{XY})}\right]^{\lambda}\right\} \\
 & \leq\left|{\cal P}_{n}({\cal Y})\right|\cdot\max_{Q_{Y}\in{\cal P}_{n}({\cal Y})}e^{n\left[H(Q_{Y})-\rho\right]}\cdot\E\left\{ \left[\sum_{Q_{UXY}}M_{y^{n}}(Q_{UXY})e^{-nL(Q_{XY})}\right]^{1-\lambda}\cdot\left[\sum_{\overline{Q}_{UXY}}M_{y^{n}}(\overline{Q}_{UXY})e^{-n\overline{L}(\overline{Q}_{XY})}\right]^{\lambda}\right\} \\
 & \leq\left|{\cal P}_{n}({\cal Y})\right|\left|{\cal P}_{n}({\cal U}\times{\cal X}\times{\cal Y})\right|^{2}\nonumber \\
 & \hphantom{=}\times\max_{Q_{Y}\in{\cal P}_{n}({\cal Y})}e^{n\left[H(Q_{Y})-\rho\right]}\cdot\E\left[\max_{Q_{UXY}}M_{y^{n}}^{1-\lambda}(Q_{UXY})e^{-n(1-\lambda)\cdot L(Q_{XY})}\cdot\max_{\overline{Q}_{UXY}}M_{y^{n}}^{\lambda}(\overline{Q}_{UXY})e^{-n\lambda\overline{L}(\overline{Q}_{XY})}\right]\\
 & \leq\left|{\cal P}_{n}({\cal Y})\right|\left|{\cal P}_{n}({\cal U}\times{\cal X}\times{\cal Y})\right|^{2}\nonumber \\
 & \hphantom{=}\times\max_{Q_{Y}\in{\cal P}_{n}({\cal Y})}e^{n\left[H(Q_{Y})-\rho\right]}\cdot\E\left[\sum_{Q_{UXY}}M_{y^{n}}^{1-\lambda}(Q_{UXY})e^{-n(1-\lambda)\cdot L(Q_{XY})}\cdot\sum_{\overline{Q}_{UXY}}M_{y^{n}}^{\lambda}(\overline{Q}_{UXY})e^{-n\lambda\overline{L}(\overline{Q}_{XY})}\right]\\
 & =\left|{\cal P}_{n}({\cal Y})\right|\left|{\cal P}_{n}({\cal U}\times{\cal X}\times{\cal Y})\right|^{2}\nonumber \\
 & \hphantom{=}\times\max_{Q_{Y}\in{\cal P}_{n}({\cal Y})}e^{n\left[H(Q_{Y})-\rho\right]}\sum_{Q_{UXY}}\sum_{\overline{Q}_{UXY}}\E\left[M_{y^{n}}^{1-\lambda}(Q_{UXY})\cdot M_{y^{n}}^{\lambda}(\overline{Q}_{UXY})\right]\cdot e^{-n\left[(1-\lambda)\cdot L(Q_{XY})+\lambda\overline{L}(\overline{Q}_{XY})\right]}\\
 & \leq\left|{\cal P}_{n}({\cal Y})\right|\left|{\cal P}_{n}({\cal U}\times{\cal X}\times{\cal Y})\right|^{4}\nonumber \\
 & \hphantom{=}\times\max_{Q_{UXY}}\max_{\overline{Q}_{UXY}}e^{n\left[H(Q_{Y})-\rho\right]}\cdot\E\left[M_{y^{n}}^{1-\lambda}(Q_{UXY})\cdot M_{y^{n}}^{\lambda}(\overline{Q}_{UXY})\right]\cdot e^{-n\left[(1-\lambda)\cdot L(Q_{XY})+\lambda\overline{L}(\overline{Q}_{XY})\right]}\\
 & \dfn c_{n}\cdot\max_{(Q_{UXY},\overline{Q}_{UXY})\in{\cal Q}_{n}}\zeta_{n}(Q_{UXY},\overline{Q}_{UXY}),
\end{align}
where $(a)$ follows since by symmetry, the expectation only depends
on the type of $y^{n}$, and by using the definitions of the enumerators
in \eqref{eq: hierarchical codewords enumerator}, and the log-likelihood
in \eqref{eq: log likelihood}. After passage $(a)$ and onward, $y^{n}$
is an arbitrary member of ${\cal T}_{n}(Q_{Y})$, and the sums and
maximization operators are over $(Q_{UXY},\overline{Q}_{UXY})$ restricted
to the set
\[
{\cal Q}_{n}\dfn\left\{ (Q_{UXY},\overline{Q}_{UXY})\in{\cal P}_{n}^{2}({\cal U}\times{\cal X}\times{\cal Y}):Q_{Y}=\overline{Q}_{Y},\;Q_{XU}=\overline{Q}_{XU}\right\} .
\]
In the last equality we have implicitly defined $c_{n}$ and $\zeta_{n}(Q_{UXY},\overline{Q}_{UXY})$.
By defining
\[
\overline{{\cal Q}}\dfn\left\{ (Q_{UXY},\overline{Q}_{UXY})\in{\cal {\cal S}}^{2}({\cal U}\times{\cal X}\times{\cal Y}):Q_{Y}=\overline{Q}_{Y},\;Q_{XU}=\overline{Q}_{XU}\right\} ,
\]
 and using standard arguments (e.g., as in the proof of Sanov's theorem
\cite[Theorem 11.4.1]{Cover:2006:EIT:1146355}) we get
\[
\liminf_{n\to\infty}-\frac{1}{n}\log\E\left\{ \sum_{y^{n}\in{\cal Y}^{n}}\left[P_{Y^{n}}^{(\mathfrak{C}_{n})}(y^{n})\right]^{1-\lambda}\cdot\left[\overline{P}_{Y^{n}}^{(\mathfrak{C}_{n})}(y^{n})\right]^{\lambda}\right\} \geq\min_{(Q_{UXY},\overline{Q}_{UXY})\in\overline{\mathcal{Q}}}\liminf_{n\to\infty}-\frac{1}{n}\log\zeta_{n}(Q_{UXY},\overline{Q}_{UXY}).
\]
The result \eqref{eq: expected Chernoff distance} will follow by
minimizing 
\begin{align}
\Lambda_{\lambda}(Q_{UXY},\overline{Q}_{UXY}) & \dfn\liminf_{n\to\infty}-\frac{1}{n}\log\zeta_{n}(Q_{UXY},\overline{Q}_{UXY})\\
 & =-H(Q_{Y})+\rho-\liminf_{n\to\infty}-\frac{1}{n}\log\E\left[M_{y^{n}}^{1-\lambda}(Q_{UXY})M_{y^{n}}^{\lambda}(\overline{Q}_{UXY})\right]\nonumber \\
 & \hphantom{=}+\left[(1-\lambda)\cdot L(Q_{XY})+\lambda\overline{L}(\overline{Q}_{XY})\right]
\end{align}
over $(Q_{UXY},\overline{Q}_{UXY})\in\overline{\mathcal{Q}}$. We
now evaluate this expression in three cases, which correspond to the
three cases of Proposition \ref{prop: Correlation of hierarchical enumeartors}.
In each one, we substitute for $\liminf_{n\to\infty}-\frac{1}{n}\log\E[M_{y^{n}}^{1-\lambda}(Q_{UXY})M_{y^{n}}^{\lambda}(\overline{Q}_{UXY})]$
the appropriate term,\footnote{In fact, Proposition \ref{prop: Correlation of hierarchical enumeartors}
implies that the limit inferior of this sequence is a proper limit.} as follows:
\begin{casenv}
\item For $Q_{UXY}=\overline{Q}_{UXY}$, 
\begin{align}
 & \Lambda_{\lambda}(Q_{UXY},\overline{Q}_{UXY})\nonumber \\
 & =-H(Q_{Y})+\rho-\left[\rho-I_{Q}(U,X;Y)\right]+(1-\lambda)\cdot L(Q_{XY})+\lambda\cdot\overline{L}(Q_{XY})\\
 & =(1-\lambda)\cdot\left[-H_{Q}(Y|X,U)+L(Q_{XY})\right]+\lambda\cdot\left[-H_{Q}(Y|X,U)+\overline{L}(Q_{XY})\right]\\
 & \trre[=,a](1-\lambda)\cdot D(Q_{Y|UX}||P_{Y|X}|Q_{UX})+\lambda\cdot D(Q_{Y|UX}||\overline{P}_{Y|X}|\overline{Q}_{UX}),
\end{align}
where $(a)$ follows from the identity 
\begin{equation}
-H_{Q}(Y|X,U)+L(Q_{XY})=D(Q_{Y|UX}||P_{Y|X}|Q_{UX}).\label{eq: identity  -coditional entropy + likelihood =00003D divergence}
\end{equation}
By defining the distribution
\[
P_{Y}^{(\lambda,x)}(y)\dfn\frac{P_{Y|X}^{1-\lambda}(y|x)\overline{P}_{Y|X}^{\lambda}(y|x)}{\sum_{y'\in{\cal Y}}P_{Y|X}^{1-\lambda}(y'|x)\overline{P}_{Y|X}^{\lambda}(y'|x)}
\]
and observing that 
\begin{align}
 & \min_{Q_{Y|UX}}\left[(1-\lambda)\cdot D(Q_{Y|UX}||P_{Y|X}|Q_{UX})+\lambda\cdot D(Q_{Y|UX}||\overline{P}_{Y|X}|Q_{UX})\right]\nonumber \\
 & =\min_{Q_{Y|X}}\left\{ D(Q_{Y|X}||P_{Y}^{(\lambda,x)}|Q_{X})-\sum_{x\in{\cal X}}Q_{X}(x)\log\left[\sum_{y\in{\cal Y}}P_{Y|X}^{1-\lambda}(y|x)\overline{P}_{Y|X}^{\lambda}(y|x)\right]\right\} \\
 & =-\sum_{x\in{\cal X}}Q_{X}(x)\log\left[\sum_{y\in{\cal Y}}P_{Y|X}^{1-\lambda}(y|x)\overline{P}_{Y|X}^{\lambda}(y|x)\right],
\end{align}
it is evident that
\begin{align}
\Lambda_{\lambda,1}(Q_{UXY},\overline{Q}_{UXY}) & =\min_{Q_{Y|X}}\left[(1-\lambda)\cdot D(Q_{Y|UX}||P_{Y|X}|Q_{UX})+\lambda\cdot D(Q_{Y|UX}||\overline{P}_{Y|X}|Q_{UX})\right]\\
 & =-\sum_{x\in{\cal X}}Q_{X}(x)\log\left[\sum_{y\in{\cal Y}}P_{Y|X}^{1-\lambda}(y|x)\overline{P}_{Y|X}^{\lambda}(y|x)\right]\\
 & =d_{\lambda}(Q_{X}).\label{eq: divergence charctarization of Chernoff}
\end{align}
\item For $Q_{UXY}\neq\overline{Q}_{UXY}$ and $Q_{UY}\neq\overline{Q}_{UY}$
\begin{align}
 & \Lambda_{\lambda}(Q_{UXY},\overline{Q}_{UXY})\nonumber \\
 & =-H(Q_{Y})+\rho+(1-\lambda)\cdot L(Q_{XY})+\lambda\cdot\overline{L}(\overline{Q}_{XY})\\
 & \hphantom{=}-(1-\lambda)\cdot\left[\rho-I_{Q}(U,X;Y)\right]+\lambda\cdot\max\left\{ \left|I_{Q}(U;Y)-\rho_{\s[c]}\right|_{+},\;I_{Q}(U,X;Y)-\rho\right\} \nonumber \\
 & \hphantom{=}-\lambda\left[\rho-I_{\overline{Q}}(U,X;Y)\right]+(1-\lambda)\cdot\max\left\{ \left|I_{\overline{Q}}(U;Y)-\rho_{\s[c]}\right|_{+},\;I_{\overline{Q}}(U,X;Y)-\rho\right\} \\
 & \trre[=,a](1-\lambda)\cdot D(Q_{Y|UX}||P_{Y|X}|Q_{UX})+\lambda\cdot D(\overline{Q}_{Y|UX}||\overline{P}_{Y|X}|\overline{Q}_{UX})\nonumber \\
 & \hphantom{=}+\lambda\cdot\max\left\{ \left|I_{Q}(U;Y)-\rho_{\s[c]}\right|_{+},\;I_{Q}(U,X;Y)-\rho\right\} \nonumber \\
 & \hphantom{=}+(1-\lambda)\cdot\max\left\{ \left|I_{\overline{Q}}(U;Y)-\rho_{\s[c]}\right|_{+},\;I_{\overline{Q}}(U,X;Y)-\rho\right\} \\
 & \dfn\Lambda_{\lambda,2}(Q_{UXY},\overline{Q}_{UXY}),
\end{align}
where $(a)$ follows from \eqref{eq: identity  -coditional entropy + likelihood =00003D divergence}
again and rearrangement.
\item For $Q_{UXY}\neq\overline{Q}_{UXY}$ and $Q_{UY}=\overline{Q}_{UY}$
\begin{align}
 & \Lambda_{\lambda}(Q_{UXY},\overline{Q}_{UXY})\nonumber \\
 & =\Lambda_{\lambda,2}(Q_{UXY},\overline{Q}_{UXY})-|I_{Q}(U;Y)-\rho_{\s[c]}|_{+}\\
 & \dfn\Lambda_{\lambda,3}(Q_{UXY},\overline{Q}_{UXY}).\label{eq: Lambda third case}
\end{align}
\end{casenv}
Hence, the required bound on the Chernoff parameter is given by 
\begin{multline}
\min_{(Q_{UXY},\overline{Q}_{UXY})\in\overline{\mathcal{Q}}}\Lambda_{\lambda}(Q_{UXY},\overline{Q}_{UXY})\\
=\min\left\{ d_{\lambda}(Q_{X}),\;\min_{(Q_{UXY},\overline{Q}_{UXY})\in\overline{{\cal Q}}:Q_{UY}\neq\overline{Q}_{UY}}\Lambda_{\lambda,2}(Q_{UXY},\overline{Q}_{UXY}),\;\min_{(Q_{UXY},\overline{Q}_{UXY})\in\overline{{\cal Q}}:Q_{UY}=\overline{Q}_{UY}}\Lambda_{\lambda,3}(Q_{UXY},\overline{Q}_{UXY})\right\} .\label{eq: minimization over three cases}
\end{multline}
Observing \eqref{eq: Lambda third case}, we note that the third term
in \eqref{eq: minimization over three cases} satisfies
\begin{align}
 & \min_{(Q_{UXY},\overline{Q}_{UXY})\in\overline{{\cal Q}}:Q_{UY}=\overline{Q}_{UY}}\Lambda_{\lambda,3}(Q_{UXY},\overline{Q}_{UXY})\nonumber \\
 & =\min\Bigg\{\min_{(Q_{UXY},\overline{Q}_{UXY})\in\overline{{\cal Q}}:Q_{UY}=\overline{Q}_{UY},\;I_{Q}(U;Y)\leq\rho_{\s[c]}}\Lambda_{\lambda,2}(Q_{UXY},\overline{Q}_{UXY}),\;\nonumber \\
 & \hphantom{=}\min_{(Q_{UXY},\overline{Q}_{UXY})\in\overline{{\cal Q}}:Q_{UY}=\overline{Q}_{UY},\;I_{Q}(U;Y)>\rho_{\s[c]}}\left\{ \Lambda_{\lambda,2}(Q_{UXY},\overline{Q}_{UXY})-I_{Q}(U;Y)+\rho_{\s[c]}\right\} \Bigg\}\\
 & \trre[=,a]\min\Bigg\{\min_{(Q_{UXY},\overline{Q}_{UXY})\in\overline{{\cal Q}}:Q_{UY}=\overline{Q}_{UY},}\Lambda_{\lambda,2}(Q_{UXY},\overline{Q}_{UXY}),\;\nonumber \\
 & \hphantom{=}\min_{(Q_{UXY},\overline{Q}_{UXY})\in\overline{{\cal Q}}:Q_{UY}=\overline{Q}_{UY},\;I_{Q}(U;Y)>\rho_{\s[c]}}\left\{ \Lambda_{\lambda,2}(Q_{UXY},\overline{Q}_{UXY})-I_{Q}(U;Y)+\rho_{\s[c]}\right\} \Bigg\},\label{eq: minimization of pair for the third case}
\end{align}
where in $(a)$ we have removed the constraint $I_{Q}(U;Y)\leq\rho_{\s[c]}$
in the first term of the outer minimization, since for $Q_{UY}$ with
$I_{Q}(U;Y)>\rho_{\s[c]}$ the second term will dominate the minimization.
Thus, the first term in \eqref{eq: minimization of pair for the third case}
may be unified with the second term of \eqref{eq: minimization over three cases}.
Doing so, the constraint $Q_{UY}\neq\overline{Q}_{UY}$ may be removed
in the second term of \eqref{eq: minimization over three cases}.
Consequently, the required bound on the Chernoff parameter is given
by 
\begin{multline}
\min\Bigg\{ d_{\lambda}(Q_{X}),\;\min_{(Q_{UXY},\overline{Q}_{UXY})\in\overline{{\cal Q}}}\Lambda_{2}(Q_{UXY},\overline{Q}_{UXY}),\;\\
\min_{(Q_{UXY},\overline{Q}_{UXY})\in\overline{{\cal Q}}:Q_{UY}=\overline{Q}_{UY},\;I_{Q}(U;Y)>\rho_{\s[c]}}\left\{ \Lambda_{2}(Q_{UXY},\overline{Q}_{UXY})-I_{Q}(U;Y)+\rho_{\s[c]}\right\} \Bigg\}.\label{eq: minimization of three cases improved}
\end{multline}
The second term in \eqref{eq: minimization of three cases improved}
corresponds to $A_{\s[rc]}'$ defined in \eqref{eq: A_rc_tag}. The
third term corresponds to $A_{\s[rc]}''$ defined in \eqref{eq: A_rc_tag2},
when using $\rho_{\s[s]}=\rho-\rho_{\s[c]}$, and noting that when
$I_{Q}(U;Y)>\rho_{\s[c]}$, it is equal to
\begin{align}
 & \Lambda_{\lambda,2}(Q_{UXY},\overline{Q}_{UXY})-I_{Q}(U;Y)+\rho_{\s[c]}\nonumber \\
 & =(1-\lambda)\cdot D(Q_{Y|UX}||P_{Y|X}|Q_{UX})+\lambda\cdot D(\overline{Q}_{Y|UX}||\overline{P}_{Y|X}|\overline{Q}_{UX})\nonumber \\
 & \hphantom{=}+\lambda\cdot\left|I_{Q}(X;Y|U)-\rho_{\s[s]}\right|_{+}+(1-\lambda)\cdot\left|I_{\overline{Q}}(X;Y|U)-\rho_{\s[s]}\right|_{+}.
\end{align}
 Using the definition \eqref{eq: A_rc} of $A_{\s[rc]}$ as the minimum
of the last two cases, \eqref{eq: expected Chernoff distance} is
obtained.
\end{IEEEproof}
Next, recall that by its definition, a valid CD code of rate $\rho$
is comprised of\emph{ $e^{n\rho}$ distinct }codewords. However, when
the codewords are independently drawn, some of them might be identical.
Nonetheless, the next lemma shows that the average number of distinct
codewords of a randomly chosen code is asymptotically close to $e^{n\rho}$.
\begin{lem}
\label{lem: average of distinct codewords}Let $\mathfrak{C}_{n}$
be drawn randomly from the hierarchical ensemble of Definition \ref{def:A-fixed-composition-hierarchical-Codebook},
with conditional distribution $Q_{U|X}$, cloud-center rate $\rho_{\s[c]}$,
and satellite rate $\rho_{\s[s]}$ (which satisfy $\rho=\rho_{\s[c]}+\rho_{\s[s]}$).
If $\rho_{\s[c]}+H_{Q}(X|U)\geq\rho$ then
\[
\E\left[\left|\mathfrak{C}_{n}\right|\right]\doteq e^{n\rho}.
\]
\end{lem}
\begin{IEEEproof}
Let us enumerate the random cloud centers as $\{U^{n}(i)\}_{i=1}^{e^{n\rho_{\s[c]}}}$
and the random satellite codebooks as $\{\mathfrak{C}_{n,\s[s]}(U^{n}(i))\}_{i=1}^{e^{n\rho_{\s[c]}}}$.
For any given $(u^{n},x^{n})\in{\cal T}_{n}(Q_{UX})$ 
\begin{align}
\P\left[x^{n}\in\mathfrak{C}_{n,\s[s]}(u^{n})\right] & =1-\left(1-\frac{1}{\left|{\cal T}_{n}(Q_{U|X},u^{n})\right|}\right)^{e^{n\rho_{\s[s]}}}\\
 & \geq\frac{1}{2}\cdot\min\left\{ 1,\frac{e^{n\rho_{\s[s]}}}{\left|{\cal T}_{n}(Q_{X|U},u^{n})\right|}\right\} \\
 & \geq\exp\left(-n\cdot\left\{ \left|H_{Q}(X|U)-\rho_{\s[s]}\right|_{+}+\delta\right\} \right),
\end{align}
using $1-(1-t)^{K}\geq\frac{1}{2}\cdot\min\left\{ 1,tK\right\} $
\cite[Lemma 1]{Somekh2007_private}. Further, for a random cloud center
$U^{n}$ and $x^{n}\in{\cal T}_{n}(Q_{X})$
\begin{align}
\P\left[x^{n}\in\mathfrak{C}_{n,\s[s]}(U^{n})\right] & =\P\left[(x^{n},U^{n})\in{\cal T}_{n}(Q_{UX})\right]\cdot\P\left[x^{n}\in\mathfrak{C}_{n,\s[s]}(U^{n})\vert(x^{n},U^{n})\in{\cal T}_{n}(Q_{UX})\right]\\
 & \trre[=,a]\P\left[(x^{n},U^{n})\in{\cal T}_{n}(Q_{UX})\right]\cdot\P\left[x^{n}\in\mathfrak{C}_{n,\s[s]}(u^{n})\right]\\
 & \geq\frac{\left|{\cal T}_{n}(Q_{U|X},x^{n})\right|}{\left|{\cal T}_{n}(Q_{U})\right|}\cdot\exp\left(-n\cdot\left\{ \left|H_{Q}(X|U)-\rho_{\s[s]}\right|_{+}+\delta\right\} \right)\\
 & \geq\exp\left(-n\cdot\left\{ I_{Q}(U;X)+\left|H_{Q}(X|U)-\rho_{\s[s]}\right|_{+}+2\delta\right\} \right)\\
 & \dfn e^{-n(\xi+2\delta)},
\end{align}
where $(a)$ is due to symmetry, and the definition
\begin{align}
\xi & \dfn I_{Q}(U;X)+\left|H_{Q}(U|X)-\rho_{\s[s]}\right|_{+}\\
 & =\max\left\{ I_{Q}(U;X),\;H(Q_{X})-\rho_{\s[s]}\right\} .
\end{align}
Therefore, the average number of distinct codewords in the random
CD code ${\cal \mathfrak{C}}_{n}$ is lower bounded as
\begin{align}
\E\left[|\mathfrak{C}_{n}|\right] & =\E\left[\sum_{x^{n}\in{\cal T}_{n}(Q_{X})}\I\left(x^{n}\in\mathfrak{C}_{n}\right)\right]\\
 & =\sum_{x^{n}\in{\cal T}_{n}(Q_{X})}\P\left(x^{n}\in\mathfrak{C}_{n}\right)\\
 & =\sum_{x^{n}\in{\cal T}_{n}(Q_{X})}\P\left\{ \bigcup_{i=1}^{e^{n\rho_{\s[c]}}}\left[x^{n}\in\mathfrak{C}_{n,\s[s]}(U^{n}(i))\right]\right\} \\
 & \trre[\geq,a]\sum_{x^{n}\in{\cal T}_{n}(Q_{X})}\frac{1}{2}\cdot\min\left\{ e^{n\rho_{\s[c]}}\cdot\P\left[x^{n}\in{\cal \mathfrak{C}}_{n,\s[s]}(U^{n}(1))\right],1\right\} \\
 & \geq\exp\left\{ n\cdot\left[\min\left\{ H(Q_{X})+\rho_{\s[c]}-\xi,\;H(Q_{X})\right\} -3\delta\right]\right\} \\
 & =\exp\left\{ n\cdot\left[\min\left\{ H(Q_{X})+\rho_{\s[c]}-I_{Q}(U;X),\rho_{\s[c]}+\rho_{\s[s]},\;H(Q_{X})\right\} -3\delta\right]\right\} \\
 & \trre[=,b]e^{n(\rho-3\delta)},\label{eq: number of unique codewords}
\end{align}
where $(a)$ holds since for a given set of $K$ pairwise independent
events $\{{\cal A}_{k}\}_{k=1}^{K}$ \cite[Lemma A.2]{shulman2003communication}
\begin{equation}
\P\left[\bigcup_{k=1}^{K}{\cal A}_{k}\right]\geq\frac{1}{2}\min\left\{ 1,\sum_{k=1}^{K}\P\left({\cal A}_{k}\right)\right\} .\label{shulman}
\end{equation}
The passage $(b)$ follows from the assumptions $\rho<H(Q_{X})$ and
$\rho_{\s[c]}+H_{Q}(X|U)\geq\rho$. Thus, on the average, a randomly
chosen $\mathfrak{C}_{n}$ has more than $e^{n(\rho-3\delta)}$ distinct
codewords. The results follows since clearly $|\mathfrak{C}_{n}|\leq e^{n\rho}$. 
\end{IEEEproof}
We are now ready to prove Theorem \ref{thm: random coding bound}.
The main argument is to show that by randomly drawing a set of $e^{n\rho}$
codewords from the hierarchical ensemble, and then removing its duplicates,
i.e., keeping only a single instance of codewords which were drawn
more than once (thus making it a valid CD code), may only cause a
negligible loss in the achieved exponent. 
\begin{IEEEproof}[Proof of Theorem \ref{thm: random coding bound}]
 Let $\delta>0$ be given, let $\lambda^{*}$ be the achiever of
the maximum on the right-hand side of \eqref{eq: CD random coding bound},
and let $Q_{U|X}^{*}$ and $\rho_{c}^{*}$ be the achievers of the
supremum, up to $\delta$. As noted in Section \ref{sec:Channel-Detection-Codes},
given a CD code ${\cal C}_{n}$, the detector faces an ordinary HT
problem between the distributions $P_{Y^{n}}^{({\cal C}_{n})}$ and
$\overline{P}_{Y^{n}}^{({\cal C}_{n})}$, and thus the bounds of Section
\ref{subsec:Ordinary-Hypothesis-Testing} can be used. Specifically,
\eqref{eq: ordinary hypothesis testing Chernoff information} (with
$\tau=\frac{1-\lambda}{\lambda}$) implies that
\begin{align}
 & \max_{{\cal C}_{n}\subseteq{\cal T}_{n}(Q_{X}):\;|{\cal C}_{n}|\geq e^{n\rho},\;p_{1}({\cal {\cal C}}_{n},\phi_{n})\leq e^{-nF_{1}}}-\frac{1}{n}\log p_{2}({\cal {\cal C}}_{n},\phi_{n})\nonumber \\
 & \geq\max_{{\cal C}_{n}\subseteq{\cal T}_{n}(Q_{X}):\;|{\cal C}_{n}|\geq e^{n\rho}}\max_{0\leq\lambda\leq1}\left\{ -\frac{1-\lambda}{\lambda}\cdot F_{1}-\frac{1}{\lambda}\cdot\frac{1}{n}\log\left\{ \sum_{y^{n}\in{\cal Y}^{n}}\left[P_{Y^{n}}^{({\cal C}_{n})}(y^{n})\right]^{1-\lambda}\cdot\left[\overline{P}_{Y^{n}}^{({\cal C}_{n})}(y^{n})\right]^{\lambda}\right\} \right\} \\
 & \geq\max_{{\cal C}_{n}\subseteq{\cal T}_{n}(Q_{X}):\;|{\cal C}_{n}|\geq e^{n\rho}}\left\{ -\frac{1-\lambda^{*}}{\lambda^{*}}\cdot F_{1}-\frac{1}{\lambda^{*}}\cdot\frac{1}{n}\log\left\{ \sum_{y^{n}\in{\cal Y}^{n}}\left[P_{Y^{n}}^{({\cal C}_{n})}(y^{n})\right]^{1-\lambda^{*}}\cdot\left[\overline{P}_{Y^{n}}^{({\cal C}_{n})}(y^{n})\right]^{\lambda^{*}}\right\} \right\} .\label{eq: Chernoff representation for CD codes blocklength n random coding}
\end{align}
Instead of maximizing over ${\cal C}_{n}$, we use ensemble averages.
To this end, let us consider a sequence of conditional type-classes
$Q_{U|X}^{(n)}\in{\cal P}_{n}({\cal U},Q_{X})$ such that $Q_{U|X}^{(n)}\to Q_{U|X}^{*}$
as $n\to\infty$ (as the types are dense in the simplex, such a sequence
always exists). Furthermore, there exists a sequence $\rho_{\s[c]}^{(n)}$
with $\rho_{\s[c]}^{(n)}\to\rho_{\s[c]}^{*}$ such that $\rho_{\s[c]}^{(n)}\geq\rho-H_{Q^{(n)}}(X|U)$
where $Q_{UX}^{(n)}=Q_{X}\times Q_{U|X}^{(n)}$ . Then, using Lemma
\ref{lem: average of Chernoff} for the hierarchical ensemble defined
by rates $(\rho,\rho_{c}^{(n)})$ and types $(Q_{X},Q_{U|X}^{(n)})$
we obtain that
\[
\E\left\{ \sum_{y^{n}\in{\cal Y}^{n}}\left[P_{Y^{n}}^{(\mathfrak{C}_{n})}(y^{n})\right]^{1-\lambda^{*}}\cdot\left[\overline{P}_{Y^{n}}^{(\mathfrak{C}_{n})}(y^{n})\right]^{\lambda^{*}}\right\} \doteq\exp\left[-n\cdot\min\left\{ d_{\lambda^{*}}(Q_{X}),\;A_{\s[rc]}(\rho,\rho_{\s[c]}^{(n)},Q_{UX}^{(n)},\lambda^{*})\right\} \right],
\]
and using Lemma \ref{lem: average of distinct codewords}, we obtain
that the average number of distinct codewords in a randomly chosen
codebook is $|{\cal C}_{n}|\doteq e^{n\rho}$. It remains to prove
the existence of a CD code, whose codewords are all distinct, and
its Chernoff parameter exponent is close to the ensemble average.
To this end, consider the events
\[
{\cal A}_{1}\dfn\left\{ \left|\mathfrak{C}_{n}\right|\geq\frac{1}{2}\E\left[\left|{\cal \mathfrak{C}}_{n}\right|\right]\right\} ,
\]
and 
\[
{\cal A}_{2}\dfn\left\{ \sum_{y^{n}\in{\cal Y}^{n}}\left[P_{Y^{n}}^{(\mathfrak{C}_{n})}(y^{n})\right]^{1-\lambda^{*}}\cdot\left[\overline{P}_{Y^{n}}^{(\mathfrak{C}_{n})}(y^{n})\right]^{\lambda^{*}}\leq e^{4n\delta}\cdot\exp\left[-n\cdot\min\left\{ d_{\lambda^{*}}(Q_{X}),\;A_{\s[rc]}^{(n)}\right\} \right]\right\} ,
\]
where, for brevity, we denote $A_{\s[rc]}^{(n)}\dfn A_{\s[rc]}(\rho,\rho_{\s[c]}^{(n)},Q_{UX}^{(n)},\lambda^{*})$.
Note that since $\P(|{\cal \mathfrak{C}}_{n}|\leq e^{n\delta}\cdot\E[|\mathfrak{C}_{n}|])=\P(|\mathfrak{C}_{n}|\leq e^{n\rho})=1$,
for all $n$ sufficiently large, the reverse Markov inequality\footnote{The reverse Markov inequality states that if $\P(0\leq X\leq\alpha\E[X])=1$
for some $\alpha>1$. Then, for any $\beta<1$,$\P\left(X>\beta\E[X]\right)\geq\frac{1-\beta}{\alpha-\beta}.$} \cite[Section 9.3, p. 159]{loeve} implies that 
\begin{equation}
\P\left[{\cal A}_{1}\right]\geq\frac{1-\nicefrac{1}{2}}{e^{n\delta}-\nicefrac{1}{2}}\geq e^{-2n\delta}.\label{eq: random coding existence of good rate}
\end{equation}
Further, Markov's inequality implies that for all $n$ sufficiently
large
\begin{equation}
\P\left[{\cal A}_{2}\right]\geq1-e^{-n3\delta}.\label{eq: random coding existence of good Chernoff distance}
\end{equation}
Then, we note that 
\begin{align}
\P\left[{\cal A}_{1}\cap{\cal A}_{2}\right] & \geq1-\P[{\cal A}_{1}^{c}]-\P\left[{\cal A}_{2}^{c}\right]\\
 & \geq1-e^{-3n\delta}-\left(1-e^{-2n\delta}\right)\\
 & =e^{-2n\delta}-e^{-3n\delta}\\
 & >0,
\end{align}
and thus deduce that there exists a CD code ${\cal C}_{n}^{*}$ such
that $|{\cal C}_{n}^{*}|\geq\frac{1}{4}e^{n(\rho-\delta)}\geq e^{n(\rho-2\delta)}$
and ${\cal A}_{2}$ holds for all $n$ sufficiently large. Let the
CD code obtained after keeping only the unique codewords of ${\cal C}_{n}^{*}$
be denoted as ${\cal C}_{n}^{**}$. It remains to show that the exponent
of the Chernoff parameter of ${\cal C}_{n}^{**}$ is asymptotically
equal to that of ${\cal C}_{n}^{*}$. Indeed, 
\begin{align}
 & \sum_{y^{n}\in{\cal Y}^{n}}\left[P_{Y^{n}}^{({\cal C}_{n}^{*})}(y^{n})\right]^{1-\lambda^{*}}\cdot\left[\overline{P}_{Y^{n}}^{({\cal C}_{n}^{*})}(y^{n})\right]^{\lambda^{*}}\nonumber \\
 & =\sum_{y^{n}\in{\cal Y}^{n}}\left[\sum_{x^{n}\in{\cal C}_{n}^{*}}\frac{1}{e^{n\rho}}P_{Y|X}(y^{n}|x^{n})\right]^{1-\lambda^{*}}\cdot\left[\sum_{\overline{x}^{n}\in{\cal C}_{n}^{*}}\frac{1}{e^{n\rho}}\overline{P}_{Y|X}(y^{n}|\overline{x}^{n})\right]^{\lambda^{*}}\\
 & \trre[\geq,a]\sum_{y^{n}\in{\cal Y}^{n}}\left[\sum_{x^{n}\in{\cal C}_{n}^{**}}\frac{1}{e^{n\rho}}P_{Y|X}(y^{n}|x^{n})\right]^{1-\lambda^{*}}\cdot\left[\sum_{\overline{x}^{n}\in{\cal C}_{n}^{**}}\frac{1}{e^{n\rho}}\overline{P}_{Y|X}(y^{n}|\overline{x}^{n})\right]^{\lambda^{*}}\\
 & \trre[\geq,b]e^{-2n\delta}\cdot\sum_{y^{n}\in{\cal Y}^{n}}\left[\sum_{x^{n}\in{\cal C}_{n}^{**}}\frac{1}{\left|{\cal C}_{n}^{**}\right|}P_{Y|X}(y^{n}|x^{n})\right]^{1-\lambda^{*}}\cdot\left[\sum_{\overline{x}^{n}\in{\cal C}_{n}^{**}}\frac{1}{\left|{\cal C}_{n}^{**}\right|}\overline{P}_{Y|X}(y^{n}|\overline{x}^{n})\right]^{\lambda^{*}},
\end{align}
where $(a)$ follows since ${\cal C}_{n}^{**}\subseteq{\cal C}_{n}^{*}$,
and $(b)$ follows since $|{\cal C}_{n}^{**}|\geq e^{n(\rho-2\delta)}$,
and therefore 
\[
\sum_{y^{n}\in{\cal Y}^{n}}\left[P_{Y^{n}}^{({\cal C}_{n}^{**})}(y^{n})\right]^{1-\lambda^{*}}\cdot\left[\overline{P}_{Y^{n}}^{({\cal C}_{n}^{**})}(y^{n})\right]^{\lambda^{*}}\leq e^{6n\delta}\cdot\exp\left[-n\cdot\min\left\{ d_{\lambda^{*}}(Q_{X}),\;A_{\s[rc]}^{(n)}\right\} \right].
\]

With the above, the derivation of \eqref{eq: Chernoff representation for CD codes blocklength n random coding}
may be continued as 
\begin{align}
 & \max_{{\cal C}_{n}\subseteq{\cal T}_{n}(Q_{X}):\;|{\cal C}_{n}|\geq e^{n\rho},\;p_{1}({\cal {\cal C}}_{n},\phi_{n})\leq e^{-nF_{1}}}-\frac{1}{n}\log p_{2}({\cal {\cal C}}_{n},\phi_{n})\nonumber \\
 & \geq-\frac{1-\lambda^{*}}{\lambda^{*}}\cdot F_{1}-\frac{1}{\lambda^{*}}\cdot\frac{1}{n}\log\left\{ \sum_{y^{n}\in{\cal Y}^{n}}\left[P_{Y^{n}}^{({\cal C}_{n}^{**})}(y^{n})\right]^{1-\lambda^{*}}\cdot\left[\overline{P}_{Y^{n}}^{({\cal C}_{n}^{**})}(y^{n})\right]^{\lambda^{*}}\right\} \\
 & \geq-\frac{1-\lambda^{*}}{\lambda^{*}}\cdot F_{1}+\frac{1}{\lambda^{*}}\cdot\left[\min\left\{ d_{\lambda^{*}}(Q_{X}),\;A_{\s[rc]}(\rho,\rho_{\s[c]}^{(n)},Q_{UX}^{(n)},\lambda^{*})\right\} -6\delta\right].
\end{align}
Now, taking the limit $n\to\infty$ over $n$'s such that ${\cal T}_{n}(Q_{X})$
is not empty, 
\begin{align}
 & \lim_{n\to\infty}\max_{{\cal C}_{n}\subseteq{\cal T}_{n}(Q_{X}):\;|{\cal C}_{n}|\geq e^{n\rho},\;p_{1}({\cal {\cal C}}_{n},\phi_{n})\leq e^{-nF_{1}}}-\frac{1}{n}\log p_{2}({\cal {\cal C}}_{n},\phi_{n})\nonumber \\
 & \geq-\frac{1-\lambda^{*}}{\lambda^{*}}\cdot F_{1}+\frac{1}{\lambda^{*}}\cdot\left[\min\left\{ d_{\lambda^{*}}(Q_{X}),\;\lim_{n\to\infty}A_{\s[rc]}(\rho,\rho_{\s[c]}^{(n)},Q_{UX}^{(n)},\lambda^{*})\right\} -6\delta\right]\\
 & =-\frac{1-\lambda^{*}}{\lambda^{*}}\cdot F_{1}+\frac{1}{\lambda^{*}}\cdot\left[\min\left\{ d_{\lambda^{*}}(Q_{X}),\;A_{\s[rc]}(\rho,\rho_{\s[c]}^{*},Q_{UX}^{*},\lambda^{*})\right\} -6\delta\right]\\
 & \geq\max_{0\leq\lambda\leq1}\left\{ -\frac{1-\lambda}{\lambda}\cdot F_{1}+\frac{1}{\lambda}\cdot\min\left[d_{\lambda}(Q_{X}),\;\sup_{Q_{U|X}}\sup_{\rho_{\s[c]}:\;\rho_{\s[c]}\geq\left|\rho-H_{Q}(X|U)\right|_{+}}A_{\s[rc]}(\rho,\rho_{\s[c]},Q_{UX},\lambda)\right]\right\} -7\delta,
\end{align}
where the first equality follows from the continuity of $A_{\s[rc]}(\rho,\rho_{\s[c]},Q_{UX},\lambda)$
in $(Q_{UX},\rho_{\s[c]})$ (which can be readily verified from \eqref{eq: A_rc_tag}
and \eqref{eq: A_rc_tag2}), and the second inequality from the definition
of $(\lambda^{*},Q_{UX}^{*},\rho_{\s[c]}^{*})$. The proof is completed
by taking $\delta\downarrow0$.
\end{IEEEproof}
\begin{rem}
\label{rem:good for all type 1 exponent}As mentioned after Theorem
\ref{thm: random coding bound}, the random-coding bound can be achieved
by a single sequence of CD codes, simultaneously for all type 1 error
exponent constraint $F_{1}$. This can be proved by showing that there
exists a CD code such that the event ${\cal A}_{2}(\lambda)$ defined
in \eqref{eq: random coding existence of good Chernoff distance}
holds for all $\lambda\in[0,1]$. To show the latter, we uniformly
quantize the interval $[0,1]$ to $\{\lambda_{i}\}_{i=0}^{K}$ with
$\lambda_{i}=\frac{i}{K}$ and a fixed $K$. Then, using the union
bound, for all $n$ sufficiently large
\begin{align}
\P\left[\bigcap_{i=0}^{K}{\cal A}_{2}(\lambda_{i})\right] & \geq1-\sum_{i=0}^{K}\P\left[{\cal A}_{2}^{c}(\lambda_{i})\right]\\
 & \geq1-(K+1)\cdot e^{-3n\delta}\\
 & \geq1-e^{-2n\delta},
\end{align}
and this bound can be used in lieu of \eqref{eq: random coding existence of good Chernoff distance}
in the proof. This will prove the simultaneous achievability of ${\cal A}_{2}(\lambda_{i})$
for all $0\leq i\leq K$. Then, utilizing the continuity of $A_{\s[rc]}(\rho,\rho_{\s[c]},Q_{UX},\lambda)$,
by taking $K$ to increase sub-exponentially in $n$ the same result
can be established to the entire $[0,1]$ interval.
\end{rem}
\begin{IEEEproof}[Proof of Theorem \ref{thm: Expurgated bound}]
 As in the proof of Theorem \ref{thm: random coding bound} we begin
with \eqref{eq: Chernoff representation for CD codes blocklength n random coding}.
Then, we use the property shown in \cite[Appendix E]{Weinberger_merhav15},
which states that for any $\delta>0$ and all $n$ sufficiently large,
there exists a CD code ${\cal C}_{n}^{*}$ (of rate $\rho$) such
that 
\[
\sum_{y^{n}\in{\cal Y}^{n}}\left[P_{Y^{n}}^{({\cal C}_{n}^{*})}(y^{n})\right]^{1-\lambda}\left[\overline{P}_{Y^{n}}^{({\cal C}_{n}^{*})}(y^{n})\right]^{\lambda}\leq\exp\left[-n\cdot\min\left\{ d_{\lambda}(Q_{X}),\;A_{\s[ex]}(\rho,Q_{X},\lambda)\right\} \right].
\]
Substituting this bound to \eqref{eq: Chernoff representation for CD codes blocklength n random coding},
taking $n\to\infty$ and $\delta\downarrow0$ completes the proof
of the theorem.
\end{IEEEproof}

\section{The Type-Enumeration Method and the Proof of Proposition \ref{prop: Correlation of hierarchical enumeartors}\label{sec:The-Type-Enumeration-Method}}

We begin with a short review of the \emph{type-enumeration method}
\cite[Sec. 6.3]{Merhav09}. To begin, let us define type-class enumerators
for the cloud centers by
\begin{equation}
N_{y^{n}}(Q_{UY})\dfn\left|\left\{ u^{n}\in\mathfrak{C}_{\s[c],n}:(u^{n},y^{n})\in{\cal T}_{n}(Q_{UY})\right\} \right|.\label{eq: cloud centers enumerators}
\end{equation}
To wit, $N_{y^{n}}(Q_{UY})$ counts the random number of cloud centers
which have joint type $Q_{UY}\in{\cal P}_{n}({\cal U}\times{\cal Y})$
with $y^{n}$. While $M_{y^{n}}(Q_{UXY})$ defined in \eqref{eq: hierarchical codewords enumerator}
is an enumerator of a hierarchical ensemble, $N_{y^{n}}(Q_{UY})$
is an enumerator of an ordinary ensemble, and thus simpler to analyze.
Furthermore, the analysis of $M_{y^{n}}(Q_{UXY})$ depends on the
properties of $N_{y^{n}}(Q_{UY})$, and thus we begin by analyzing
the latter.

As the cloud centers in the ensemble are drawn independently, $N_{y^{n}}(Q_{UY})$
is a binomial random variable. It pertains to $e^{n\rho_{\s[c]}}$
trials and probability of success of the exponential order of $\exp[-n\cdot I_{Q}(U;Y)]$,
and consequently, $\E[N_{y^{n}}(Q_{UY})]\doteq\exp\{n\cdot[\rho_{\s[c]}-I_{Q}(U;Y)]\}$.
In the sequel, we will need refined properties of the enumerator,
and specifically, its large-deviations behavior and the moments $\E[N_{y^{n}}^{\lambda}(Q_{UY})]$.
To this end, we note that as $N_{y^{n}}(Q_{UY})$ is just an enumerator
for a code drawn from an ordinary ensemble, the analysis of \cite[Sec. 6.3]{Merhav09}
\cite[Appendix A.2]{Merhav2009_relations} holds. As was shown there,
when $I_{Q}(U;Y)\leq\rho_{\s[c]}$, $\E[N_{y^{n}}(Q_{UY})]$ increases
exponentially with $n$ as $\exp\{n\cdot[\rho_{\s[c]}-I_{Q}(U;Y)]\}$,
and $N_{y^{n}}(Q_{UY})$ concentrates double-exponentially rapidly
around this average. Specifically, letting
\begin{equation}
{\cal B}_{n}(Q_{UY},\delta)\dfn\left\{ e^{-n\delta}\cdot\E\left[N_{y^{n}}(Q_{UY})\right]\leq N_{y^{n}}(Q_{UY})\leq e^{n\delta}\cdot\E\left[N_{y^{n}}(Q_{UY})\right]\right\} ,\label{eq: enumerator concentrates at average}
\end{equation}
then for any $\delta>0$ sufficiently small 
\begin{equation}
\P\left[{\cal B}_{n}^{c}(Q_{UY},\delta)\right]\leq\exp\left[-e^{n\delta}\right].\label{eq: enumerator concntration, high rate}
\end{equation}
When $I_{Q}(U;Y)>\rho_{\s[c]}$ holds, $\E[N_{y^{n}}(Q_{UY})]$ decreases
exponentially with $n$ as $\exp\{-n\cdot[I_{Q}(U;Y)-\rho_{\s[c]}]\}$,
and $N_{y^{n}}(Q_{UY})=0$ almost surely. Furthermore, the probability
that even a single codeword has joint type $Q_{UY}$ with $y^{n}$
is exponentially small, and the probability that there is an exponential
number of such codewords is double-exponentially small. Specifically,
for all sufficiently large $n$
\begin{equation}
\P\left\{ N_{y^{n}}(Q_{UY})\ge1\right\} \leq\exp\left\{ -n\cdot\left[I_{Q}(U;Y)-\rho_{\s[c]}\right]\right\} ,\label{eq: enumerator large deviation, high rate}
\end{equation}
and for any given $\delta>0$
\begin{equation}
\P\left\{ N_{y^{n}}(Q_{UY})\ge e^{2n\delta}\right\} \leq\exp\left[-e^{n\delta}\right].\label{eq: enumerator concntration, low rate}
\end{equation}
Using the properties above, it can be easily deduced that
\begin{align}
\E\left[N_{y^{n}}^{\lambda}(Q_{UY})\right] & \doteq\begin{cases}
\exp\left\{ n\lambda\cdot\left[\rho_{\s[c]}-I_{Q}(U;Y)\right]\right\} , & I_{Q}(U;Y)\leq\rho_{\s[c]}\\
\exp\left\{ -n\cdot\left[I_{Q}(U;Y)-\rho_{\s[c]}\right]\right\} , & I_{Q}(U;Y)>\rho_{\s[c]}
\end{cases},\label{eq: enumerator moment}
\end{align}
or, in an equivalent and more compact form,
\begin{align}
\E\left[N_{y^{n}}^{\lambda}(Q_{UY})\right] & \doteq\exp\left(n\cdot\left[\lambda\left[\rho_{\s[c]}-I_{Q}(U;Y)\right]-(1-\lambda)\cdot\left|I_{Q}(U;Y)-\rho_{\s[c]}\right|_{+}\right]\right).\label{eq: enumeartor moment no cases}
\end{align}

We can now turn to analyze the behavior of the more complicated enumerator
$M_{y^{n}}(Q_{UXY})$. To this end, note that conditioned on the event
$N_{y^{n}}(Q_{UY})=e^{n\nu},$ $M_{y^{n}}(Q_{UXY})$ is a binomial
random variable pertaining to $\exp[n(\nu+\rho_{\s[s]})]$ trials
and probability of success of the exponential order of $\exp[-n\cdot I_{Q}(X;Y|U)]$.
Thus,
\begin{equation}
\E\left[M_{y^{n}}^{\lambda}(Q_{UXY})\Big|N_{y^{n}}(Q_{UY})=e^{n\nu}\right]\doteq\begin{cases}
\exp\left\{ n\lambda\cdot\left[\nu+\rho_{\s[s]}-I_{Q}(X;Y|U)\right]\right\} , & I_{Q}(X;Y|U)\leq\nu+\rho_{\s[s]}\\
\exp\left\{ -n\cdot\left[I_{Q}(X;Y|U)-\nu-\rho_{\s[s]}\right]\right\} , & I_{Q}(X;Y|U)>\nu+\rho_{\s[s]}
\end{cases},\label{eq: hierarchical moment conditional on cloud center moment}
\end{equation}
and the conditional large-deviations behavior of $M_{y^{n}}(Q_{UXY})$
is identical to the large-deviations behavior of $N_{y^{n}}(Q_{UY})$,
with $\nu+\rho_{\s[s]}$ and $I_{Q}(X;Y|U)$ replacing $\rho_{\s[c]}$
and $I_{Q}(U;Y)$, respectively. The next lemma provides an asymptotic
expression for the unconditional moments of $M_{y^{n}}(Q_{UXY})$.
It can be easily seen that \eqref{eq: enumeartor moment no cases}
is obtained as a special case, when setting $U=X$ and $\rho=\rho_{\s[c]}$. 
\begin{lem}
\label{lem: enumerators fractional moments}For $\lambda>0$
\begin{equation}
\E\left[M_{y^{n}}^{\lambda}(Q_{UXY})\right]\doteq\exp\left(n\cdot\left[\lambda\left[\rho-I_{Q}(U,X;Y)\right]-(1-\lambda)\cdot\max\left\{ \left|I_{Q}(U;Y)-\rho_{\s[c]}\right|_{+},\;I_{Q}(U,X;Y)-\rho\right\} \right]\right).\label{eq: moments of enumerator M}
\end{equation}
\end{lem}
\begin{IEEEproof}
Let 
\[
\delta\in\left(0,\liminf_{n\to\infty}\frac{1}{n}\log\E[N_{y^{n}}(Q_{UY})]\right),
\]
define the events
\begin{equation}
{\cal A}_{n}^{=0}(Q_{UY})\dfn\left\{ N_{y^{n}}(Q_{UY})=0\right\} ,\label{eq: enumeartor is zero}
\end{equation}
\begin{equation}
{\cal A}_{n}^{=1}(Q_{UY},\delta)\dfn\left\{ 1\leq N_{y^{n}}(Q_{UY})\leq e^{2n\delta}\right\} ,\label{eq: enumeartor is 1}
\end{equation}
\begin{equation}
{\cal A}_{n}^{\geq1}(Q_{UY},\delta)\dfn\left\{ N_{y^{n}}(Q_{UY})\geq e^{2n\delta}\right\} ,\label{eq: enumeartor is exponential}
\end{equation}
and recall the definition of the event ${\cal B}_{n}(Q_{UY},\delta)$
in \eqref{eq: enumerator concentrates at average}. We will consider
four cases depending on the relations between the rates and mutual
information values. For the sake of brevity, only the first case will
be analyzed with a strictly positive $\delta>0$ and then the limit
$\delta\downarrow0$ will be taken. In all other three cases, we shall
derive the expressions for the moments assuming $\delta=0$, with
the understanding that upper and lower bounds can be derived in a
similar manner to the first case. For notational convenience, when
the expressions are derived assuming $\delta=0$ we will omit $\delta$
from the notation of the events defined above {[}e.g., ${\cal A}_{n}^{=1}(Q_{UY})${]}.
We will use the moments \eqref{eq: enumerator moment}, the strong
concentration relation \eqref{eq: enumerator concntration, high rate}
and the large-deviations bound \eqref{eq: enumerator large deviation, high rate},
for both $N_{y^{n}}(Q_{UY})$ and $M_{y^{n}}(Q_{UXY})$ {[}when the
latter is conditioned on the value of $N_{y^{n}}(Q_{UY})${]}. 
\begin{casenv}
\item If $I_{Q}(U;Y)>\rho_{\s[c]}$ and $I_{Q}(X;Y|U)>\rho_{\s[s]}$, then
for all $\delta>0$ sufficiently small, $I_{Q}(U;Y)>\rho_{\s[c]}+\delta$
and $I_{Q}(X;Y|U)>\rho_{\s[s]}+\delta$ and thus,
\begin{align}
 & \E\left[M_{y^{n}}^{\lambda}(Q_{UXY})\right]\nonumber \\
 & \leq\P\left[{\cal A}_{n}^{=0}(Q_{UY})\right]\cdot0+\P\left[{\cal A}_{n}^{=1}(Q_{UY},\delta)\right]\cdot\E\left[M_{y^{n}}^{\lambda}(Q_{UXY})\vert{\cal A}_{n}^{=1}(Q_{UY},\delta)\right]\nonumber \\
 & \hphantom{=}+\P\left[{\cal A}_{n}^{\geq1}(Q_{UY},\delta)\right]\cdot e^{n\lambda\rho}\\
 & \dotleq0+\exp\left\{ -n\cdot\left[I_{Q}(U;Y)-\rho_{\s[c]}-\delta\right]\right\} \cdot\exp\left\{ -n\cdot\left[I_{Q}(X;Y|U)-\rho_{\s[s]}-2\delta\right]\right\} +\exp\left[-e^{n\delta}\right]\cdot e^{n\lambda\rho}\\
 & \doteq\exp\left\{ n\cdot\left[\rho-I_{Q}(U,X;Y)+3\delta\right]\right\} .
\end{align}
Similarly, it can be shown that 
\[
\E\left[M_{y^{n}}^{\lambda}(Q_{UXY})\right]\dotgeq\exp\left\{ n\cdot\left[\rho-I_{Q}(U,X;Y)-3\delta\right]\right\} .
\]
As $\delta\geq0$ is arbitrary, we obtain
\[
\E\left[M_{y^{n}}^{\lambda}(Q_{UXY})\right]\doteq\exp\left\{ n\cdot\left[\rho-I_{Q}(U,X;Y)\right]\right\} .
\]
\item If $I_{Q}(U;Y)>\rho_{\s[c]}$ and $I_{Q}(X;Y|U)<\rho_{\s[s]}$ then
\begin{align}
 & \E\left[M_{y^{n}}^{\lambda}(Q_{UXY})\right]\nonumber \\
 & \leq\P\left[{\cal A}_{n}^{=0}(Q_{UY})\right]\cdot0+\P\left[{\cal A}_{n}^{=1}(Q_{UY})\right]\cdot\E\left[M_{y^{n}}^{\lambda}(Q_{UXY})\vert{\cal A}_{n}^{=1}(Q_{UY})\right]\nonumber \\
 & \hphantom{=}+\P\left[{\cal A}_{n}^{\geq1}(Q_{UY})\right]\cdot e^{n\lambda\rho}\\
 & \doteq0+\exp\left\{ -n\cdot\left[I_{Q}(U;Y)-\rho_{\s[c]}\right]\right\} \cdot\exp\left\{ n\cdot\lambda\cdot\left[\rho_{\s[s]}-I_{Q}(X;Y|U)\right]\right\} +0\\
 & =\exp\left\{ n\cdot\left[\rho_{\s[c]}-I_{Q}(U;Y)+\lambda\cdot\left(\rho_{\s[s]}-I_{Q}(X;Y|U)\right)\right]\right\} \\
 & =\exp\left\{ n\cdot\left[\rho_{\s[c]}-I_{Q}(U;Y)+\rho_{\s[s]}-I_{Q}(X;Y|U)-(1-\lambda)\cdot\left(\rho_{\s[s]}-I_{Q}(X;Y|U)\right)\right]\right\} \\
 & =\exp\left\{ n\cdot\left[\rho-I_{Q}(U,X;Y)-(1-\lambda)\left(\rho_{\s[s]}-I_{Q}(X;Y|U)\right)\right]\right\} .
\end{align}
\item If $I_{Q}(U;Y)<\rho_{\s[c]}$ and $I_{Q}(X;Y|U)>\rho_{\s[c]}-I_{Q}(U;Y)+\rho_{\s[s]}$
then
\begin{align}
 & \E\left[M_{y^{n}}^{\lambda}(Q_{UXY})\right]\nonumber \\
 & =\P\left\{ {\cal B}_{n}(Q_{UY})\right\} \cdot\E\left[M_{y^{n}}^{\lambda}(Q_{UXY})\vert{\cal B}_{n}(Q_{UY})\right]\nonumber \\
 & \hphantom{=}+\P\left\{ {\cal B}_{n}^{c}(Q_{UY})\right\} \cdot\E\left[M_{y^{n}}^{\lambda}(Q_{UXY})\vert{\cal B}_{n}^{c}(Q_{UY})\right]\\
 & \doteq1\cdot\exp\left\{ n\cdot\left[\rho_{\s[c]}-I_{Q}(U;Y)+\rho_{\s[s]}-I_{Q}(X;Y|U)\right]\right\} +0\\
 & =\exp\left\{ n\cdot\left[\rho-I_{Q}(U,X;Y)\right]\right\} .
\end{align}
\item If $I_{Q}(U;Y)<\rho_{\s[c]}$ and $I_{Q}(X;Y|U)<\rho_{\s[c]}-I_{Q}(U;Y)+\rho_{\s[s]}$
then
\begin{align}
 & \E\left[M_{y^{n}}^{\lambda}(Q_{UXY})\right]\nonumber \\
 & =\P\left\{ {\cal B}_{n}(Q_{UY})\right\} \cdot\E\left[M_{y^{n}}^{\lambda}(Q_{UXY})\vert{\cal B}_{n}(Q_{UY})\right]\nonumber \\
 & \hphantom{=}+\P\left\{ {\cal B}_{n}^{c}(Q_{UY})\right\} \cdot\E\left[M_{y^{n}}^{\lambda}(Q_{UXY})\vert{\cal B}_{n}^{c}(Q_{UY})\right]\\
 & \doteq1\cdot\exp\left\{ n\cdot\lambda\cdot\left[\rho_{\s[c]}-I_{Q}(U;Y)+\rho_{\s[s]}-I_{Q}(X;Y|U)\right]\right\} +0\\
 & =\exp\left\{ n\cdot\lambda\cdot\left[\rho-I_{Q}(U,X;Y)\right]\right\} \\
 & =\exp\left\{ n\cdot\left[\rho-I_{Q}(U,X;Y)-(1-\lambda)\left[\rho-I_{Q}(U,X;Y)\right]\right]\right\} .
\end{align}
\end{casenv}
Noting that $I_{Q}(X;Y|U)>\rho_{\s[c]}-I_{Q}(U;Y)+\rho_{\s[s]}$ is
equivalent to $I_{Q}(U,X;Y)>\rho$, it is easy to verify that
\[
\E\left[M_{y^{n}}^{\lambda}(Q_{UXY})\right]\doteq\exp\left(n\cdot\left\{ \rho-I_{Q}(U,X;Y)-(1-\lambda)\left|\rho_{\s[s]}-I_{Q}(X;Y|U)+\left|\rho_{\s[c]}-I_{Q}(U;Y)\right|_{+}\right|_{+}\right\} \right)
\]
matches all four cases. The final expression is obtained from the
identities
\begin{align}
 & \rho-I_{Q}(U,X;Y)-(1-\lambda)\cdot\left|\rho_{\s[s]}-I_{Q}(X;Y|U)+\left|\rho_{\s[c]}-I_{Q}(U;Y)\right|_{+}\right|_{+}\nonumber \\
 & \trre[=,a]\rho-I_{Q}(U,X;Y)-(1-\lambda)\left|\rho-I_{Q}(U,X;Y)+\left|I_{Q}(U;Y)-\rho_{\s[c]}\right|_{+}\right|_{+}\\
 & \trre[=,b]\lambda\left[\rho-I_{Q}(U,X;Y)\right]-(1-\lambda)\left|I_{Q}(U;Y)-\rho_{\s[c]}\right|_{+}-(1-\lambda)\left|I_{Q}(U,X;Y)-\rho-\left|I_{Q}(U;Y)-\rho_{\s[c]}\right|_{+}\right|_{+}\\
 & \trre[=,c]\lambda\left[\rho-I_{Q}(U,X;Y)\right]-(1-\lambda)\cdot\max\left\{ \left|I_{Q}(U;Y)-\rho_{\s[c]}\right|_{+},\;I_{Q}(U,X;Y)-\rho\right\} ,
\end{align}
where $(a)$ follows from the identity $\left|t\right|_{+}=t+\left|-t\right|_{+}$
with $t=\rho_{\s[c]}-I_{Q}(U;Y)$, $(b)$ follows from the same identity
with $t=\rho-I_{Q}(U,X;Y)+|I_{Q}(U;Y)-\rho_{\s[c]}|_{+}$ , and $(c)$
follows from $t+\left|s-t\right|_{+}=\max\left\{ t,\;s\right\} $
with $t=|I_{Q}(U;Y)-\rho_{\s[c]}|_{+}$ and $s=I_{Q}(U,X;Y)-\rho$.
\end{IEEEproof}
To continue, we will need to separate the analysis according to whether
$Q_{UY}\neq\overline{Q}_{UY}$ (Lemma \ref{lem: correlation of enumerators- different clouds})
or $Q_{UY}=\overline{Q}_{UY}$ (Lemma \ref{lem: correlation of enumerators- same clouds}).
In the former case $M_{y^{n}}(Q_{UXY})$ and $M_{y^{n}}(\overline{Q}_{UXY})$
count codewords which pertain to different cloud centers, and as we
shall next see, $M_{y^{n}}^{1-\lambda}(Q_{UXY})$ and $M_{y^{n}}^{\lambda}(\overline{Q}_{UXY})$
are asymptotically \emph{uncorrelated}. In the later case, these enumerators
may count codewords which pertain to the same cloud center, and correlation
between $M_{y^{n}}^{1-\lambda}(Q_{UXY})$ and $M_{y^{n}}^{\lambda}(\overline{Q}_{UXY})$
is possible. We will need two auxiliary lemmas.
\begin{lem}
\label{lem: avarge of monotone function w.r.t. binomials - number of trials}Let
$f(t)$ be a monotonically non-decreasing function, let $B\sim\binomial(N,p)$
and $\tilde{B}\sim\binomial(\tilde{N},p)$ with $\tilde{N}>N$. Then,
\[
\E\left[f(B)\right]\leq\E\left[f(\tilde{B})\right].
\]
\end{lem}
\begin{IEEEproof}
Let $L\dfn N-\tilde{N}$ and $A\sim\binomial(L,p)$, independent of
$B$. As the sum of independent binomial random variables with the
same success probability is also binomially distributed, we have $B+A$
is equal in distribution to $\tilde{B}\sim\binomial(\tilde{N},p)$.
Thus,
\begin{align}
\E\left[f(\tilde{B})\right] & =\E\left[f(B+A)\right]\\
 & \geq\E\left[f(B)\right],
\end{align}
where the inequality holds pointwise for any given $A=a$, and thus
also under expectation. 
\end{IEEEproof}
As is well known from the method of types, $\P[(U^{n},y^{n})\in{\cal T}_{n}(Q_{UY})]\doteq\exp[-nI_{Q}(U;Y)]$.
The next lemma shows that this probability can be upper bounded asymptotically
even when $I_{Q}(U;Y)=0$, and the large-deviations behavior does
not hold. 
\begin{lem}
\label{lem: probability of being exactly in type} Let $Q_{UY}\in{\cal P}({\cal U}\times{\cal Y})$
be given such that $\supp(Q_{U})\geq2$ and $\supp(Q_{Y})\geq2$.
Also let $y^{n}\in{\cal T}_{n}(Q_{Y})$ and assume that $U^{n}$ is
distributed uniformly over ${\cal T}_{n}(Q_{U})$. Then, for any given
$\epsilon>0$ there exists $n_{0}(Q_{U},Q_{Y})$ such that for all
$n\geq n_{0}(Q_{U},Q_{Y})$ 
\[
\P\left[\left(U^{n},y^{n}\right)\in{\cal T}_{n}(Q_{UY})\right]\leq\epsilon.
\]
\end{lem}
\begin{IEEEproof}
Using Robbins' sharpening of Stirling's formula (e.g., \cite[Problem 2.2]{csiszar2011information}),
it can be shown that the size of a type class satisfies 
\begin{equation}
\left|{\cal T}_{n}(Q_{X})\right|\cong\exp\left\{ n\cdot\left[H(Q_{X})-\left(\frac{\supp(Q_{X})-1}{2}\right)\frac{\log n}{n}\right]\right\} n^{-\frac{1}{2}\left[\supp(Q_{X})-1\right]}.\label{eq: exact size of type class}
\end{equation}
Hence,
\begin{align}
 & \P\left[\left(U^{n},y^{n}\right)\in{\cal T}_{n}(Q_{UY})\right]\nonumber \\
 & \trre[=,a]\P\left[\left(U^{n},Y^{n}\right)\in{\cal T}_{n}(Q_{UY})\right]\\
 & =\frac{\left|{\cal T}_{n}(Q_{UY})\right|}{\left|{\cal T}_{n}(Q_{U})\right|\left|{\cal T}_{n}(Q_{Y})\right|}\\
 & \cong e^{-n\cdot I_{Q}(U;Y)}\cdot n^{-\frac{1}{2}\cdot\left[\supp(Q_{UY})-\supp(Q_{U})-\supp(Q_{Y})+1\right]},\label{eq: bound on exact type probability Robbins}
\end{align}
where $(a)$ holds by symmetry {[}assuming that $Y^{n}$ is drawn
uniformly over ${\cal T}_{n}(Q_{Y})${]}. 

Now, by the Kullback-Csisz\'{a}r-Kemperman-Pinsker inequality \cite[Lemma 11.6.1]{Cover:2006:EIT:1146355}\cite[Exercise 3.18]{csiszar2011information}
\begin{align}
I_{Q}(U;Y) & =D(Q_{UY}||Q_{U}\times Q_{Y})\\
 & \geq\frac{1}{2\log2}\left\Vert Q_{UY}-Q_{U}\times Q_{Y}\right\Vert ^{2},
\end{align}
and thus for any given $\delta>0$
\begin{align}
 & \left\{ \tilde{Q}_{UY}:\tilde{Q}_{U}=Q_{U},\;\tilde{Q}_{Y}=Q_{Y},\;I_{\tilde{Q}}(U;Y)\leq\delta\right\} \nonumber \\
 & \subseteq\left\{ \tilde{Q}_{UY}:\tilde{Q}_{U}=Q_{U},\;\tilde{Q}_{Y}=Q_{Y},\;\left\Vert \tilde{Q}_{UY}-Q_{U}\times Q_{Y}\right\Vert \leq\eta\right\} \\
 & \dfn{\cal J}(\eta,Q_{U},Q_{Y}),
\end{align}
where $\eta\dfn\sqrt{2\delta\log2}$. Choose $\eta(Q_{U,}Q_{Y})$
such that $\supp(\tilde{Q}_{UY})=\supp(Q_{U}\times Q_{Y})=\supp(Q_{U})\cdot\supp(Q_{Y})$
for all $\tilde{Q}_{UY}\in{\cal J}(\delta_{0},Q_{U},Q_{Y})$. We consider
two cases:
\begin{casenv}
\item If $Q_{UY}\in{\cal J}(\delta,Q_{U},Q_{Y})$ then it is elementary
to verify that in this event, as $\supp(Q_{U})\geq2$ and $\supp(Q_{Y})\geq2$
was assumed, 
\begin{align}
 & \supp(Q_{UY})-\supp(Q_{U})-\supp(Q_{Y})\nonumber \\
 & =\supp(Q_{U})\cdot\supp(Q_{Y})-\supp(Q_{U})-\supp(Q_{Y})\\
 & \geq0,
\end{align}
and thus \eqref{eq: bound on exact type probability Robbins} implies
that
\[
\P\left[\left(U^{n},y^{n}\right)\in{\cal T}_{n}(Q_{UY})\right]\leq\frac{1}{\sqrt{n}}.
\]
\item If $Q_{UY}\not\in{\cal J}(\delta,Q_{U},Q_{Y})$ then $I_{Q}(U;Y)\geq\delta=\frac{1}{2\log2}\eta^{2}$
and 
\[
\P\left[\left(U^{n},y^{n}\right)\in{\cal T}_{n}(Q_{UY})\right]\leq e^{-n\frac{1}{2\log2}\eta^{2}}.
\]
\end{casenv}
This completes the proof. 
\end{IEEEproof}
We are now ready to state and prove the asymptotic uncorrelation lemma
for $Q_{UY}\neq\overline{Q}_{UY}$. 
\begin{lem}
\label{lem: correlation of enumerators- different clouds} Let $(Q_{UXY},\overline{Q}_{UXY})$
be given such that $Q_{UY}\neq\overline{Q}_{UY}$. Then, 
\[
\E\left[M_{y^{n}}^{1-\lambda}(Q_{UXY})M_{y^{n}}^{\lambda}(\overline{Q}_{UXY})\right]\doteq\E\left[M_{y^{n}}^{1-\lambda}(Q_{UXY})\right]\cdot\E\left[M_{y^{n}}^{\lambda}(\overline{Q}_{UXY})\right].
\]
\end{lem}
\begin{IEEEproof}
We begin by lower bounding the correlation. To this end, let us decompose
\[
M_{y^{n}}(Q_{UXY})=M_{y^{n}}(Q_{UXY},1)+M_{y^{n}}(Q_{UXY},2),
\]
where $M_{y^{n}}(Q_{UXY},1)$ is the enumerator of to the subcode
of ${\cal C}_{n}$ of codewords which pertain to half of the cloud
centers (say, for cloud centers with odd indices) and $M_{y^{n}}(Q_{UXY},2)$
is the enumerator pertaining to the rest of the codewords. Note that
$M_{y^{n}}(Q_{UXY},1)$ and $M_{y^{n}}(\overline{Q}_{UXY},2)$ are
independent for any given $(Q_{UXY},\overline{Q}_{UXY})$. Hence,
\begin{align}
 & \E\left[M_{y^{n}}^{1-\lambda}(Q_{UXY})M_{y^{n}}^{\lambda}(\overline{Q}_{UXY})\right]\nonumber \\
 & =\E\left\{ \left[M_{y^{n}}(Q_{UXY},1)+M_{y^{n}}(Q_{UXY},2)\right]^{1-\lambda}\cdot\left[M_{y^{n}}(\overline{Q}_{UXY},1)+M_{y^{n}}(\overline{Q}_{UXY},2)\right]^{\lambda}\right\} \\
 & \trre[\geq,a]\E\left[M_{y^{n}}^{1-\lambda}(Q_{UXY},1)\cdot M_{y^{n}}^{\lambda}(\overline{Q}_{UXY},2)\right]\\
 & \trre[=,b]\E\left[M_{y^{n}}^{1-\lambda}(Q_{UXY},1)\right]\cdot\E\left[M_{y^{n}}^{\lambda}(\overline{Q}_{UXY},2)\right]\\
 & \trre[\doteq,c]\E\left[M_{y^{n}}^{1-\lambda}(Q_{UXY})\right]\cdot\E\left[M_{y^{n}}^{\lambda}(\overline{Q}_{UXY})\right],\label{eq: fractional moments correlation lower bound}
\end{align}
where $(a)$ follows since enumerators are positive, and $(b)$ follows
since $M_{y^{n}}(Q_{UXY},1)$ and $M_{y^{n}}(\overline{Q}_{UXY},2)$
are independent for any given $(Q_{UXY},\overline{Q}_{UXY})$. $(c)$
holds true since $M_{y^{n}}(Q_{UXY},i)$ ($i=1,2$) pertain to codebooks
of cloud-center size $\frac{1}{2}e^{n\rho_{\s[c]}}\doteq e^{n\rho_{\s[c]}}$
and satellite rate $\rho_{\s[s]}$, and thus clearly $\E[M_{y^{n}}^{\lambda}(Q_{UXY},i)]\doteq\E[M_{y^{n}}^{\lambda}(Q_{UXY})].$

To derive an upper bound on the correlation, we first note two properties.
First, as the codewords enumerated by $M_{y^{n}}(\overline{Q}_{UXY})$
necessarily correspond to different cloud centers from the codewords
enumerated by $M_{y^{n}}(Q_{UXY})$, the following Markov relation
holds: 
\begin{equation}
M_{y^{n}}(Q_{UXY})-N_{y^{n}}(Q_{UY})-N_{y^{n}}(\overline{Q}_{UY})-M_{y^{n}}(\overline{Q}_{UXY}).\label{eq: Markov chain of enumerators}
\end{equation}
Second, conditioned on $N_{y^{n}}(Q_{UY})$, $N_{y^{n}}(\overline{Q}_{UY})$
is a binomial random variable pertaining to $e^{n\rho_{\s[c]}}-N_{y^{n}}(Q_{UY})\leq e^{n\rho_{\s[c]}}$
trials with a probability of success given by
\[
\P\left[\left(U^{n},y^{n}\right)\in{\cal T}_{n}(Q_{UY})\Big|\left(U^{n},y^{n}\right)\not\in{\cal T}_{n}(\overline{Q}_{UY})\right].
\]
However, this probability is not significantly different than the
unconditional probability. More rigorously, for any $\epsilon\in(0,1),$
and all $n>n_{0}(Q_{U},Q_{Y})$ sufficiently large
\begin{equation}
\P\left[\left(U^{n},y^{n}\right)\in{\cal T}_{n}(Q_{UY})\right]\leq\P\left[\left(U^{n},y^{n}\right)\in{\cal T}_{n}(Q_{UY})\Big|\left(U^{n},y^{n}\right)\not\in{\cal T}_{n}(\overline{Q}_{UY})\right]\leq\frac{1}{1-\epsilon}\cdot\P\left[\left(U^{n},y^{n}\right)\in{\cal T}_{n}(Q_{UY})\right].\label{eq: conditional success probability does not change much}
\end{equation}
To see this, note that since $Q_{UY}\neq\overline{Q}_{UY}$ we must
have $\supp(Q_{Y})\geq2$, and thus Lemma \ref{lem: probability of being exactly in type}
implies that for any $\epsilon\in(0,1),$ and all $n$ sufficiently
large
\begin{align}
 & \P\left[\left(U^{n},y^{n}\right)\in{\cal T}_{n}(Q_{UY})\Big|\left(U^{n},y^{n}\right)\not\in{\cal T}_{n}(\overline{Q}_{UY})\right]\nonumber \\
 & =\frac{\P\left[\left(U^{n},y^{n}\right)\in{\cal T}_{n}(Q_{UY})\cap\left(U^{n},y^{n}\right)\not\in{\cal T}_{n}(\overline{Q}_{UY})\right]}{\P\left[\left(U^{n},y^{n}\right)\not\in{\cal T}_{n}(\overline{Q}_{UY})\right]}\\
 & \leq\frac{\P\left[\left(U^{n},y^{n}\right)\in{\cal T}_{n}(Q_{UY})\right]}{1-\epsilon}\\
 & \leq\frac{1}{1-\epsilon}\cdot\P\left[\left(U^{n},y^{n}\right)\in{\cal T}_{n}(Q_{UY})\right],\label{eq: derving upper bound on being in type while not in other type}
\end{align}
and, similarly,
\[
\P\left[\left(U^{n},y^{n}\right)\in{\cal T}_{n}(Q_{UXY})\Big|\left(U^{n},X^{n},y^{n}\right)\not\in{\cal T}_{n}(\overline{Q}_{UXY})\right]\geq\P\left[\left(U^{n},X^{n},y^{n}\right)\in{\cal T}_{n}(Q_{UXY})\right].
\]
Equipped with the Markov relation \eqref{eq: Markov chain of enumerators}
and the bound on the conditional probability \eqref{eq: conditional success probability does not change much},
we can derive an upper bound on the correlation that asymptotically
matches the lower bound \eqref{eq: fractional moments correlation lower bound},
and thus prove the lemma. Indeed, 
\begin{align}
 & \E\left[M_{y^{n}}^{1-\lambda}(Q_{UXY})\cdot M_{y^{n}}^{\lambda}(\overline{Q}_{UXY})\right]\nonumber \\
 & \trre[=,a]\E\left\{ \E\left[M_{y^{n}}^{1-\lambda}(Q_{UXY})\cdot M_{y^{n}}^{\lambda}(\overline{Q}_{UXY})\bigg|N_{y^{n}}(Q_{UY}),N_{y^{n}}(\overline{Q}_{UY})\right]\right\} \\
 & \trre[=,b]\E\left\{ \E\left[M_{y^{n}}^{1-\lambda}(Q_{UXY})\bigg|N_{y^{n}}(Q_{UY})\right]\cdot\E\left[M_{y^{n}}^{\lambda}(\overline{Q}_{UXY})\bigg|N_{y^{n}}(\overline{Q}_{UY})\right]\right\} \\
 & \trre[=,c]\E\left[\E\left(\E\left[M_{y^{n}}^{1-\lambda}(Q_{UXY})\bigg|N_{y^{n}}(Q_{UY})\right]\cdot\E\left[M_{y^{n}}^{\lambda}(\overline{Q}_{UXY})\bigg|N_{y^{n}}(\overline{Q}_{UY})\right]\Bigg|N_{y^{n}}(Q_{UY})\right)\right]\\
 & =\E\left[\E\left[M_{y^{n}}^{1-\lambda}(Q_{UXY})\bigg|N_{y^{n}}(Q_{UY})\right]\cdot\E\left(\E\left[M_{y^{n}}^{\lambda}(\overline{Q}_{UXY})\bigg|N_{y^{n}}(\overline{Q}_{UY})\right]\Bigg|N_{y^{n}}(Q_{UY})\right)\right]\\
 & \trre[\leq,d]\E\left[\E\left[M_{y^{n}}^{1-\lambda}(Q_{UXY})\bigg|N_{y^{n}}(Q_{UY})\right]\cdot\E\left(\E\left[M_{y^{n}}^{\lambda}(\overline{Q}_{UXY})\bigg|N_{y^{n}}(\overline{Q}_{UY}),N_{y^{n}}(Q_{UY})=0\right]\right)\right]\\
 & =\E\left[M_{y^{n}}^{1-\lambda}(Q_{UXY})\right]\cdot\E\left(\E\left[M_{y^{n}}^{\lambda}(\overline{Q}_{UXY})\bigg|N_{y^{n}}(\overline{Q}_{UY}),N_{y^{n}}(Q_{UY})=0\right]\Bigg|\right),\label{eq: correlation between enumerators upper bound derivation}
\end{align}
where $(a)$ and $(c)$ follows from the law of total expectation,
and $(b)$ follows from the Markov relation \eqref{eq: Markov chain of enumerators}.
To see $(d)$ note that $\E[M_{y^{n}}^{\lambda}(\overline{Q}_{UXY})|N_{y^{n}}(\overline{Q}_{UY})=\overline{s}]$
is a non-decreasing function of\textbf{ $\overline{s}$} {[}see \eqref{eq: hierarchical moment conditional on cloud center moment}{]}.
In addition, conditioned on $N_{y^{n}}(Q_{UY})=s$, $N_{y^{n}}(\overline{Q}_{UY})$
is a binomial random variable pertaining to less $e^{n\rho_{\s[c]}}-N_{y^{n}}(Q_{UY})\leq e^{n\rho_{\s[c]}}$
trials. Thus, $(d)$ follows from Lemma \ref{lem: avarge of monotone function w.r.t. binomials - number of trials}.
We now note that 
\begin{equation}
\E\left(\E\left[M_{y^{n}}^{\lambda}(\overline{Q}_{UXY})\bigg|N_{y^{n}}(\overline{Q}_{UY}),N_{y^{n}}(Q_{UY})=0\right]\right)\label{eq: enumerator expectation with zero competing cloud centers}
\end{equation}
and 
\[
\E\left(\E\left[M_{y^{n}}^{\lambda}(\overline{Q}_{UXY})\bigg|N_{y^{n}}(\overline{Q}_{UY})\right]\right)=\E\left[M_{y^{n}}^{\lambda}(\overline{Q}_{UXY})\right]
\]
are both moments of binomial random variables with the same number
of trials, but the former has a success probability 
\[
\P\left[\left(U^{n},y^{n}\right)\in{\cal T}_{n}(Q_{UY})\right]\doteq e^{-nI_{\overline{Q}}(U;Y)},
\]
and the latter has a success probability 
\[
\P\left[\left(U^{n},y^{n}\right)\in{\cal T}_{n}(Q_{UY})\big|\left(U^{n},y^{n}\right)\not\in{\cal T}_{n}(\overline{Q}_{UY})\right].
\]
However, \eqref{eq: conditional success probability does not change much}
shows that the latter success probability has the same exponential
order. In turn, the proof of Lemma \ref{lem: enumerators fractional moments}
shows that the exponential order of this expectation only depends
on the exponential order of the success probability. Consequently,
\begin{align}
 & \E\left(\E\left[M_{y^{n}}^{\lambda}(\overline{Q}_{UXY})\bigg|N_{y^{n}}(\overline{Q}_{UY}),N_{y^{n}}(Q_{UY})=0\right]\right)\nonumber \\
 & \doteq\E\left(\E\left[M_{y^{n}}^{\lambda}(\overline{Q}_{UXY})\bigg|N_{y^{n}}(\overline{Q}_{UY})\right]\Bigg|\right)\\
 & =\E\left[M_{y^{n}}^{\lambda}(\overline{Q}_{UXY})\right].
\end{align}
Using this in \eqref{eq: correlation between enumerators upper bound derivation}
completes the proof.
\end{IEEEproof}
Next, we move to the case where the cloud centers may be the same,
i.e., $Q_{UY}=\overline{Q}_{UY}$. In this case, correlation between
$M_{y^{n}}^{1-\lambda}(Q_{UXY})$ and $M_{y^{n}}^{\lambda}(\overline{Q}_{UXY})$
is possible even asymptotically. Apparently, this is due to the fact
that $N_{y^{n}}(Q_{UY})=0$ with high probability whenever $I_{Q}(U;Y)>\rho_{\s[c]}$,
and thus, in this case, $M_{y^{n}}(Q_{UXY})=M_{y^{n}}(\overline{Q}_{UXY})=0$
with high probability. However, as we will next show, $M_{y^{n}}^{1-\lambda}(Q_{UXY})$
and $M_{y^{n}}^{\lambda}(\overline{Q}_{UXY})$ are asymptotically
uncorrelated when \emph{conditioned} \emph{on} $N_{y^{n}}(Q_{UY})$. 

To show this, we first need a result analogous to Lemma \ref{lem: probability of being exactly in type}.
To this end, we first need to exclude possible $Q_{UY}$ from the
discussion. Let us say that $(Q_{UX},Q_{UY})$ is a \emph{joint-distribution-dictator
}(JDD) pair if it determines $Q_{UXY}$ unambiguously. For example,
when $|{\cal U}|=|{\cal X}|=|{\cal Y}|=2$, $Q_{X|U}$ corresponds
to a Z-channel and $Q_{Y|U}$ corresponds to an S-channel,\footnote{That is $Q_{Y|U}(0|1)=Q_{X|U}(1|0)=0$ and all other transition probabilities
are non-zero.} $(Q_{UX},Q_{UY})$ is a JDD pair. Clearly, in this case no $Q_{UXY}\neq\overline{Q}_{UXY}$
exists with the same $(U,X)$ and $(U,Y)$ marginals, and thus such
$Q_{UY}$ are of no interest to the current discussion.

By carefully observing the Z-channel/S-channel example above, it is
easy to verify if for all $u\in{\cal U}$ either $\supp(Q_{X|U=u^{*}})<2$
or $\supp(Q_{Y|U=u^{*}})<2$ then $(Q_{UX},Q_{UY})$ is a JDD pair.
Therefore, if $(Q_{UX},Q_{UY})$ is not a JDD pair then there must
exist $u^{*}\in{\cal U}$ such that both $\supp(Q_{X|U=u^{*}})\geq2$
and $\supp(Q_{Y|U=u^{*}})\geq2$. This property will be used in the
proof of the following lemma. 
\begin{lem}
\label{lem: probability of being exactly in type conditional} Let
$Q_{UXY}\in{\cal P}({\cal U}\times{\cal X}\times{\cal Y})$ be given
such that $\supp(Q_{X})\geq2$, $\supp(Q_{Y})\geq2$, and $(Q_{UX},Q_{UY})$
is not a JDD pair. Also, let $(u^{n},y^{n})\in{\cal T}_{n}(Q_{UY})$
and assume that $X^{n}$ is distributed uniformly over ${\cal T}_{n}(Q_{X|U},u^{n})$.
Then, for any given $\epsilon>0$ there exists $n_{0}(Q_{UX},Q_{UY})$
such that for all $n\geq n_{0}(Q_{UX},Q_{UY})$
\[
\P\left[\left(u^{n},X^{n},y^{n}\right)\in{\cal T}_{n}(Q_{UXY})\right]\leq\epsilon.
\]
\end{lem}
\begin{IEEEproof}
Had $X^{n}$ been distributed uniformly over ${\cal T}_{n}(Q_{X})$,
the claim would follow directly from Lemma \ref{lem: probability of being exactly in type},
where $y^{n}$ and $U^{n}$ there are replaced by $(u^{n},y^{n})$
and $X^{n}$, respectively. However, since $X^{n}$ is distributed
uniformly over ${\cal T}_{n}(Q_{X|U},u^{n})$ the proof is not immediate.
Nonetheless, it follows the same lines, and thus we will only highlight
the required modifications.

Just as in \eqref{eq: exact size of type class}, the size of a conditional
type class can be shown to satisfy 
\[
\left|{\cal T}_{n}(Q_{X|U},u^{n})\right|\cong\prod_{u\in\supp(Q_{U})}\exp\left\{ nQ_{U}(u)\cdot H_{Q}(X|U=u)\right\} \cdot\left[nQ_{U}(u)\right]^{-\frac{1}{2}\left[\supp(Q_{X|U=u})-1\right]}
\]
and thus
\begin{align}
 & \P\left[\left(u^{n},X^{n},y^{n}\right)\in{\cal T}_{n}(Q_{UXY})\right]\nonumber \\
 & =\frac{\left|{\cal T}_{n}(Q_{X|UY},u^{n},y^{n})\right|}{\left|{\cal T}_{n}(Q_{X|U},u^{n})\right|}\\
 & \trre[\cong,a]\frac{\prod_{(u,y)\in\supp(Q_{UY})}\exp\left\{ nQ_{UY}(u,y)\cdot H_{Q}(X|U=u,Y=Y)\right\} \cdot\left[nQ_{UY}(u,y)\right]^{-\frac{1}{2}\left[\supp(Q_{X|U=u,Y=y})-1\right]}}{\prod_{u\in\supp(Q_{U})}\exp\left\{ nQ_{U}(u)\cdot H_{Q}(X|U=u)\right\} \cdot\left[nQ_{U}(u)\right]^{-\frac{1}{2}\left[\supp(Q_{X|U=u})-1\right]}}\\
 & =e^{-n\cdot I_{Q}(X;Y|U)}\cdot n^{\frac{1}{2}\sum_{u\in\supp(Q_{U})}\left[\supp(Q_{X|U=u})-1\right]-\frac{1}{2}\sum_{(u,y)\in\supp(Q_{UY})}\left[\supp(Q_{X|U=u,Y=y})-1\right]}\cdot c(Q_{UY}),
\end{align}
where
\[
c(Q_{UY})\dfn\frac{\prod_{(u,y)\in\supp(Q_{UY})}Q_{UY}(u,y)^{-\frac{1}{2}\left[\supp(Q_{X|U=u,Y=y})-1\right]}}{\prod_{u\in\supp(Q_{U})}Q_{U}(u)^{-\frac{1}{2}\left[\supp(Q_{X|U=u})-1\right]}}.
\]
Now, suppose that $I_{Q}(X;Y|U)=0$. Then, 
\begin{align}
 & \frac{1}{2}\sum_{u\in\supp(Q_{U})}\left[\supp(Q_{X|U=u})-1\right]-\frac{1}{2}\sum_{(u,y)\in\supp(Q_{UY})}\left[\supp(Q_{X|U=u,Y=y})-1\right]\nonumber \\
 & \trre[=,a]\frac{1}{2}\sum_{u\in\supp(Q_{U})}\left[\supp(Q_{X|U=u})-1\right]-\frac{1}{2}\sum_{(u,y)\in\supp(Q_{UY})}\left[\supp(Q_{X|U=u})-1\right]\\
 & =\frac{1}{2}\sum_{u\in\supp(Q_{U})}\left\{ \left[\supp(Q_{X|U=u})-1\right]-\sum_{y\in\supp(Q_{Y|U=u})}\left[\supp(Q_{X|U=u})-1\right]\right\} \\
 & \trre[=,b]\frac{1}{2}\sum_{u\in\supp(Q_{U})}\left[1-\supp(Q_{Y|U=u})\right]\left[\supp(Q_{X|U=u})-1\right]\\
 & \leq-\frac{1}{2},
\end{align}
where $(a)$ follows since $Q_{X|U=u}=Q_{X|U=u,Y=y}$ for all $u\in\supp(Q_{U})$,
and $(b)$ follows since $(Q_{UX},Q_{UY})$ is not a JDD pair, and
thus there must exist $u^{*}\in\supp(Q_{U})$ such that both $\supp(Q_{X|U=u^{*}})\geq2$
and $\supp(Q_{X|U=u^{*}})\geq2$ (as noted before the statement of
the lemma). Thus, when $I_{Q}(X;Y|U)=0$ we get 
\begin{equation}
\P\left[\left(u^{n},X^{n},y^{n}\right)\in{\cal T}_{n}(Q_{UXY})\right]\leq\frac{c(Q_{UY})}{\sqrt{n}}.\label{eq: probabiliy of exact conditional type, zero MI}
\end{equation}
The proof then may continue as the proof of Lemma \ref{lem: probability of being exactly in type}.
One can find $\eta>0$ sufficiently small such that any $\tilde{Q}_{UXY}\in\left\{ \tilde{Q}_{UXY}:\Vert\tilde{Q}_{UXY}-Q_{U}\times Q_{X|U}\times Q_{Y|U}\Vert_{1}\leq\eta\right\} $
has the same support as $Q_{U}\times Q_{X|U}\times Q_{Y|U}$ (at least
conditioned on $u^{*}$). Further, one can find $\delta(\eta,Q_{UX},Q_{UY})>0$
such that 
\[
\left\{ \tilde{Q}_{UXY}:I_{\tilde{Q}}(X;Y|U)\leq\delta\right\} \subseteq\left\{ \tilde{Q}_{UXY}:\Vert\tilde{Q}_{UXY}-Q_{U}\times Q_{X|U}\times Q_{Y|U}\Vert_{1}\leq\eta\right\} ,
\]
and the two cases considered in Lemma \ref{lem: probability of being exactly in type}
can be considered here as well. In the first case, $I_{Q}(X;Y|U)$
may vanish, but $\supp(Q_{UXY})=\supp(Q_{U}\times Q_{X|U}\times Q_{Y|U})$
and thus \eqref{eq: probabiliy of exact conditional type, zero MI}
holds. In the second case $I_{Q}(X;Y|U)\geq\delta$, and thus for
any given $\epsilon>0$ there exists $n_{0}(Q_{UX},Q_{UY})$ such
that
\[
\P\left[\left(u^{n},X^{n},y^{n}\right)\in{\cal T}_{n}(Q_{UXY})\right]\leq e^{-nI_{Q}(X;Y|U)}\cdot n^{|{\cal U}||{\cal X}|}\cdot c(Q_{UY})\leq\epsilon
\]
for all $n\geq n_{0}(Q_{UX},Q_{UY})$. 
\end{IEEEproof}
\begin{lem}
\label{lem: correlation of enumerators- same clouds - conditional}
Let $(Q_{UXY},\overline{Q}_{UXY})$ be given such that $Q_{UY}=\overline{Q}_{UY}$.
Then, 
\[
\E\left[M_{y^{n}}^{1-\lambda}(Q_{UXY})M_{y^{n}}^{\lambda}(\overline{Q}_{UXY})\bigg|N_{y^{n}}(Q_{UY})\right]\doteq\E\left[M_{y^{n}}^{1-\lambda}(Q_{UXY})\bigg|N_{y^{n}}(Q_{UY})\right]\cdot\E\left[M_{y^{n}}^{\lambda}(\overline{Q}_{UXY})\bigg|N_{y^{n}}(Q_{UY})\right].
\]
\end{lem}
\begin{IEEEproof}
The proof follows the same lines of the proof of Lemma \ref{lem: correlation of enumerators- different clouds},
and so we only provide a brief outline. For a lower bound on the conditional
correlation, one can decompose 
\[
M_{y^{n}}(Q_{UXY})=M_{y^{n}}(Q_{UXY},1)+M_{y^{n}}(Q_{UXY},2)
\]
where $M_{y^{n}}(Q_{UXY},1)$ {[}respectively, $M_{y^{n}}(Q_{UXY},2)${]}
corresponds to codewords of odd (even) satellite indices (say). 

For a asymptotically matching upper bound on the correlation, we note
that similarly to \eqref{eq: conditional success probability does not change much},
when $X^{n}$ is drawn uniformly over ${\cal T}_{n}(Q_{X|U},u^{n})$,
\[
\P\left[\left(u^{n},X^{n},y^{n}\right)\in{\cal T}_{n}(Q_{UXY})\Big|\left(u^{n},X^{n},y^{n}\right)\not\in{\cal T}_{n}(\overline{Q}_{UXY})\right]
\]
is close to the unconditional probability, in the sense that for any
given $\epsilon>0$,
\begin{multline}
\P\left[\left(u^{n},X^{n},y^{n}\right)\in{\cal T}_{n}(Q_{UXY})\right]\leq\P\left[\left(U^{n},y^{n}\right)\in{\cal T}_{n}(Q_{UY})\Big|\left(U^{n},y^{n}\right)\not\in{\cal T}_{n}(\overline{Q}_{UY})\right]\\
\leq\frac{1}{1-\epsilon}\cdot\P\left[\left(u^{n},X^{n},y^{n}\right)\in{\cal T}_{n}(Q_{UXY})\right].\label{eq: conditional success probability does not change much hierarchical}
\end{multline}
To prove this, a derivation similar to \eqref{eq: derving upper bound on being in type while not in other type}
can be used, while noting that $(Q_{UX},Q_{UY})$ is not a JDD pair,
and so according to Lemma \ref{lem: probability of being exactly in type conditional}
\[
\P\left[\left(u^{n},X^{n},y^{n}\right)\in{\cal T}_{n}(\overline{Q}_{UXY})\right]\leq\epsilon
\]
for all $n$ sufficiently large. Equipped with these results, we get
\begin{align}
 & \E\left[M_{y^{n}}^{1-\lambda}(Q_{UXY})M_{y^{n}}^{\lambda}(\overline{Q}_{UXY})\bigg|N_{y^{n}}(Q_{UY})\right]\nonumber \\
 & \trre[=,a]\E\left\{ M_{y^{n}}^{\lambda}(\overline{Q}_{UXY})\cdot\E\left[M_{y^{n}}^{1-\lambda}(Q_{UXY})\bigg|N_{y^{n}}(Q_{UY}),M_{y^{n}}(\overline{Q}_{UXY})\right]\bigg|N_{y^{n}}(Q_{UY})\right\} \\
 & \trre[\leq,b]\E\left\{ M_{y^{n}}^{\lambda}(\overline{Q}_{UXY})\cdot\E\left[M_{y^{n}}^{1-\lambda}(Q_{UXY})\bigg|N_{y^{n}}(Q_{UY}),M_{y^{n}}(\overline{Q}_{UXY})=0\right]\bigg|N_{y^{n}}(Q_{UY})\right\} \\
 & \trre[\doteq,c]\E\left\{ M_{y^{n}}^{\lambda}(\overline{Q}_{UXY})\bigg|N_{y^{n}}(Q_{UY})\right\} \cdot\E\left[M_{y^{n}}^{1-\lambda}(Q_{UXY})\bigg|M_{y^{n}}(\overline{Q}_{UXY})\right],
\end{align}
where $(a)$ follows from the law of total expectation. For $(b)$
note that conditioned on both $N_{y^{n}}(Q_{UY}),M_{y^{n}}(\overline{Q}_{UXY})$,
$M_{y^{n}}(Q_{UXY})$ is a binomial random variable pertaining to
$N_{y^{n}}(Q_{UY})e^{n\rho_{\s[s]}}-M_{y^{n}}(\overline{Q}_{UXY})\leq N_{y^{n}}(Q_{UY})e^{n\rho_{\s[s]}}$
trials. Thus, $\E[M_{y^{n}}^{1-\lambda}(Q_{UXY})|N_{y^{n}}(Q_{UY}),M_{y^{n}}(\overline{Q}_{UXY})=\overline{s}]$
is a non-increasing function of $\overline{s}$ {[}see \eqref{eq: hierarchical moment conditional on cloud center moment}{]},
and $(b)$ follows from Lemma \ref{lem: avarge of monotone function w.r.t. binomials - number of trials}.
For $(c)$, we note that from \eqref{eq: conditional success probability does not change much hierarchical},
the conditioning on $M_{y^{n}}(\overline{Q}_{UXY})=0$ does not change
the exponential order of the success probability of $M_{y^{n}}(\overline{Q}_{UXY})$.
As evident from \eqref{eq: hierarchical moment conditional on cloud center moment},
this conditioning can be removed without changing the exponential
order of the expression.
\end{IEEEproof}
Proceeding with the case of $Q_{UY}=\overline{Q}_{UY}$, we next evaluate
the expectation over $N_{y^{n}}(Q_{UY})$. We show that the asymptotic
uncorrelation result of Lemma \ref{lem: correlation of enumerators- different clouds}
holds, albeit with a correction term required when $I_{Q}(U;Y)>\rho_{\s[c]}$. 
\begin{lem}
\label{lem: correlation of enumerators- same clouds}Let $\delta>0$,
and $(Q_{UXY},\overline{Q}_{UXY})$ be given such that $Q_{UY}=\overline{Q}_{UY}$.
Then, 
\[
\E\left[M_{y^{n}}^{1-\lambda}(Q_{UXY})M_{y^{n}}^{\lambda}(\overline{Q}_{UXY})\right]\doteq\E\left[M_{y^{n}}^{1-\lambda}(Q_{UXY})\right]\cdot\E\left[M_{y^{n}}^{\lambda}(\overline{Q}_{UXY})\right]\cdot e^{n\left|I_{Q}(U;Y)-\rho_{\s[c]}\right|_{+}}.
\]
\end{lem}
\begin{IEEEproof}
We consider two cases separately. First suppose that $I_{Q}(U;Y)\leq\rho_{\s[c]}$.
In this case, $N_{y^{n}}(Q_{UY})$ concentrates double-exponentially
fast around its expected value, where the latter equals $\exp[n(\rho_{\s[c]}-I_{Q}(U;Y))]$
up to the first order in the exponent. Thus, the conditional expectation
and the unconditional expectation are equal up to the first order
in the exponent. More rigorously, let $\delta>0$ be given and recall
the definition of the event ${\cal B}_{n}(Q_{UY},\delta)$ in \eqref{eq: enumerator concentrates at average}.
Then, 
\begin{align}
\E\left[M_{y^{n}}^{1-\lambda}(Q_{UXY})\right] & =\P\left[N_{y^{n}}(Q_{UY})\in{\cal B}_{n}(Q_{UY},\delta)\right]\cdot\E\left[M_{y^{n}}^{1-\lambda}(Q_{UXY})\bigg|N_{y^{n}}(Q_{UY})\in{\cal B}_{n}(Q_{UY},\delta)\right]\nonumber \\
 & \hphantom{=}+\P\left[N_{y^{n}}(Q_{UY})\in{\cal B}_{n}^{c}(Q_{UY},\delta)\right]\cdot\E\left[M_{y^{n}}^{1-\lambda}(Q_{UXY})\bigg|N_{y^{n}}(Q_{UY})\in{\cal B}_{n}^{c}(Q_{UY},\delta)\right]\\
 & \trre[\doteq,a]\E\left[M_{y^{n}}^{1-\lambda}(Q_{UXY})\bigg|N_{y^{n}}(Q_{UY})\in{\cal B}_{n}(Q_{UY},\delta)\right]\\
 & \trre[\geq,b]e^{-n\delta}\cdot\E\left\{ M_{y^{n}}^{1-\lambda}(Q_{UXY})\bigg|N_{y^{n}}(Q_{UY})=\E\left[N_{y^{n}}(Q_{UY})\right]\right\} ,\label{eq: moment w/o conditioning}
\end{align}
where $(a)$ follows from the fact $\P[{\cal B}_{n}(Q_{UY},\delta)]$
decays double-exponentially {[}see \eqref{eq: enumerator concntration, high rate}{]},
and $(b)$ follows from \eqref{eq: hierarchical moment conditional on cloud center moment}.
Similarly 
\begin{equation}
\E\left[M_{y^{n}}^{\lambda}(\overline{Q}_{UXY})\right]\dotgeq e^{-n\delta}\cdot\E\left\{ M_{y^{n}}^{\lambda}(\overline{Q}_{UXY})\bigg|N_{y^{n}}(Q_{UY})=\E\left[N_{y^{n}}(Q_{UY})\right]\right\} .\label{eq: moment bar w.o conditioning}
\end{equation}
Thus, 
\begin{align}
 & \E\left[M_{y^{n}}^{1-\lambda}(Q_{UXY})M_{y^{n}}^{\lambda}(\overline{Q}_{UXY})\right]\nonumber \\
 & =\P\left[N_{y^{n}}(Q_{UY})\in{\cal B}_{n}(Q_{UY},\delta)\right]\cdot\E\left[M_{y^{n}}^{1-\lambda}(Q_{UXY})M_{y^{n}}^{\lambda}(\overline{Q}_{UXY})\bigg|N_{y^{n}}(Q_{UY})\in{\cal B}_{n}(Q_{UY},\delta)\right]\nonumber \\
 & \hphantom{=}+\P\left[N_{y^{n}}(Q_{UY})\in{\cal B}_{n}^{c}(Q_{UY},\delta)\right]\cdot\E\left[M_{y^{n}}^{1-\lambda}(Q_{UXY})M_{y^{n}}^{\lambda}(\overline{Q}_{UXY})\bigg|N_{y^{n}}(Q_{UY})\in{\cal B}_{n}^{c}(Q_{UY},\delta)\right]\\
 & \trre[\doteq,a]\E\left[M_{y^{n}}^{1-\lambda}(Q_{UXY})M_{y^{n}}^{\lambda}(\overline{Q}_{UXY})\bigg|N_{y^{n}}(Q_{UY})\in{\cal B}_{n}(Q_{UY},\delta)\right]\\
 & =\sum_{s\in{\cal B}_{n}(Q_{UY},\delta)}\P\left[N_{y^{n}}(Q_{UY})=s\right]\cdot\E\left[M_{y^{n}}^{1-\lambda}(Q_{UXY})M_{y^{n}}^{\lambda}(\overline{Q}_{UXY})\bigg|N_{y^{n}}(Q_{UY})=s\right]\\
 & \trre[\doteq,b]\sum_{s\in{\cal B}_{n}(Q_{UY},\delta)}\P\left[N_{y^{n}}(Q_{UY})=s\right]\cdot\E\left\{ M_{y^{n}}^{1-\lambda}(Q_{UXY})\bigg|N_{y^{n}}(Q_{UY})=s\right\} \cdot\E\left\{ M_{y^{n}}^{\lambda}(\overline{Q}_{UXY})\bigg|N_{y^{n}}(Q_{UY})=s\right\} \\
 & \trre[\leq,c]e^{2n\delta}\cdot\P\left[N_{y^{n}}(Q_{UY})\in{\cal B}_{n}(Q_{UY},\delta)\right]\nonumber \\
 & \hphantom{=}\times\E\left\{ M_{y^{n}}^{1-\lambda}(Q_{UXY})\bigg|N_{y^{n}}(Q_{UY})=\E\left[N_{y^{n}}(Q_{UY})\right]\right\} \cdot\E\left\{ M_{y^{n}}^{\lambda}(\overline{Q}_{UXY})\bigg|N_{y^{n}}(Q_{UY})=\E\left[N_{y^{n}}(Q_{UY})\right]\right\} \\
 & \trre[\doteq,d]e^{2n\delta}\cdot\E\left\{ M_{y^{n}}^{1-\lambda}(Q_{UXY})\bigg|N_{y^{n}}(Q_{UY})=\E\left[N_{y^{n}}(Q_{UY})\right]\right\} \cdot\E\left\{ M_{y^{n}}^{\lambda}(\overline{Q}_{UXY})\bigg|N_{y^{n}}(Q_{UY})=\E\left[N_{y^{n}}(Q_{UY})\right]\right\} \\
 & \trre[\dotleq,e]e^{4n\delta}\cdot\E\left\{ M_{y^{n}}^{1-\lambda}(Q_{UXY})\right\} \cdot\E\left\{ M_{y^{n}}^{\lambda}(\overline{Q}_{UXY})\right\} ,\label{eq: uncorrelation derivation for the same clouds and high cloud rate}
\end{align}
where $(a)$ and $(d)$ follow from \eqref{eq: enumerator concntration, high rate},
$(b)$ follows from Lemma \ref{lem: correlation of enumerators- same clouds - conditional},
$(c)$ from \eqref{eq: hierarchical moment conditional on cloud center moment},
and $(e)$ from \eqref{eq: moment w/o conditioning} and \eqref{eq: moment bar w.o conditioning}.

We next address the case $I_{Q}(U;Y)>\rho_{\s[c]}$. In this case,
$N_{y^{n}}(Q_{UY})=0$ with high probability, $1\leq N_{y^{n}}(Q_{UY})\leq e^{2n\delta}$
with probability $\exp\{-n[I_{Q}(U;Y)-\rho_{\s[c]}]\}$, and $N_{y^{n}}(Q_{UY})\geq e^{2n\delta}$
with probability double-exponentially small {[}see \eqref{eq: enumerator large deviation, high rate}
and \eqref{eq: enumerator concntration, low rate}{]}. For brevity,
we will use the definitions of ${\cal A}_{n}^{=0}(Q_{UY})$, ${\cal A}_{n}^{=1}(Q_{UY})$
and ${\cal A}_{n}^{\geq1}(Q_{UY},\delta)$ in \eqref{eq: enumeartor is zero}-\eqref{eq: enumeartor is exponential}.
Then,
\begin{align}
\E\left[M_{y^{n}}^{1-\lambda}(Q_{UXY})\right] & =\P\left[{\cal A}_{n}^{=0}(Q_{UY})\right]\cdot\E\left\{ M_{y^{n}}^{1-\lambda}(Q_{UXY})\bigg|{\cal A}_{n}^{=0}(Q_{UY})\right\} \nonumber \\
 & \hphantom{=}+\P\left[{\cal A}_{n}^{=1}(Q_{UY},\delta)\right]\cdot\E\left\{ M_{y^{n}}^{1-\lambda}(Q_{UXY})\bigg|{\cal A}_{n}^{=1}(Q_{UY},\delta)\right\} \nonumber \\
 & \hphantom{=}+\P\left[{\cal A}_{n}^{\geq1}(Q_{UY},\delta)\right]\cdot\E\left\{ M_{y^{n}}^{1-\lambda}(Q_{UXY})\bigg|{\cal A}_{n}^{\geq1}(Q_{UY},\delta)\right\} \\
 & \trre[\doteq,a]0+\P\left[{\cal A}_{n}^{=1}(Q_{UY},\delta)\right]\cdot\E\left\{ M_{y^{n}}^{1-\lambda}(Q_{UXY})\bigg|{\cal A}_{n}^{=1}(Q_{UY},\delta)\right\} \\
 & \trre[\leq,b]e^{n\delta}\cdot\P\left[{\cal A}_{n}^{=1}(Q_{UY},\delta)\right]\cdot\E\left\{ M_{y^{n}}^{1-\lambda}(Q_{UXY})\bigg|N_{y^{n}}(Q_{UY})=1\right\} ,
\end{align}
where for $(a)$ and $(b)$ we apply \eqref{eq: enumerator concntration, low rate}
and \eqref{eq: hierarchical moment conditional on cloud center moment},
respectively. Similarly, 
\[
\E\left[M_{y^{n}}^{\lambda}(\overline{Q}_{UXY})\right]\dotleq e^{n\delta}\cdot\P\left[{\cal A}_{n}^{=1}(Q_{UY},\delta)\right]\cdot\E\left\{ M_{y^{n}}^{\lambda}(\overline{Q}_{UXY})\bigg|N_{y^{n}}(Q_{UY})=1\right\} .
\]
Then,
\begin{align}
 & \E\left[M_{y^{n}}^{1-\lambda}(Q_{UXY})M_{y^{n}}^{\lambda}(\overline{Q}_{UXY})\right]\nonumber \\
 & \trre[\dotleq,a]e^{2n\delta}\cdot\P\left[{\cal A}_{n}^{=1}(Q_{UY},\delta)\right]\cdot\E\left[M_{y^{n}}^{1-\lambda}(Q_{UXY})\bigg|N_{y^{n}}(Q_{UY})=1\right]\cdot\E\left[M_{y^{n}}^{\lambda}(\overline{Q}_{UXY})\bigg|N_{y^{n}}(Q_{UY})=1\right]\\
 & \dotleq e^{4n\delta}\cdot\frac{1}{\P\left[{\cal A}_{n}^{=1}(Q_{UY},\delta)\right]}\cdot\E\left[M_{y^{n}}^{1-\lambda}(Q_{UXY})\right]\cdot\E\left[M_{y^{n}}^{\lambda}(\overline{Q}_{UXY})\right]\\
 & \trre[\doteq,b]e^{4n\delta}\cdot e^{n\left[I_{Q}(U;Y)-\rho_{\s[c]}\right]}\cdot\E\left[M_{y^{n}}^{1-\lambda}(Q_{UXY})\right]\cdot\E\left[M_{y^{n}}^{\lambda}(\overline{Q}_{UXY})\right],
\end{align}
where $(a)$ follows from a derivation similar to \eqref{eq: uncorrelation derivation for the same clouds and high cloud rate},
and $(b)$ follows from \eqref{eq: enumerator large deviation, high rate}. 

Analogous asymptotic lower bounds on $\E[M_{y^{n}}^{1-\lambda}(Q_{UXY})M_{y^{n}}^{\lambda}(\overline{Q}_{UXY})]$
for both cases can be obtained in the same manner, when $\delta$
is replaced by $(-\delta)$. The proof is then completed by taking
$\delta\downarrow0$. 
\end{IEEEproof}
Using all the above, we are now ready to prove Proposition \ref{prop: Correlation of hierarchical enumeartors}.
\begin{IEEEproof}[Proof of Prop. \ref{prop: Correlation of hierarchical enumeartors}]
For the first case of \eqref{eq: correlation of hierarchical enumerator}
\[
\E\left[M_{y^{n}}^{1-\lambda}(Q_{UXY})M_{y^{n}}^{\lambda}(\overline{Q}_{UXY})\right]=\E\left[M_{y^{n}}(Q_{UXY})\right],
\]
and the result follows from Lemma \ref{lem: enumerators fractional moments}
with $\lambda=1$. For the second case, using Lemma \ref{lem: correlation of enumerators- different clouds}
\[
\E\left[M_{y^{n}}^{1-\lambda}(Q_{UXY})M_{y^{n}}^{\lambda}(\overline{Q}_{UXY})\right]\doteq\E\left[M_{y^{n}}^{1-\lambda}(Q_{UXY})\right]\cdot\E\left[M_{y^{n}}^{\lambda}(\overline{Q}_{UXY})\right],
\]
and the result follows from Lemma \ref{lem: enumerators fractional moments}.
Similarly, the third case follows from Lemmas \ref{lem: correlation of enumerators- same clouds}
and \ref{lem: enumerators fractional moments}. Specifically, the
result is just as in the second case, except for the correction term
$|I_{Q}(U;Y)-\rho_{\s[c]}|_{+}$ to the exponent. Standard manipulations
lead to the expression shown in the third case. 
\end{IEEEproof}

\end{document}